\documentclass[12pt]{article}
\pdfoutput=1
\usepackage{amsmath}
\usepackage{amssymb}
\usepackage{graphicx}
\usepackage{esint}

\usepackage{ifpdf}
\usepackage{epsfig}
\usepackage{amsfonts}
\usepackage{amsmath}
\usepackage{amssymb}
\usepackage{color} 
\usepackage{graphicx}
\usepackage{grffile}
\usepackage{subfig}
\usepackage{notoccite}
\usepackage{hyperref}
\usepackage{geometry}

\newcommand{\be}{\begin{equation}}
\newcommand{\ee}{\end{equation}}
\newcommand{\ba}{\begin{eqnarray}}
\newcommand{\ea}{\end{eqnarray}}
\newcommand{\bi}{\begin{itemize}}
\newcommand{\ei}{\end{itemize}}

\newcommand{\sgn}{{\rm sgn}}

\newcommand{\aslash}[1]{\,\,{\raise.15ex\hbox{/}\mkern-12mu #1}}
\newcommand{\bslash}[1]{\,\,{\raise.15ex\hbox{/}\mkern-9mu #1}}
\newcommand{\of}[1]{\left( #1 \right)}
\newcommand{\sqof}[1]{\left[ #1 \right]}

\renewcommand{\bar}{\overline}
\renewcommand{\tilde}{\widetilde}
\renewcommand{\hat}{\widehat}

\begin{document}

\begin{titlepage}

\begin{center}
\vspace{1cm}
{\Large \bf  Polarised Black Holes in ABJM}
\vspace{0.8cm}

{\large  Miguel S. Costa$^{\dagger}$, Lauren Greenspan$^\dagger$,  
Jo\~ao Penedones$^{\ddagger}$, Jorge E. Santos$ ^{\Diamond}$}

\vspace{.5cm}

{\it  $ ^\dagger$Centro de F\'\i sica do Porto,
Departamento de F\'\i sica e Astronomia\\
Faculdade de Ci\^encias da Universidade do Porto\\
Rua do Campo Alegre 687,
4169--007 Porto, Portugal}
\\
\vspace{.3cm}
{\it  $ ^\ddagger$
Institute of Physics, Fields and Strings Laboratory\\ 
Ecole Polytechnique F\'ed\' erale de Lausanne (EPFL)\\
Rte de la Sorge, BSP 728, CH-1015 Lausanne, Switzerland}
\\
\vspace{.3cm}
{\it  $ ^\Diamond$Department of Applied Mathematics and Theoretical Physics\\
University of Cambridge, Wilberforce Road\\
Cambridge CB3 0WA, UK}

\end{center}
\vspace{1cm}

\begin{abstract}

We numerically construct asymptotically $AdS_4$ solutions to Einstein-Maxwell-dilaton theory. These have a dipolar electrostatic potential turned on at the
conformal boundary $S^2\times \mathbb{R}_t$. We find two classes of geometries: $AdS$ soliton solutions that encode the full backreaction of the electric field on the $AdS$ geometry without a horizon, and neutral black holes that are ``polarised" by the dipolar potential. For a certain range of the electric field $\mathcal{E}$, we find two distinct branches of the $AdS$ soliton that exist for the same value of $\mathcal{E}$. For the black hole, we find either two or four branches depending on the value of the electric field and horizon temperature. These branches meet at critical values of the electric field and impose a maximum value of $\mathcal{E}$ that should be reflected in the dual field theory. For both the soliton and black hole geometries, we study boundary data such as the stress tensor. For the black hole, we also consider horizon observables such as the entropy.  At finite temperature, we consider the Gibbs free energy for both phases and determine the phase transition between them. We find that the $AdS$ soliton dominates at low temperature for an electric field up to the maximum value. Using the gauge/gravity duality, we propose that these solutions are dual to deformed ABJM theory and compute the corresponding weak coupling phase diagram. 
\end{abstract}

\bigskip
\bigskip

\end{titlepage}

\section{Introduction}

In a recent work \cite{PBH} we showed that neutral black holes of spherical topology placed in four-dimensional Anti-de Sitter space-time are polarised when subject to an external electric field. At finite temperature, this gravitational system is described by a two-dimensional phase diagram, in terms of its temperature $T$ and electric field parameter ${\cal E}$. The system has two phases, a lower temperature phase described by an AdS soliton with a self-gravitating electric field, and a higher temperature phase described by the polarised black hole. The critical temperature decreases with the external electric field. For pure Einstein-Maxwell theory we observed that the external electric field could be made arbitrarily large, with the critical temperature converging  to zero in the limit of large electric field.

One of the motivations for the above study was to consider three-dimensional conformal theories on $\mathbb{R}_t\times S^2 $ subject to an external electric field source that couples to a global current operator. The two distinct phases are then expected to describe confining and deconfining phases, with a critical temperature that depends on the external electric field. Intuitively we expect the critical temperature to decrease with the external field because of the electric repulsion between the partonic degrees of freedom. The analysis of a free conformally coupled scalar field in the presence of such external electric field supports this intuition and shows the existence of a maximal electric field, above which the vacuum is itself unstable. However, such a maximum electric field is not seen in the gravitational analysis done with pure Einstein-Maxwell theory. This fact is not necessarily contradictory because we do not know if pure Einstein-Maxwell theory is dual to a  CFT. The goal of this paper is to clarify this point by considering a consistent truncation of the gravitational dual of ABJM theory \cite{ABJM}. This truncation includes a current operator that we can turn on in order to  deform the ABJM CFT, therefore studying this problem in a precise holographic setup. 

We consider a consistent truncation of eleven-dimensional supergravity on $AdS_4\times S^7/Z_k$. More concretely, we can break the $SU(4)\times U(1)$ R-symmetry to $U(1)^4$. The bosonic sector of this theory is given by the metric, three scalar fields and, as expected, four $U(1)$ gauge fields \cite{Cvetic:1999xp}. A further consistent truncation can be made to reduce the theory to gravity, a gauge field and a single scalar field, with the following action\footnote{In the notation of \cite{Cvetic:1999xp} we set $\Phi_1=\Phi$, $\Phi_2=\Phi_3=0$, $A=A_1=-A_2$ and $A_3=A_4=0$.}
\be
S_{bulk}=\frac{1}{16\pi G_N}\int d^4x\sqrt{g}\left(R-\frac{1}{2}\nabla_\mu\Phi\nabla^\mu\Phi+\frac{2}{l^2}(\cosh\Phi+2)-\frac{1}{2}e^\Phi F^2\right) \label{action}\,,
\ee
where $F=dA$ and $G_N$ is the Newton constant. For vanishing scalar and gauge field the theory has an  $AdS_4$ vacuum with radius $l$.
This is a simple generalisation  of pure Einstein-Maxwell gravity but, as we shall see, the response to an external electric field contains important differences.

This paper is organised as follows. In section two we construct the $AdS$ soliton in the presence of an external static electric field for the theory with action (\ref{action}). 
Our analysis is numerical, so we can choose any functional form of the source $C_a$ for the global current operator $J_a$ on the boundary theory
(with lower case latin indices running over the boundary coordinates). This simply translates into the choice of the non-normalizable mode of the bulk $U(1)$ gauge field $A$. The most natural thing to do is to decompose the electrostatic potential $C_t=C_t(\theta,\phi)$ in $S^2$ scalar harmonics. As in \cite{PBH}, we shall consider for simplicity the 
 $AdS$ soliton for the particular case of a dipolar potential
\be
C_t(\theta)= {\mathcal {\cal E}}\cos\theta \,.
\label{eq:DipolarSource}
\ee
For this theory we find that there is indeed a maximum allowed electric field.  
In section three we find the polarised neutral black hole for this theory subject to the same external electric field. Section four begins with the gravitational thermodynamics that leads to the construction of the phase diagram for ABJM theory at strong coupling. Then we consider the free ABJM theory subject to the external electrostatic potential (\ref{eq:DipolarSource}). We see that at zero coupling the theory exhibits a behaviour qualitatively similar to the gravity description. We conclude in section five.
In the appendix we present the perturbative analytical result of a small electric field in $AdS$ which matches our numerical result to a very good approximation. 

\section{AdS Soliton}\label{sec:soliton}
The  equations of motion that follow from the action (\ref{action}) are
\begin{align}	
&R_{\mu \nu}+\frac{1}{l^2}(\cosh\Phi+2)g_{\mu \nu}-\frac{1}{2}\nabla_\mu\Phi \nabla_\nu	\Phi+e^{\Phi}\left(\frac{1}{4}F^2 g_{\mu \nu}-F_{\mu \alpha}F_\nu^{\ \alpha}\right)=0\,,
\nonumber
\\
&d\left(e^\Phi \star F\right)=0\,,\ \ \ \ \ \ \ \ \ 
\nabla^2 \Phi+\frac{2}{l^2}\sinh\Phi-\frac{1}{2}e^\Phi F^2=0\,.
\label{EOM}
\end{align}
Notice that once we turn on an external electric field, setting $\Phi=0$ is not consistent. For the simple form of the source (\ref{eq:DipolarSource}) at the $AdS$ boundary 
we can consider the axially-symmetric ansatz (setting the $AdS$ radius $l=1$)
\begin{align}	
ds^2=\ &\frac{1}{\left(1-r^2\right)^2}\left(A(r,\theta )f(r)\,d\tau ^2 +\frac{ \left(1+r^2\right)^2G(r,\theta )}{f(r)}\,dr^2\right)
\nonumber
\\
&+ r^2\left(C(r,\theta ) \left(d\theta+\frac{1}{r}  H(r,\theta )\,dr\right)^2+B(r,\theta )\sin^2\theta\,d\phi ^2\right),
\label{AdSansatz}
\\
A_\tau=\ & irD(r,\theta)\,, 
\qquad\qquad
\Phi(r,\theta)=(1-r^2) \, \varphi(r,\theta)\,,
\nonumber
\end{align}
where $ f(r)=1-r^2+r^4$. The radial coordinate 
$r$  runs from the $AdS$ origin at $r=0$ to $r=1$ at the $AdS$ boundary. We will work in 
the Euclidean setting with the time coordinate $\tau=it$ periodically identified.
Global $AdS$ corresponds to $A=G=B=C=1$ and $H=D=\varphi=0$. The radial coordinate $r$ can be related to the canonical radial coordinate $y$ of $AdS$  by $y=r/(1-r^{2})$. The dipolar source we will consider imposes a reflection symmetry $\theta \to \pi-\theta$ on the ansatz. Thus, we will use the reflection properties of the functions $A$, $G$, $C$, $B$, $H$, $\varphi$, and $D$ to discretise the equations of motion on a domain bounded by $r=0, 1$ and $\theta=0,\pi/2$. 

The above ansatz must by smooth at the fixed points of the coordinate system. These include the origin at $r=0$, the equator at $\theta=\pi/2$, and the pole at $\theta=0$. Regularity at the origin is achieved by imposing that the first derivatives along $r$ vanish at $r=0$. At the pole, smoothness implies that the first derivatives along $\theta$ vanish at $\theta=0$ as well as requiring that $B(r,0)=C(r,0)$ and $H(r,0)=0$. 
Finally, at the equator we require that the first derivatives along $\theta$ vanish, except for the functions $D(r,\theta)$ and $H(r,\theta)$ which vanish at that point. 

The $AdS$ boundary at $r=1$ is the only real boundary of this coordinate system. Here we require the metric to approach that of global $AdS$ by setting
\be
A(1,\theta)=B(1,\theta)=C(1,\theta)=G(1,\theta)=1\,, \ \ \  H(1,\theta)=0\,.
\ee
The boundary condition for the gauge field can be used to turn on the dipolar potential (\ref{eq:DipolarSource}) at $r=1$. In other words, we turn on a non-normalizable mode by setting
\be
D(1,\theta)= C_t(\theta) = {\mathcal {\cal E}}\cos\theta\,.
\label{eq:BounCondD}
\ee
Finally we need to impose the boundary condition to the scalar field $\Phi$. From (\ref{action}) it is simple to see that this field has $m^2=-2$, corresponding for the ABJM theory to a dual operator of dimension $\Delta=1$. 
We must choose a boundary condition to ensure that the non-normalizable mode corresponding to turning on a source for this operator at the boundary is zero, leaving only its VEV to be determined by the equations of motion. This corresponds to setting the second radial derivative, in Fefferman-Graham coordinates,  of the scalar field to zero. In our ansatz, this becomes  
\be
\varphi(1,\theta)+\partial_r\varphi(r,\theta)\big|_{r=1}=0\, .
\label{eq:BounCondVarphi}
\ee
We will solve the harmonic Einstein equations of motion for this ansatz using the deTurck trick for gauge fixing the Einstein equations. The details of this method were developed in \cite{Headrick:2009pv,Wiseman:2011by} and reviewed in \cite{Dias:2015nua}. 

\subsection{Results}

\begin{figure}[t!]
\centering
\subfloat[ ]{
\includegraphics[width=50mm]{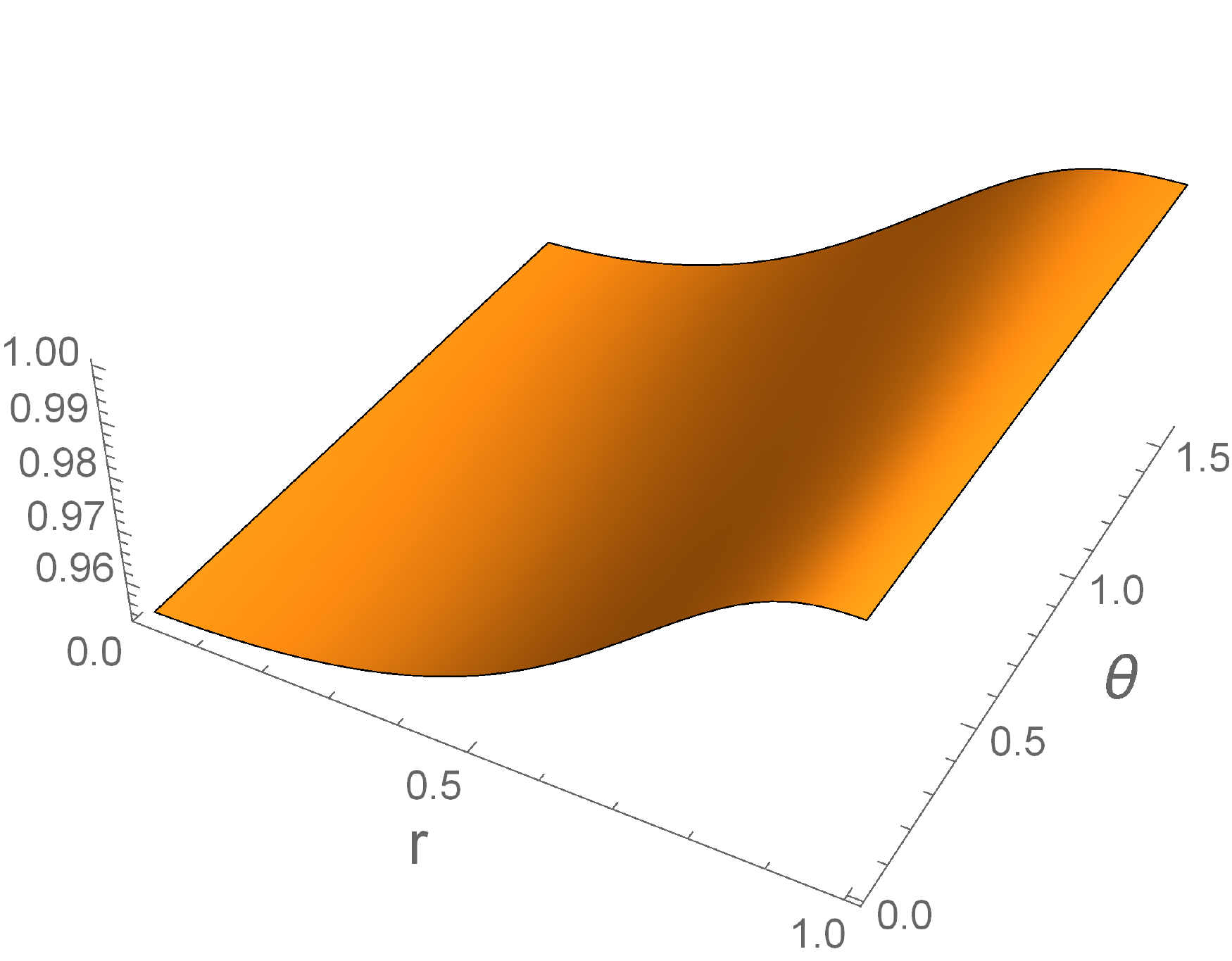}
}
\hspace{2cm}
\subfloat[ ]{
\includegraphics[width=50mm]{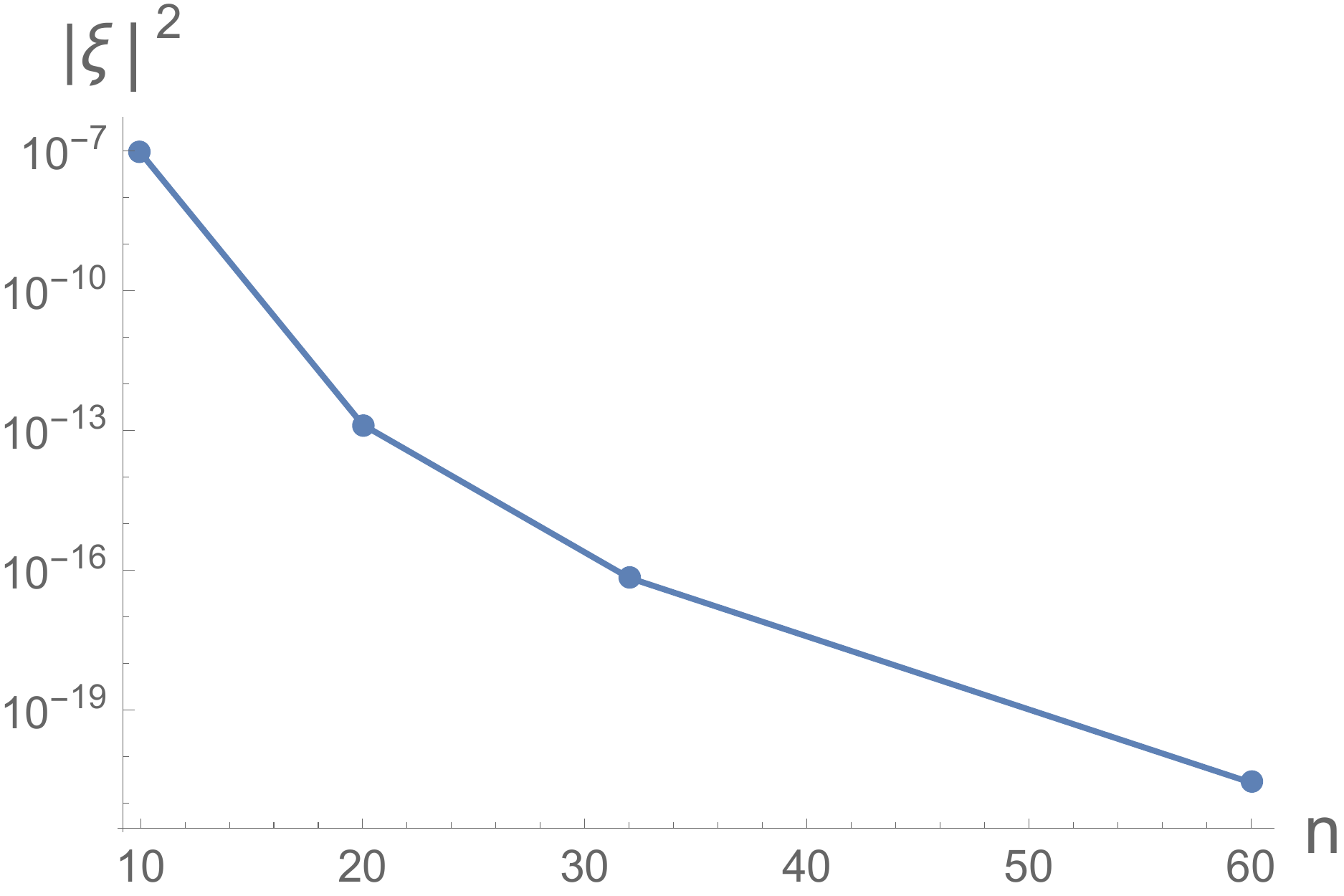}
}
\caption{(a) Metric function $A$ for ${\cal E}=1$ and a numerical grid of $60\times 60$ points.
(b) A linear-log plot of the deTurck vector norm, a measure of convergence, as a function of the number of gridpoints $n$.
\label{metnoBH}}
\end{figure}
\begin{figure}[t!]
\centering
\subfloat[]{
\includegraphics[width=65mm]{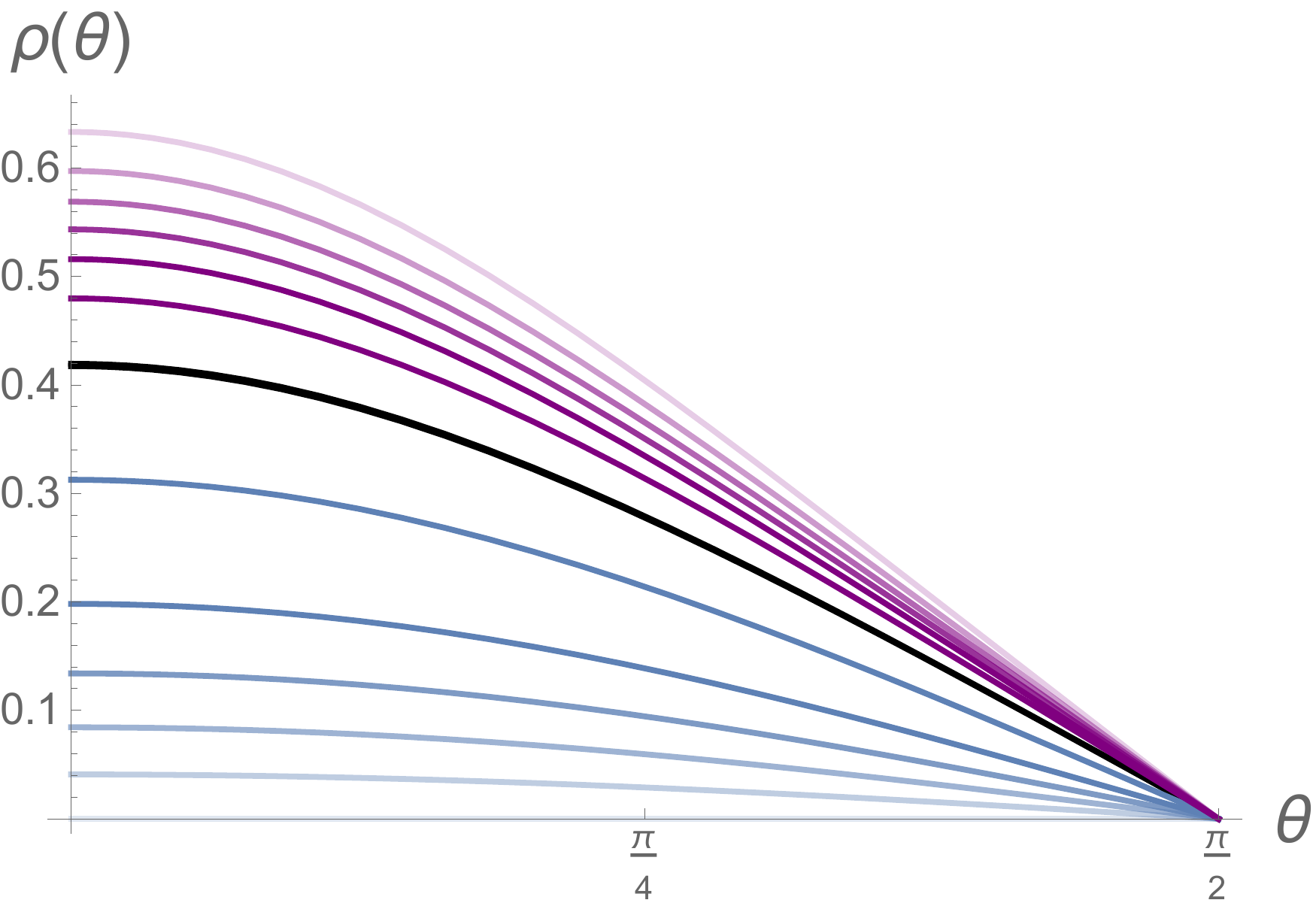}
\label{chargedensAdS}
}
\subfloat[]{
\includegraphics[width=65mm]{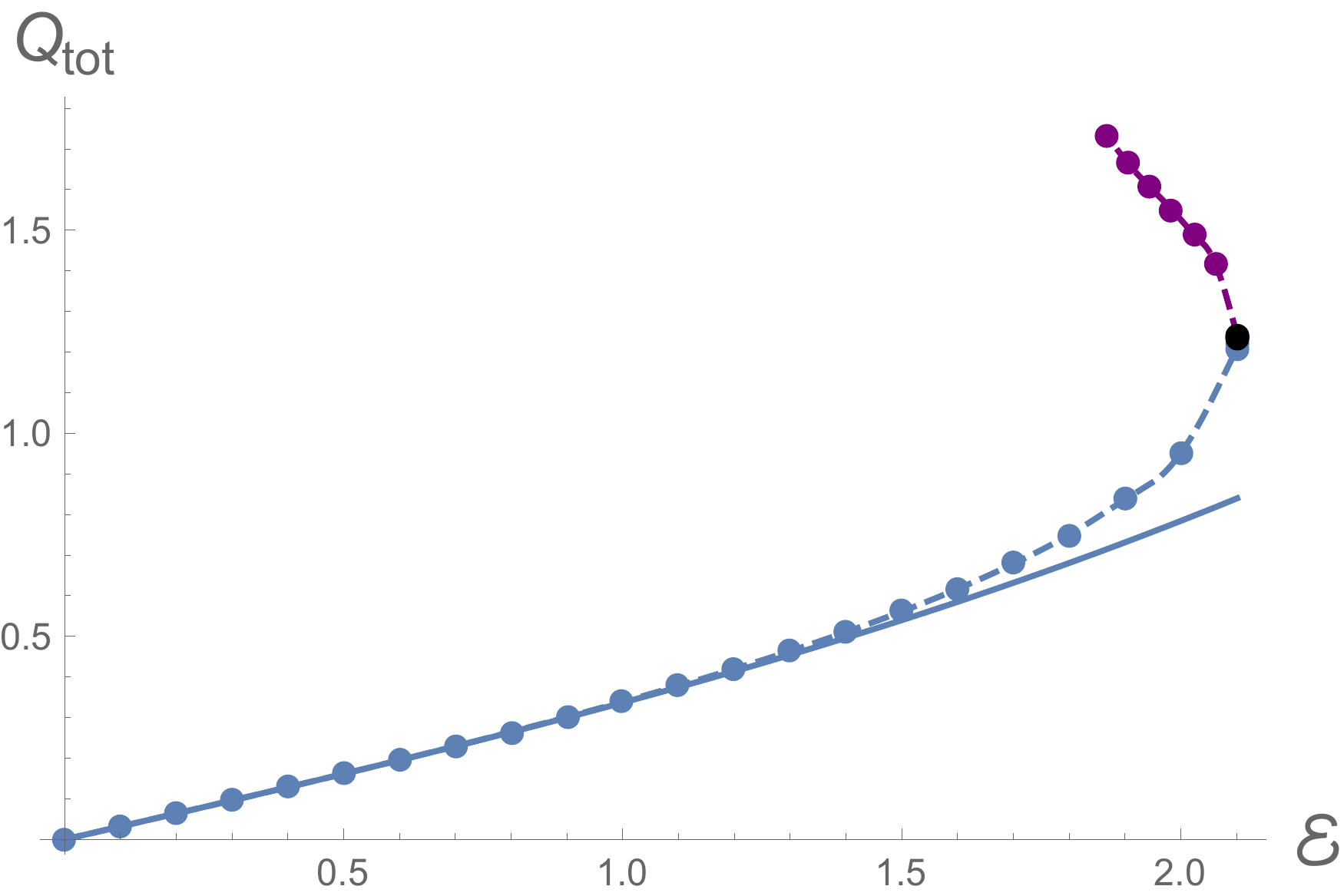}
\label{QtotAdS}
}
\caption{(a) The charge density and
(b) the total charge on the $AdS$ boundary for several values of 
${\cal E} \in \left[0, \,{\cal E}^{Sol}_c\right] $ with $G_N$ set to one. Blue and purple correspond to the two branches of the soliton, with the maximum value of the electric field shown in black. The solid curve in (b) represents the analytical perturbative result in ${\cal E}$, as derived in appendix A.}
\end{figure}

In figure \ref{metnoBH}(a) we show the function $A$ of the ansatz (\ref{AdSansatz}) for a value of the electric field ${\cal E}=1$.  A good measure of convergence when solving the Harmonic Einstein equations is the norm of the deTurck vector $\xi$. This is plotted in figure \ref{metnoBH}(b) as a function of $n$ for ${\cal E}=1$. The results have new features when compared to 
the scalar-free case described in \cite{PBH}, with the notable difference that we now find a maximum allowed value of the electric field at ${\cal E}^{Sol}_c=2.101$,
beyond which this soliton solution does not exist. This maximum value is where two branches of the solution meet. 
This can be seen by calculating several boundary observables, including the charge density $\rho(\theta)$, which can be written as 
\be
\frac{1}{4\pi G_N} \left.\left(e^{\Phi }\star F\right)\right|_{r=1}=\rho(\theta)\, d\Omega_{2}\,,
\label{eq:carge_dens}
\ee
where $ d\Omega_2$ is the volume form on the unit $S^2$. This charge density is plotted in figure \ref{chargedensAdS} for several values of the electric field magnitude. The blue curves correspond to the first branch of soliton solutions up to the maximum value ${\cal E}^{Sol}_c$, while the purple curves correspond to decreasing the electric field from that maximum. The black curve corresponds to the maximum value of the electric field, and the opacity the curves are is proportional to the value of the electric field. We will use this key in all plots of the $AdS$ soliton that follow. In all cases, the charge density is maximal at the pole and vanishes at the equator, as expected from the choice of boundary condition. The total charge in one hemisphere at infinity is also plotted in figure \ref{QtotAdS}. Here we see a qualitative difference between the two branches related to the electric field. The total charge increases with the electric field up to the maximum, then decreases along the second branch of solutions.

We can also compute the stress tensor of the boundary field theory \cite{gr-qc/9209012,Skenderis:2002wp}. For this we need to consider the total action
$S=S_{bulk}+S_{bdy}+S_{CT}$, where $S_{bulk}$ is given by (\ref{action}) and $S_{bdy}$ is the Gibbons-Hawking-York term 
\be
S_{bdy}=\frac{1}{8\pi G_N}\int d^3x\sqrt{h}K\,,
\ee
where $K$ is the trace of the extrinsic curvature and $h$ the induced boundary metric.
In $AdS$, the cosmological constant term in the action leads to a divergence proportional to the volume of spacetime. Such a divergence can be canceled by adding a counter term of the form \cite{hep-th/9902121,deHaro:2000xn} 
\be
S_{CT}=-\frac{1}{8\pi G_N}\int_{\partial M} d^{3}x\sqrt{h}\left(1-\frac{l^{2}}{12}\,R+\frac{1}{4}\, \Phi^2\right).
\label{counter}
\ee
From this renormalized on-shell action, we can derive the stress tensor in the usual way. In units such that $l=1$, this is
\be
T_{\mu\nu}=\frac{2}{\sqrt{h}}\frac{\delta S}{\delta h^{\mu\nu}}=
\frac{1}{8\pi G_N}  \left(K_{\mu\nu}-K\, h_{\mu\nu}+G_{\mu\nu}-2h_{\mu\nu}-\frac{1}{4}\, \Phi^2 h_{\mu\nu}\, \right) .\label{stress}
\ee
The first two terms come from the Gibbons-Hawking-York boundary term in the on-shell action, while the last three terms come from the counter-term action (\ref{counter}). In the above expression, $K_{\mu\nu}$ is the extrinsic curvature and $G_{\mu\nu}$ is the Einstein tensor on the boundary. To evaluate the stress tensor at the boundary, we will use the
asymptotic expansion of the metric functions
up to $O(1-r)^{5}$ including logs, for example, 
\ba
A(r,\theta)=\underset{i=0}{\sum}(1-r)^{i}\alpha_{i}(\theta) +\log(1-r) \,\underset{i=4}{\sum} (1-r)^{i}a_{i}(\theta)\,.
\ea
Most of the expansion coefficients will be fixed by the equations of motion. Those that remain are normalizable modes that depend on the behavior of the numerical solution into the bulk. They are computed by taking derivatives of the appropriate numerical solutions and evaluating them at the boundary. 
\begin{figure}[t!]
\centering
\subfloat[]{
\includegraphics[height=40mm]{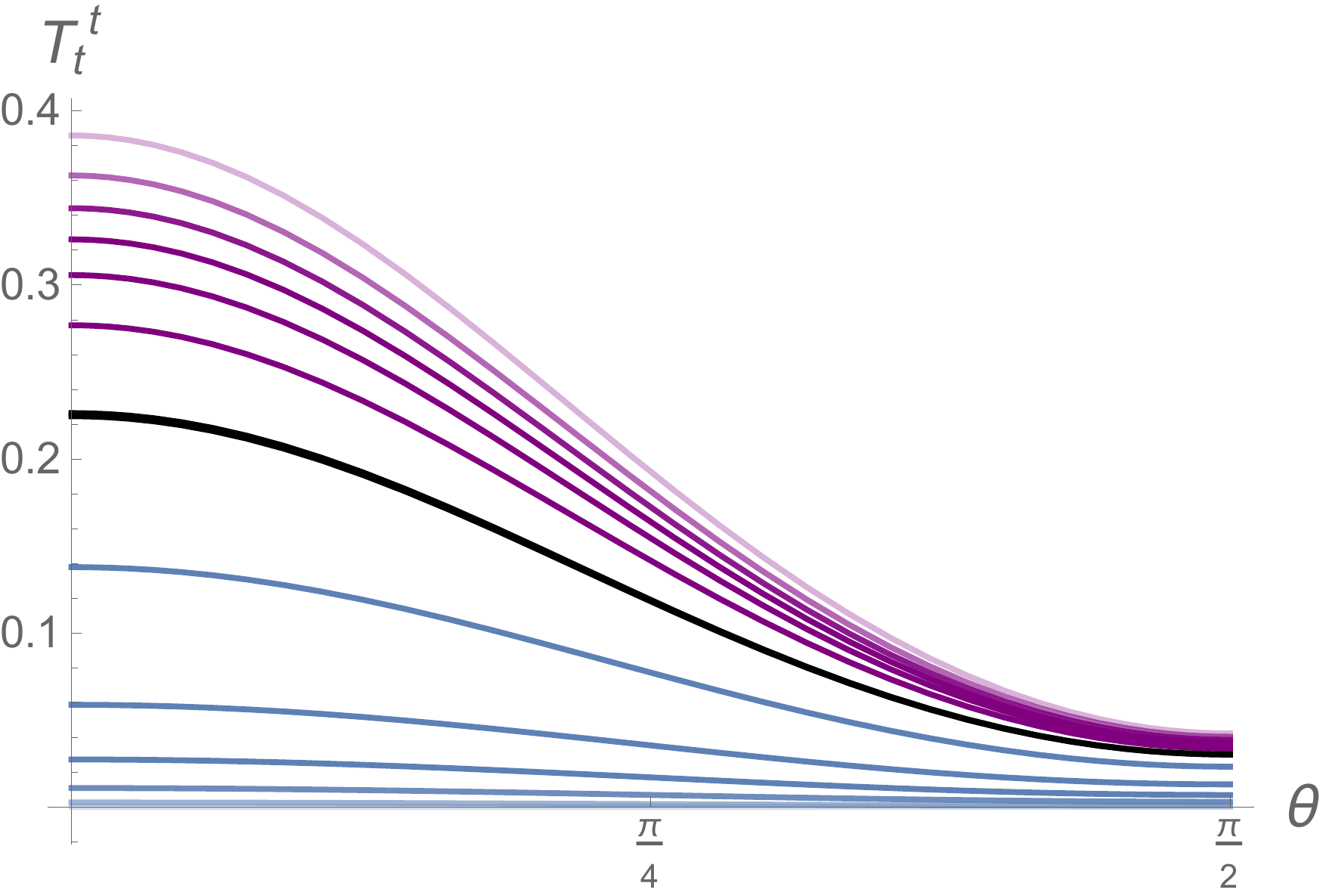}
\label{endensAdS}
}
\subfloat[]{
\includegraphics[height=40mm]{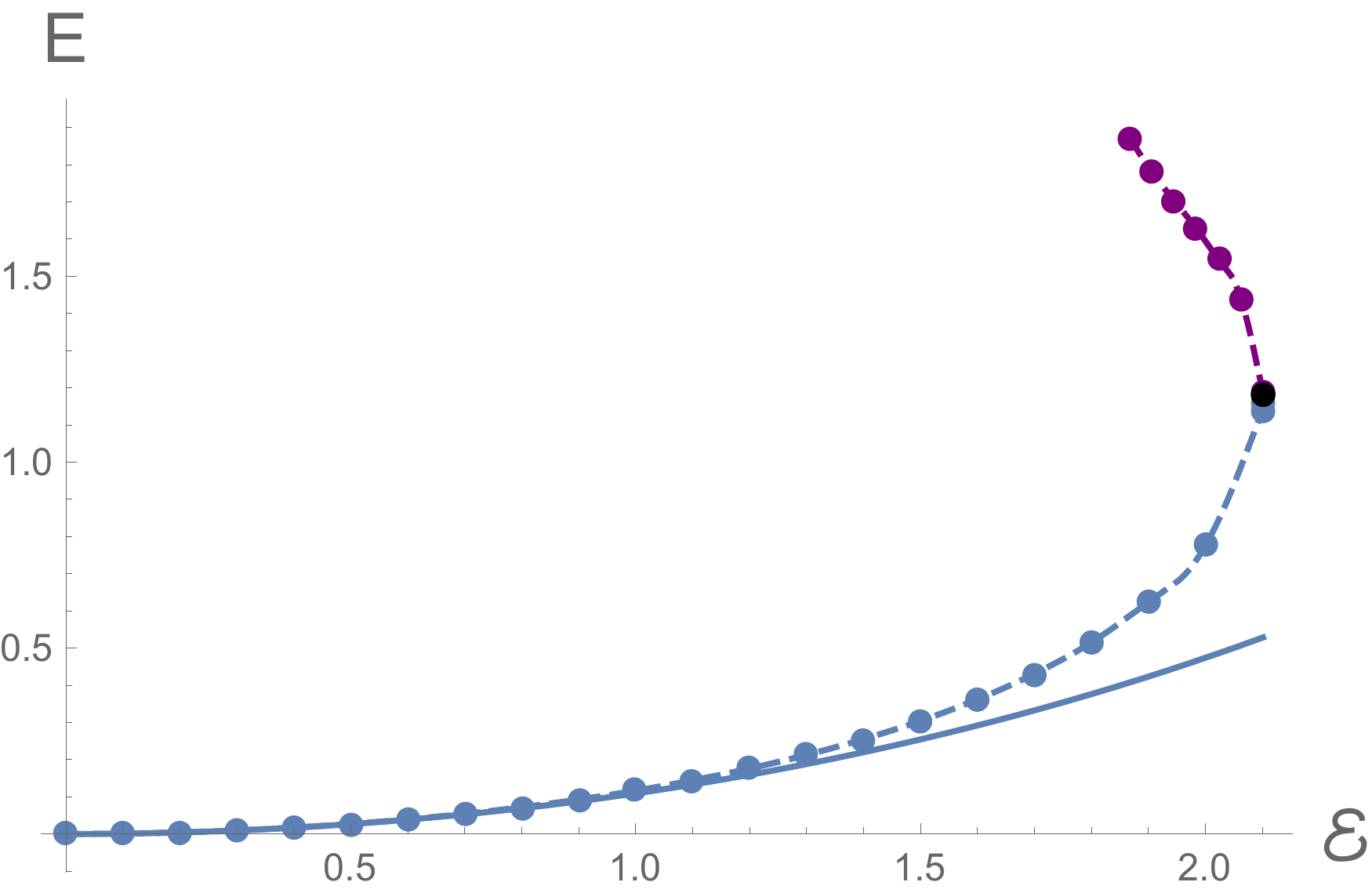}
\label{energyAdS}}
\caption{(a) Energy density on the $AdS$ boundary and (b) total energy 
of the boundary theory as a function of the electric field ${\cal E}$ (setting $G_N=1$).
The solid curve in (b) is the analytical result found from perturbation theory around ${\cal E} =0$.}
\end{figure}
\begin{figure}[t!]
\centering
\subfloat[]{
\includegraphics[height=40mm]{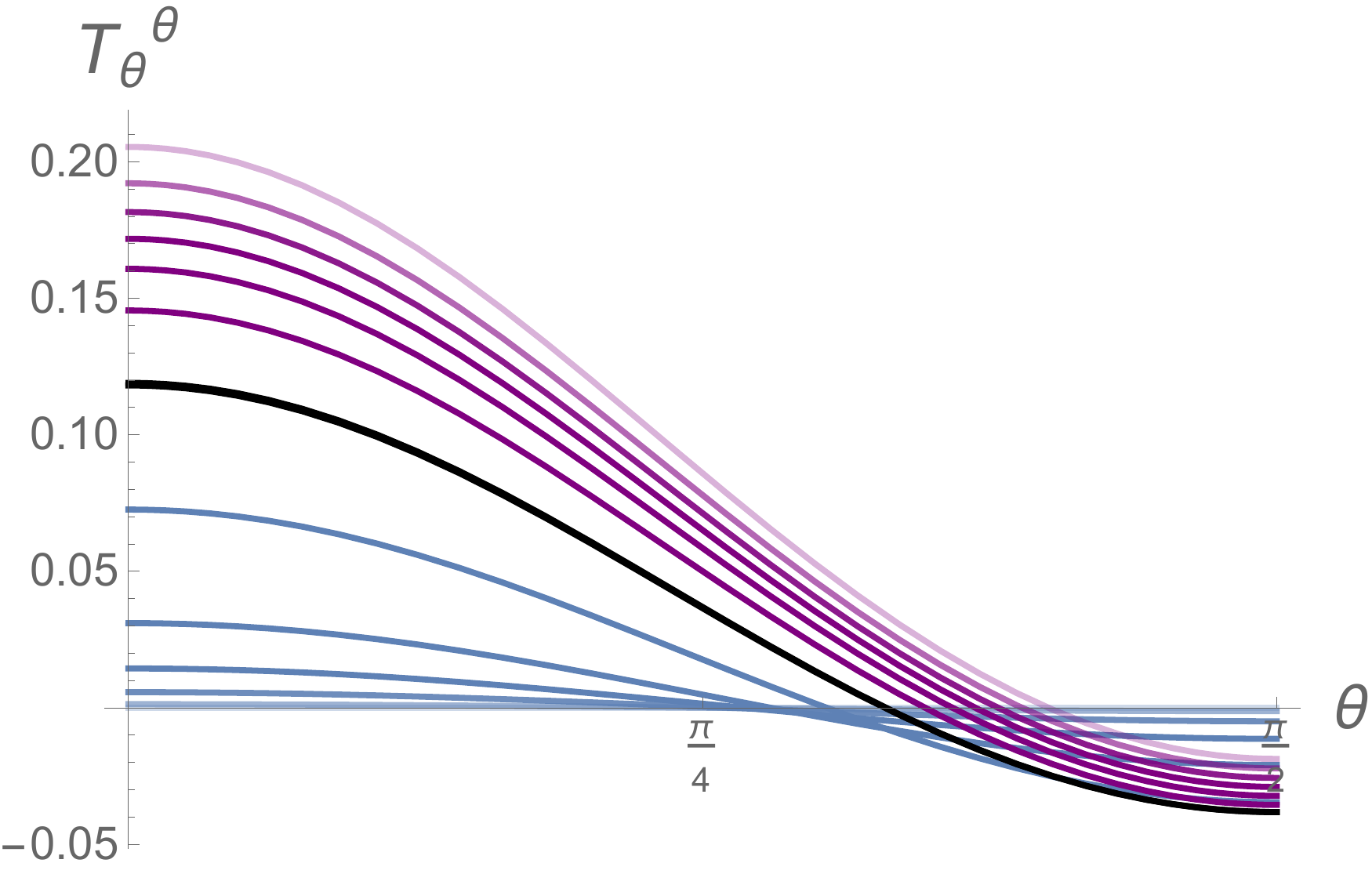}
\label{TththAdS}
}
\subfloat[]{
\includegraphics[height=40mm]{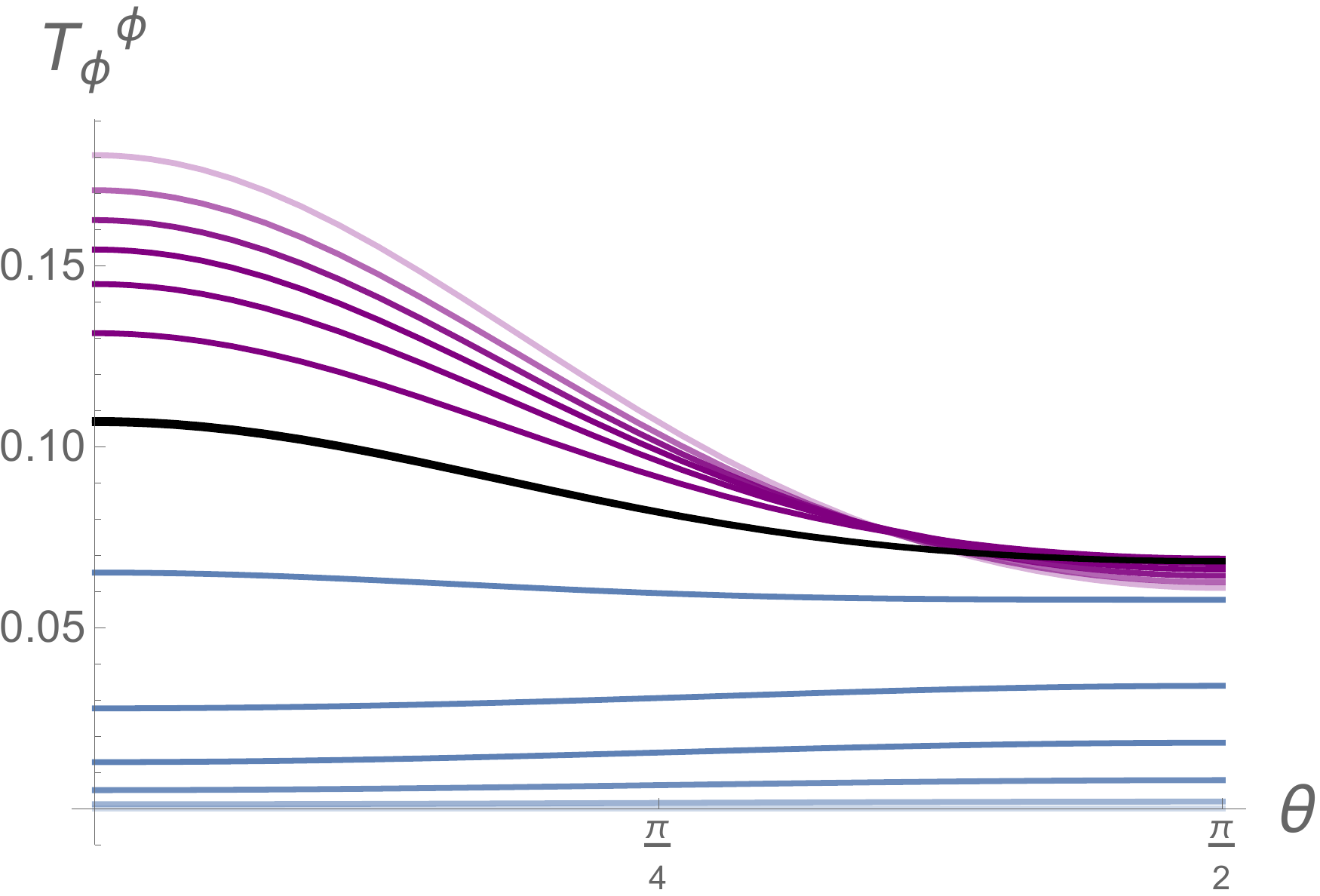}
\label{TffAdS}
}
\caption{Spatial components of the boundary stress tensor for ${\cal E}\in [0,{\cal E}^{Sol}_c]$.
We set $G_N=1$.}
\end{figure}
The energy density is
\be
T_{t}^{\,t}=\frac{3}{256\pi G_N}\left(2\alpha_3+\phi_0^2\right) ,
\ee
where $\phi_0$ is $\varphi$ evaluated at the boundary, that is, $\phi_0(\theta)=\varphi(1,\theta)$. This is plotted in figure \ref{endensAdS}. Like the charge density, $T_{t}^{\,t}$ is maximal at the pole and  minimal at the equator. For the first branch of solutions it increases for increasing ${\cal E}$, while the other branch has the opposite behavior. 
In figure \ref{energyAdS} we plot, as a function of ${\cal E}$, the boundary theory total energy
\be
E=\int d\Omega_2 T_t^{\,t} \,.
\ee
The slope of the energy curve becomes singular at the maximum value of the electric field, where the two branches meet.
The  spatial components of the stress tensor are
\ba
T_{\theta}^{\,\theta}&=&\frac{3}{256\pi G_N}\left(2\chi_3+\phi_0^2\right) ,
\\
 T_{\phi}^{\,\phi}&=&\frac{-3}{128\pi G_N}\left(\chi_3+\alpha_3+\phi_0^2 \right),
\notag
\ea
where $\chi_3(\theta)$ is the third-order radial power-law mode for the metric function $C(r,\theta)$. If we think of the stress tensor as describing a fluid of the boundary theory, the $\theta \theta$ component, plotted in figure \ref{TththAdS}, shows that the pressure along $\theta$ is positive up to a critical point dependent on  ${\cal E}$ and negative thereafter. Since there is no net flow of momenta in the $\phi$ direction, the $\phi \phi$ component, which measures the pressure along that direction, is independent of $\phi$. This is plotted in figure \ref{TffAdS}, and decreases from the poles to the equator. These are the only non-zero components of the stress tensor, which is traceless, as can be seen from the expressions written above. 

As a consequence of the Ward identities, the equation governing the conservation of energy and momentum in a background electric field is
\be
\nabla_a T^{ab} +\frac{1}{2} J_a F^{ab}=0\,,
\ee
where $J^a=(\rho,J^i)$ and $J^i$ is the current density on the sphere at the boundary. The only nontrivial component in this ansatz corresponds to $b=\theta$ and leads to the expression
\be
\partial_\theta \left( \sin \theta \,T^{\,\theta}_\theta \right) -\cos \theta \, T^{\,\phi}_\phi=-\frac{{\cal E}}{2}\rho(\theta) \sin^2 \theta\,.
\label{conservationT}
\ee
Our numerical solutions satisfy this equation to within a precision of $10^{-2}$ relative to $T^{\,\phi}_\phi$.

\begin{figure}[t!]
\centering
\subfloat[\label{fluxAdSABJMa}]{\includegraphics[width=70mm]{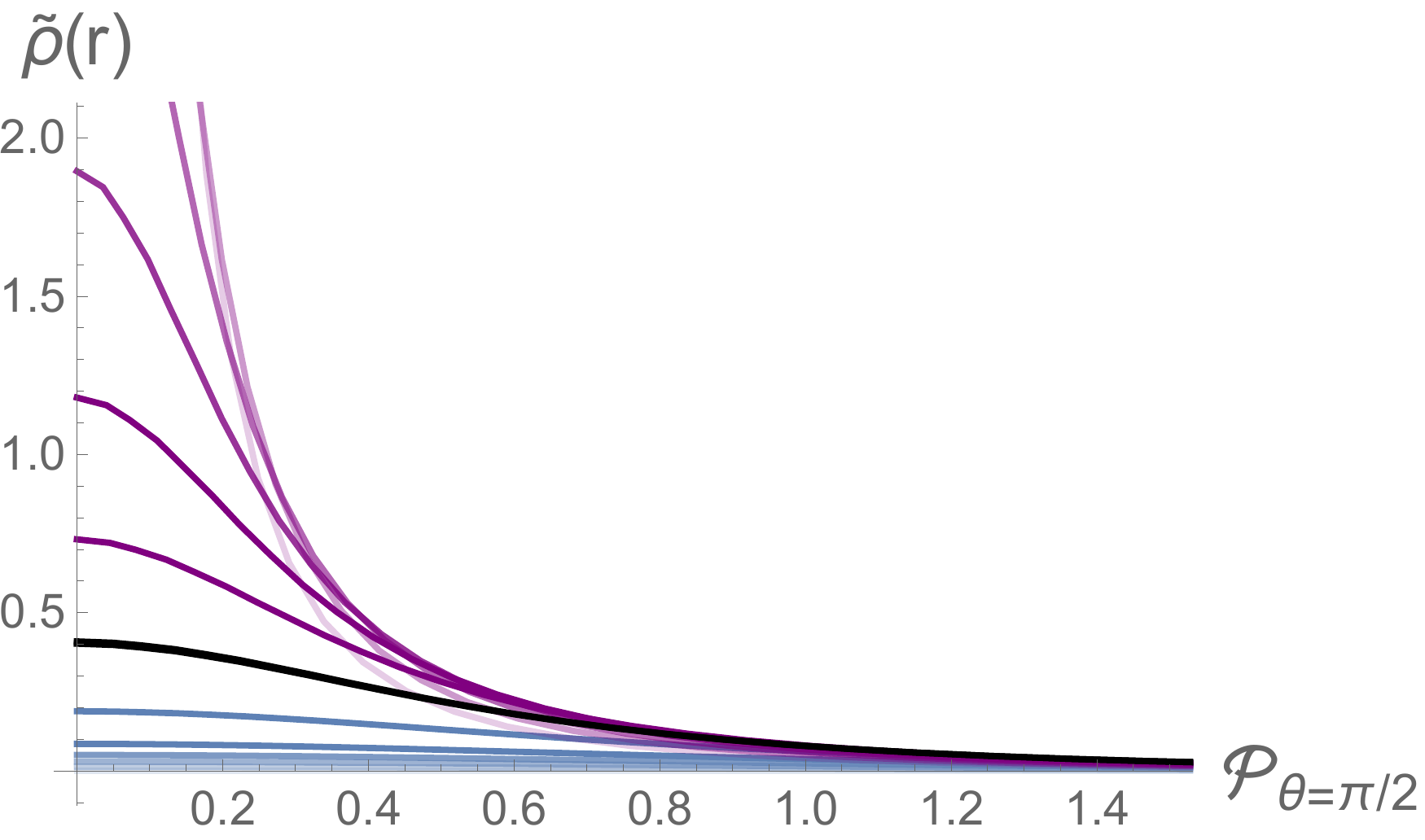}}
\subfloat[\label{fluxAdSABJMb}]{
\includegraphics[width=65mm]{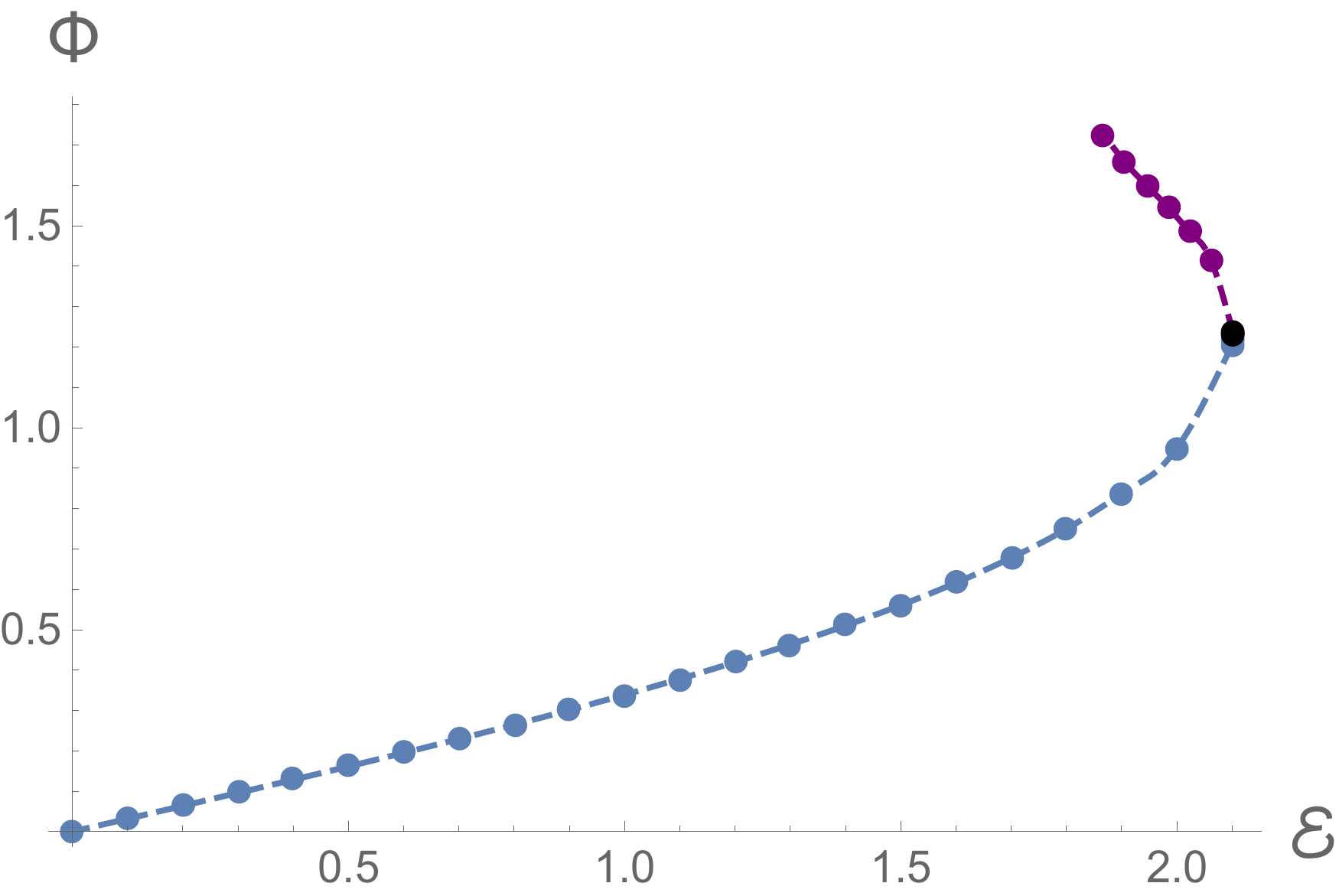}
}
\caption{(a) The flux density through the equatorial plane as a function of the proper distance for several values of the electric field. (b) Total flux through the equator.}
\end{figure}

There are also several bulk observables that will allow us to develop intuition of the $AdS$ geometry. The flux density $\tilde{\rho}$ through the $\theta=\pi/2$ plane is defined by
\be
\frac{1}{4\pi G_N}\star \left. \left( e^{\Phi(r,\theta)}\!F\right) \right |_{\theta=\pi/2}=\tilde{\rho}(r)\sqrt{g_{rr}g_{\phi\phi}}\, dr\wedge  d\phi\,.\label{fluxeq}
\ee
This is greatest at the origin and goes to zero at $r=1$. The flux density in terms of the proper radial distance from the $AdS$ center along the equatorial plane is plotted in figure \ref{fluxAdSABJMa}. As before, darker curves correspond to higher values of the electric field. For $\theta=0,\pi$  this proper distance is given by
\be
{\cal P}_\theta(r) = \int_0^r  \sqrt{g_{rr} (r',\theta)} \,dr'\,.
\ee
The total flux through the equator is plotted in figure \ref{fluxAdSABJMb}. By conservation, the total flux through the equatorial plane should be equal to the total charge at one hemisphere. We have checked that this is true with an error of $10^{-3}$.
The flux increases up to the maximum electric field and then keeps growing in the other branch of the solution, as the electric field decreases to another critical value. 

\begin{figure}[b!]
\centering
\subfloat[]{
\includegraphics[width=65mm]{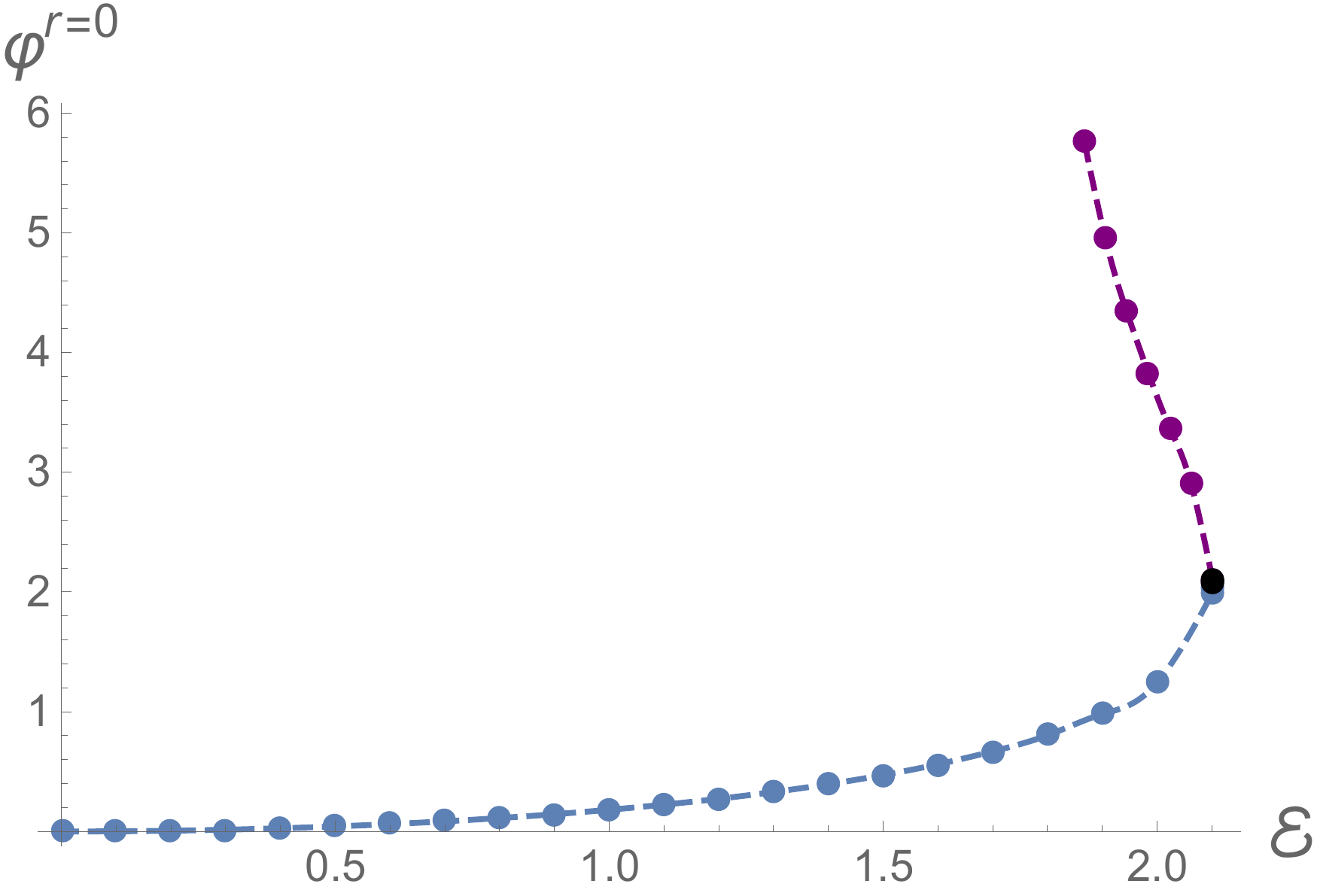}\label{phiAdSABJM}
}
\subfloat[]{
\includegraphics[width=60mm]{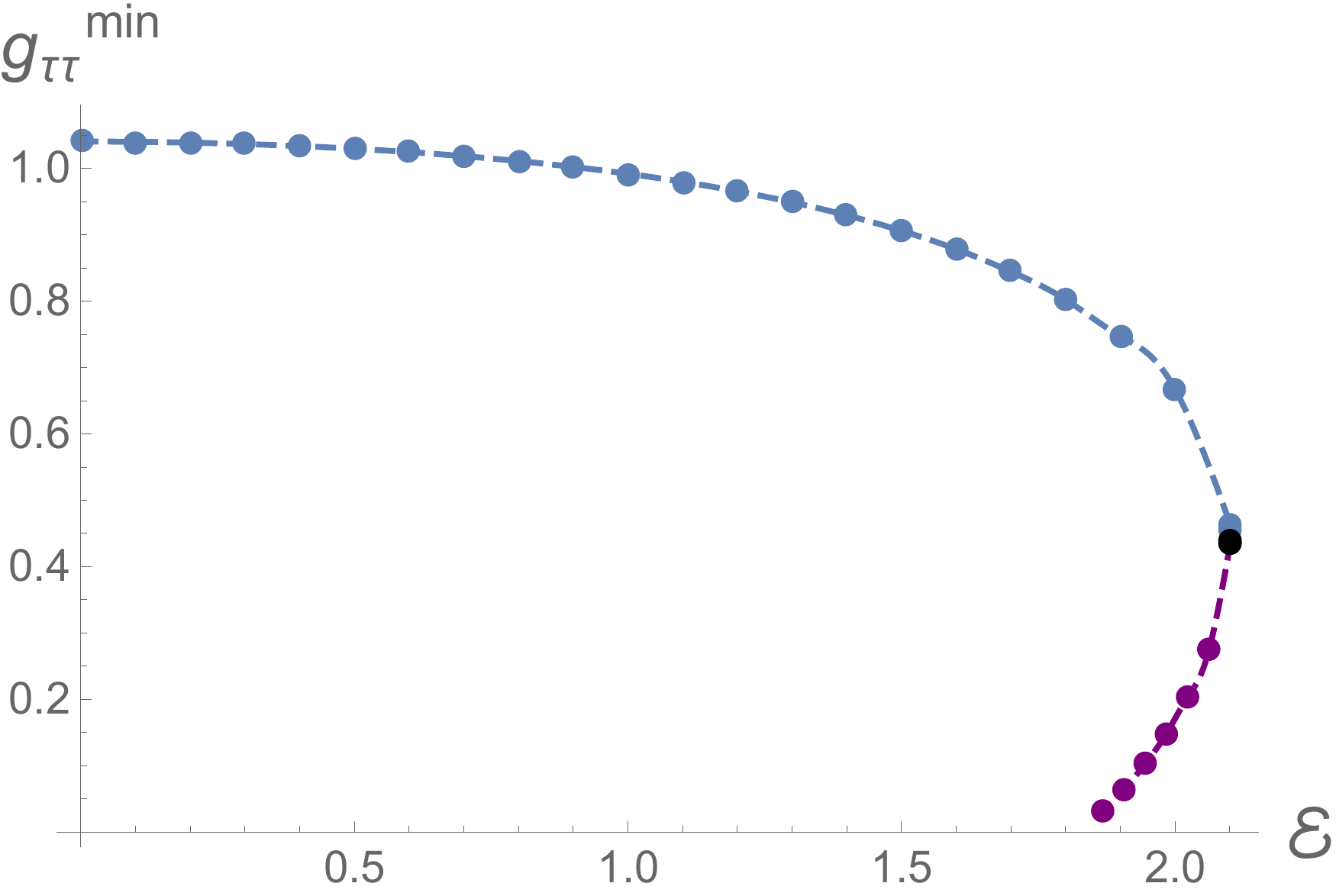}\label{gttAdSABJM}
}
\caption{(a) The value of $\varphi$ at the AdS center and (b) the $tt$ component of the metric up to maximum value of  ${\cal E}$ for the two branches of the $AdS$ soliton, denoted in blue and purple, respectively. 
}
\end{figure}
\begin{figure}[t!]
\centering
\includegraphics[width=70mm]{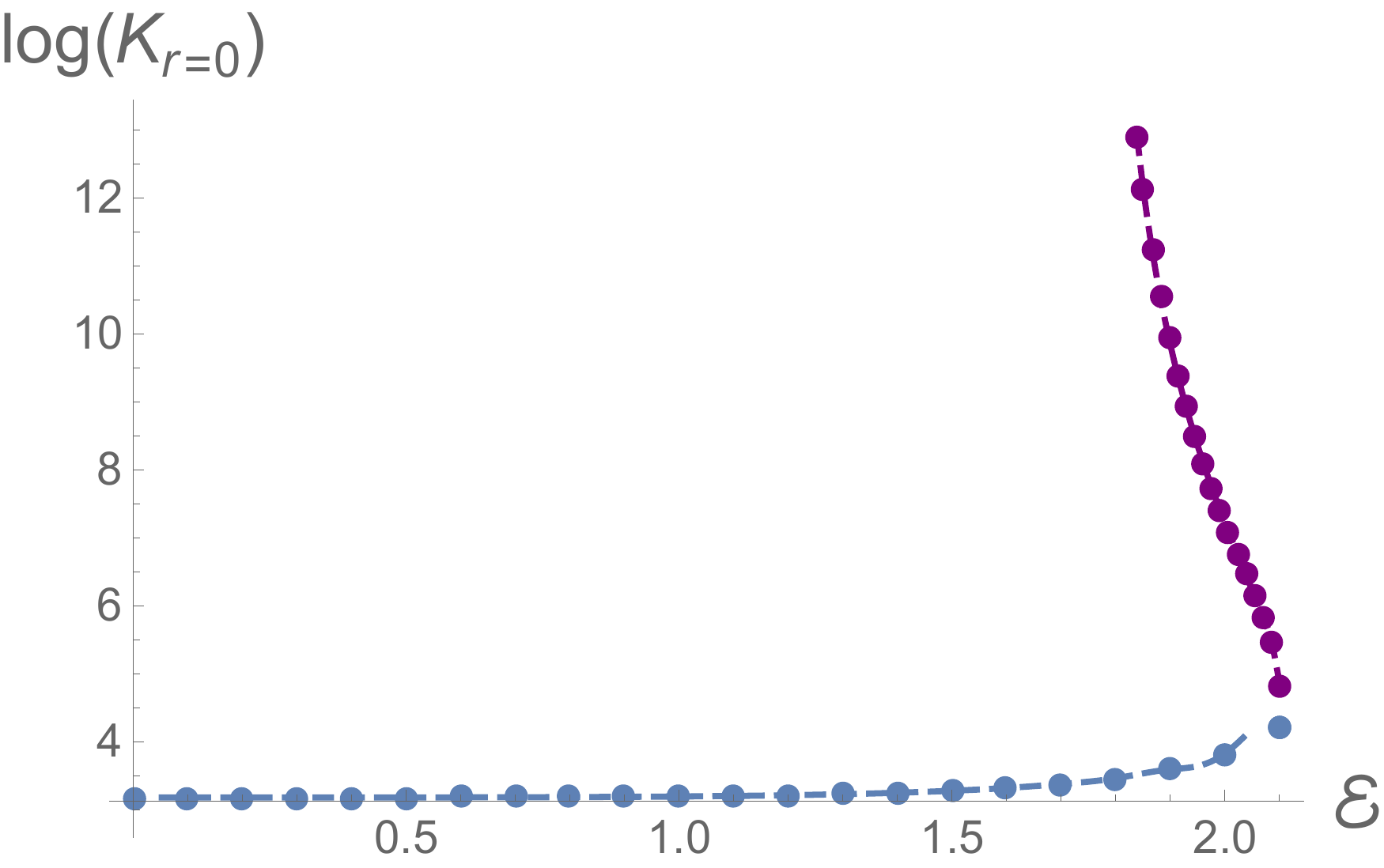}
\caption{A log plot of the Kretschman invariant at the origin $r=0$.
\label{kretschsol}}
\end{figure}

The value of $\varphi$ at the AdS center  is plotted in figure \ref{phiAdSABJM} and the metric component $g_{\tau \tau}$ in figure  \ref{gttAdSABJM}. The two branches are marked in blue and purple, respectively, and in both plots these meet at the maximum allowed  electric field. The fact that the $g_{\tau \tau}$ metric component tends to zero at some value of $\mathcal{E}$ could mean that a horizon develops at the origin when the electric field for the second soliton branch is reaches a minimum.
However, the Kretschman curvature invariant $K=R_{\mu \nu \rho \sigma}R^{\mu \nu \rho \sigma}$ blows up at this value of the electric field, therefore a singularity will form at this point. The Kretschman curvature at the origin is  plotted in figure \ref{kretschsol} as a function of $\mathcal{E}$. 

\section{Black Hole}\label{sec:BH}

Next we construct the solution describing a polarised black hole immersed  in the external electric field. 
A convenient ansatz  to find this geometry is
\begin{align}	
ds^2=\ &\frac{r^2}{\left(1-r^2\right)^2} \,A(r,\theta ) f(r)\,  d\tau ^2 
\nonumber\\
&+\frac{y_0^2}{\left(1-r^2\right)^2}\left(\frac{4 G(r,\theta )}{f(r)}\,dr^2+C(r,\theta ) \big(d\theta+2r  H(r,\theta ) \,dr \big)^2+B(r,\theta )\sin^2\theta \,d\phi^2\right),
\label{BHansatz}\\
A_\tau=\ &ir^2D(r,\theta)\,,
\qquad\qquad
\Phi(r,\theta)=(1-r^2) \, \varphi(r,\theta)\,,
\nonumber
\end{align}
where $f(r)=(1-r^2)^2-q_0^2(1-r^2)^3+y_0^2(3-3r^2+r^4)$.
The radial coordinate $r$ runs from the black hole horizon at $r=0$ to the $AdS$ boundary at $r=1$. For $r=1$ the metric functions obey the 
boundary conditions 
\be
A(1,\theta)=G(1,\theta)=C(1,\theta)=B(1,\theta)=1\,,
\qquad\qquad
H(1,\theta)=0\,.
\ee
For the scalar field, vanishing of the second derivative  in Fefferman-Graham coordinates now gives the condition
\be
\partial_r\varphi(r,\theta)\big|_{r=1}=0 \,.
\ee
For the gauge field we require a  dipolar potential as for the soliton, imposing condition  (\ref{eq:BounCondD}).
Thus the $AdS$ soliton and black hole solutions have the same asymptotics, and will be thermodynamically competing solutions. 

Setting $A=B=C=G=1$, $H=0$ and $D=q_0$ we would obtain the metric and gauge field for the Reissner-Nordstrom-$AdS$ black hole of charge $q_0$, with usual radial coordinate  $y=y_0/(1-r^{2})$. Therefore, parametrizing the metric ansatz with $y_0$ and $q_0$ will allow us to search for solutions with temperatures below the minimum temperature of the $AdS$-Schwarzschild solution as explained in detail in \cite{PBH}. We will, however, only consider neutral black holes, as follows from the boundary condition
(\ref{eq:BounCondD}) imposed on the gauge field.
%

At $r=0$, the first derivative with respect to $r$ vanish on all functions. These are readily imposed by the ansatz (\ref{BHansatz}) as it requires all metric functions,  and the gauge and scalar fields to be smooth functions of $r^2$. The condition $A(0,\theta)=G(0,\theta)$ is then fixed implicitly by the equations of motion, which guarantee that the geometry closes off smoothly at the fixed point $r=0$. Given that these conditions are satisfied, we may use 
the parameters $y_0$ and $q_0$ to fix the temperature of the solution.
 The symmetries about $\theta =0\, ,\pi$ are the same as for the $AdS$ soliton of the last section, as are the corresponding boundary conditions. 

\subsection{Results}\label{results}

\begin{figure}[b!]
\centering
\subfloat[]{
\includegraphics[width=55mm]{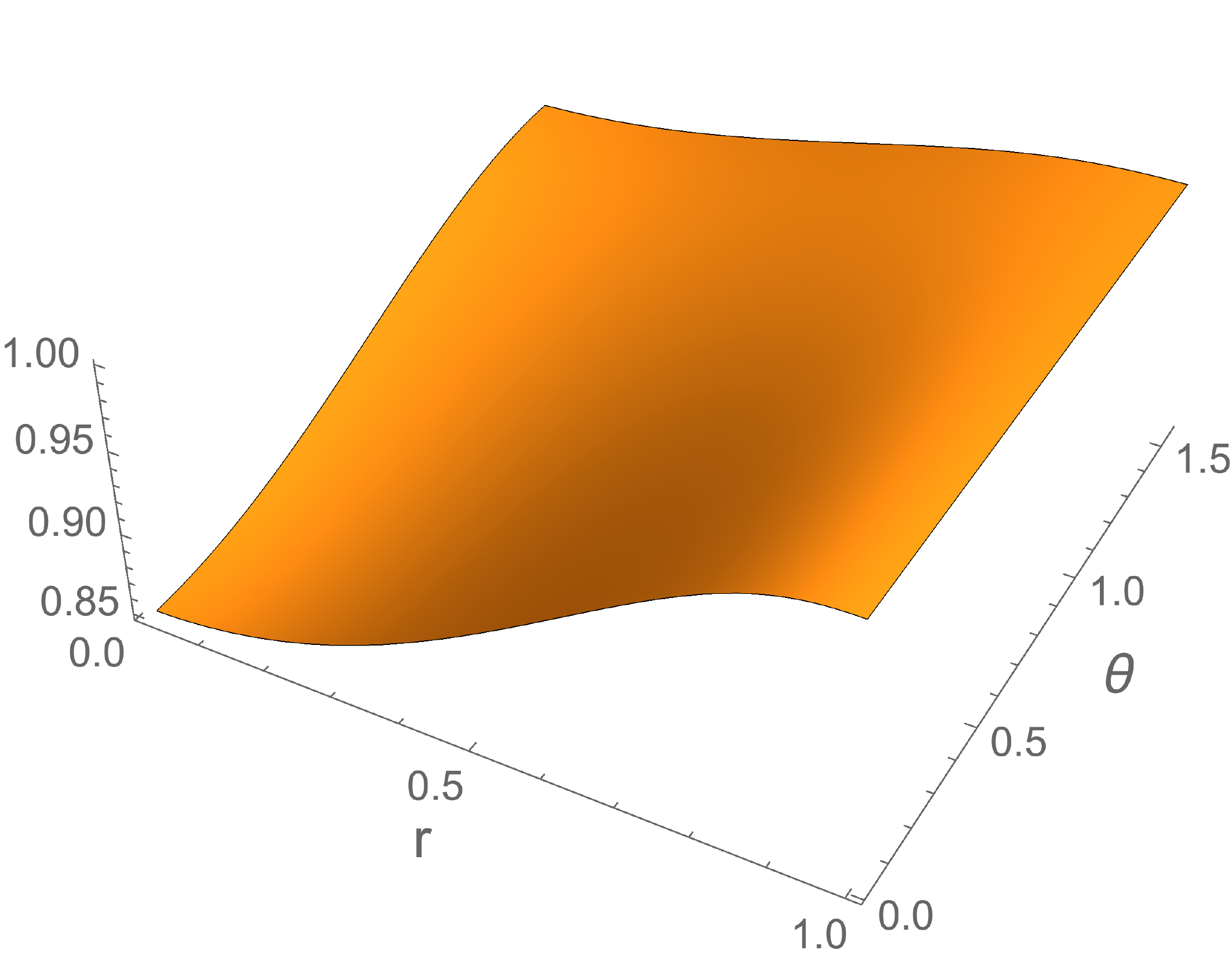}
}
\hspace{2cm}
\subfloat[ ]{
\includegraphics[width=55mm]{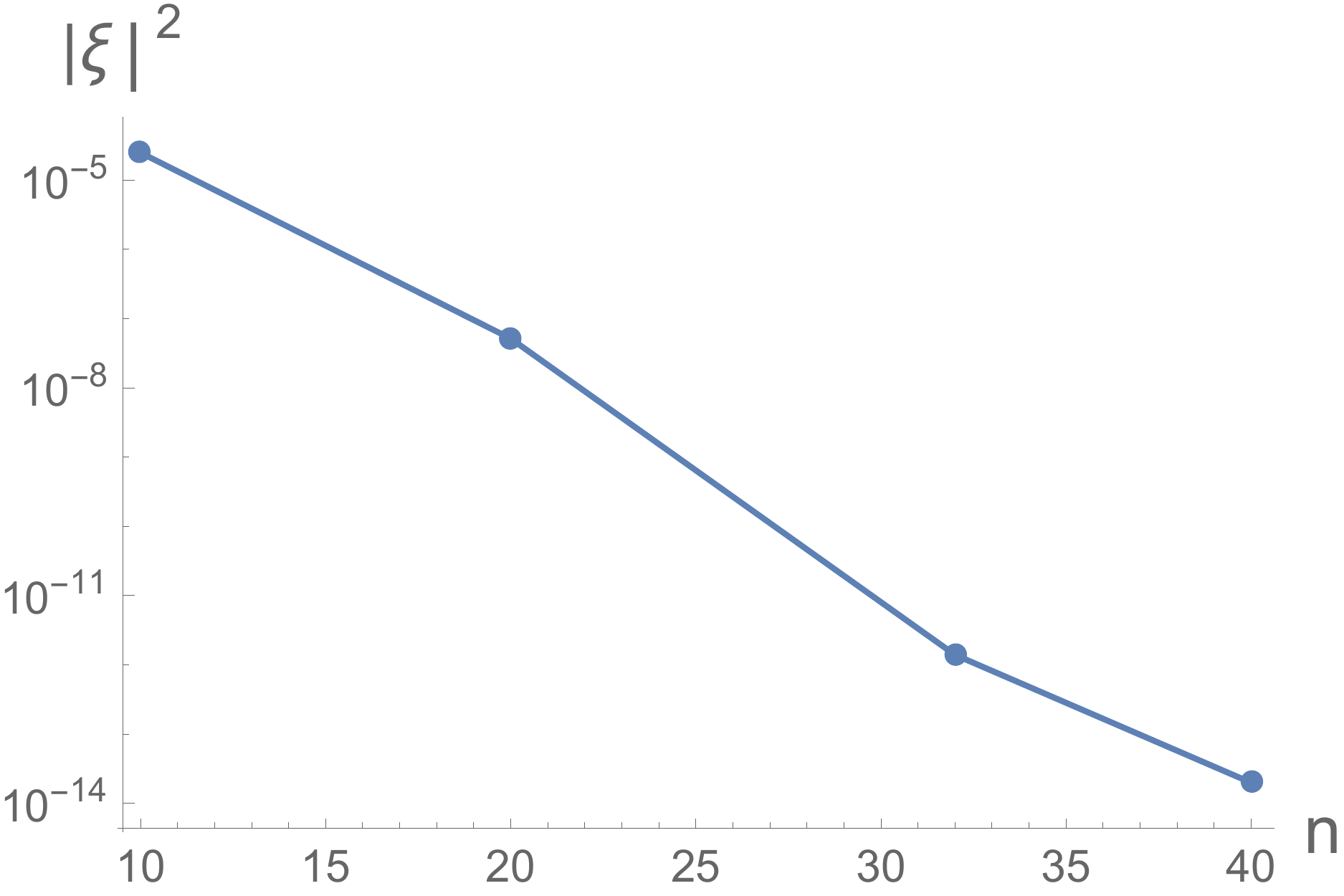}
}
\caption{(a) Examples of numerical solutions for the metric function $A$ of a large polarised black hole for an electric field ${\cal E}=1$ and temperature $T=1/\pi$ for a numerical grid of $40\times 40$ points.(b) Plot of convergence for the polarised black hole for a value of the electric field ${\cal E}=1$ and temperature $T=1/\pi$. \label{convergenceABJM}
}
\end{figure}

Figure \ref{convergenceABJM}(a) shows the ansatz function $A$ for a value of temperature and electric field. A linear-logarithmic plots of the deTurck vector norm as a function of the number of grid points $n$ is plotted in figure \ref{convergenceABJM}(b).

For a given temperature there will be four branches of solutions. Two are analogous to the so-called large and small black holes of Schwarzscild-AdS without a scalar or a source, so we will refer to these solutions as ``L1" and ``S1" black holes. These solution are continuously connected to the large and small black holes of Schwarzscild-AdS by turning off the external electric field. 
In the case of black holes in $AdS$ with an electric field but no scalar, as discussed in $\cite{PBH}$, the large and small black hole branches connect at a minimum value of the temperature that depends on the electric field, which can be made arbitrarily large. No solutions exist at temperatures below these minimum values.  
The same behaviour occurs in the presence of the scalar fields, however now there is a 
maximum value allowed for the electric field on the $AdS$ black hole geometry that changes itself with temperature.
This gives another degenerate point at which the large and small black holes branch again. We will call these branches the ``L2" and ``S2" black holes. 

\begin{figure}[t!]
\centering
\subfloat[]{
\includegraphics[width=45mm]{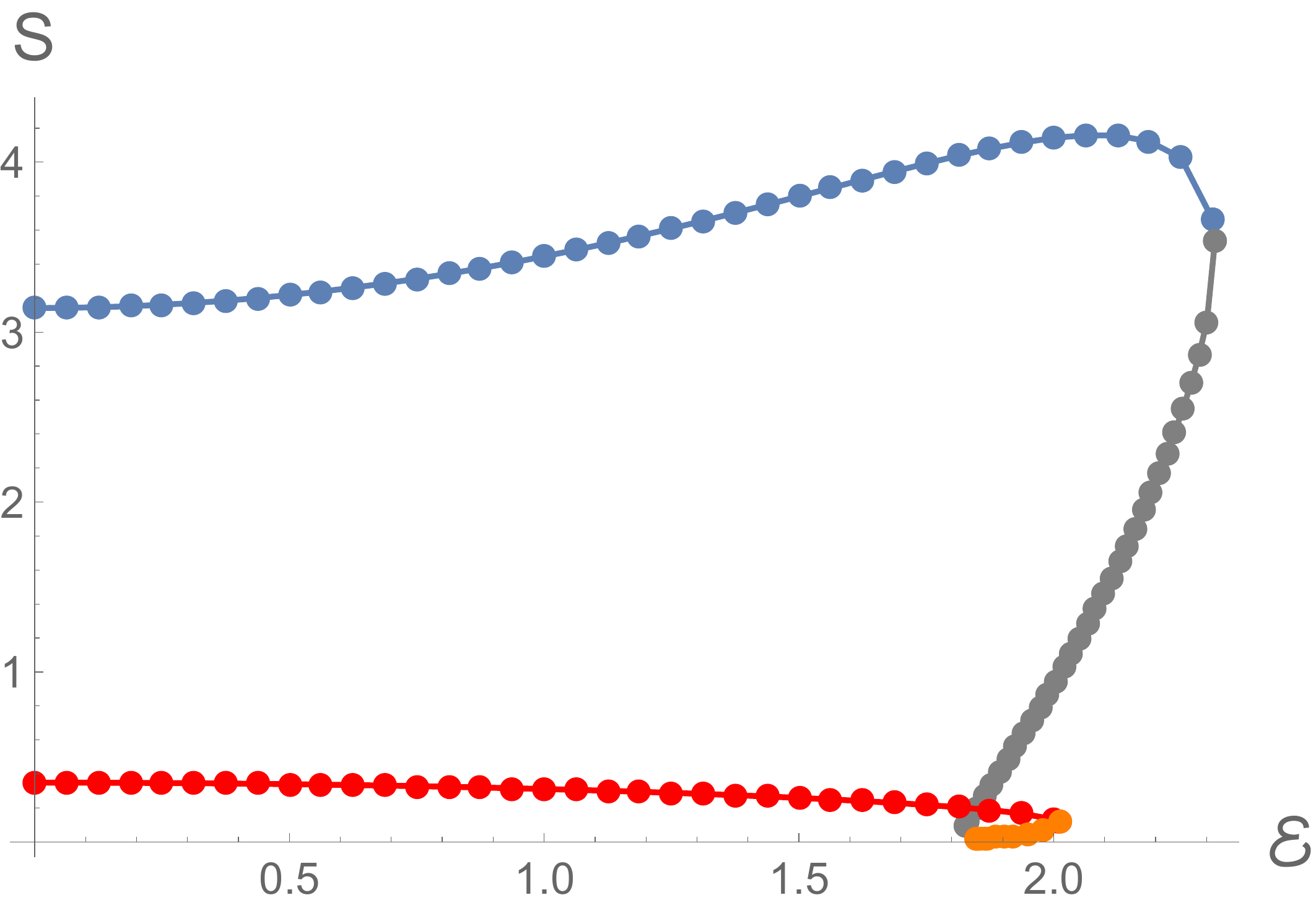}
}
\subfloat[ ]{
\includegraphics[width=45mm]{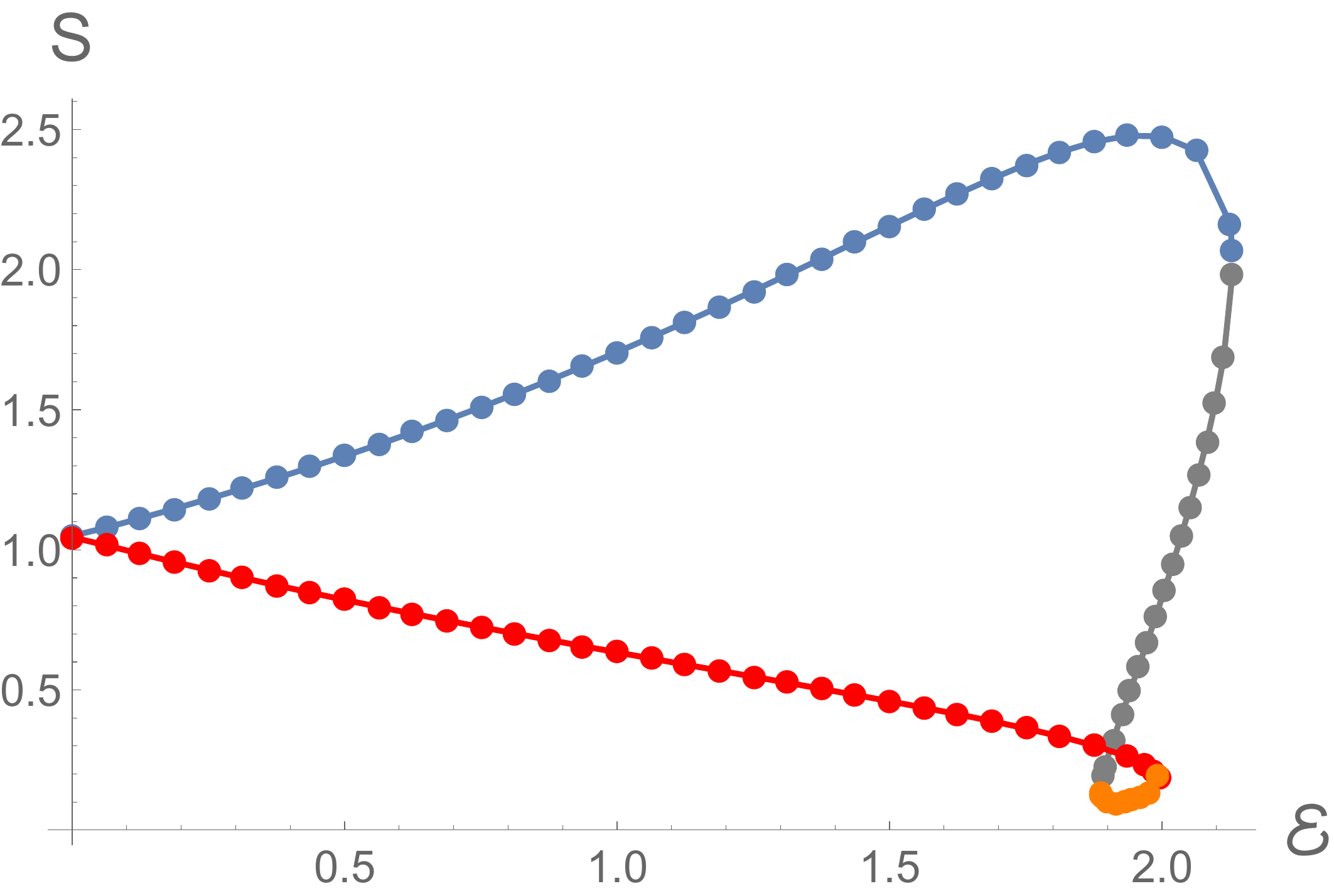}
}
\subfloat[ ]{
\includegraphics[width=45mm]{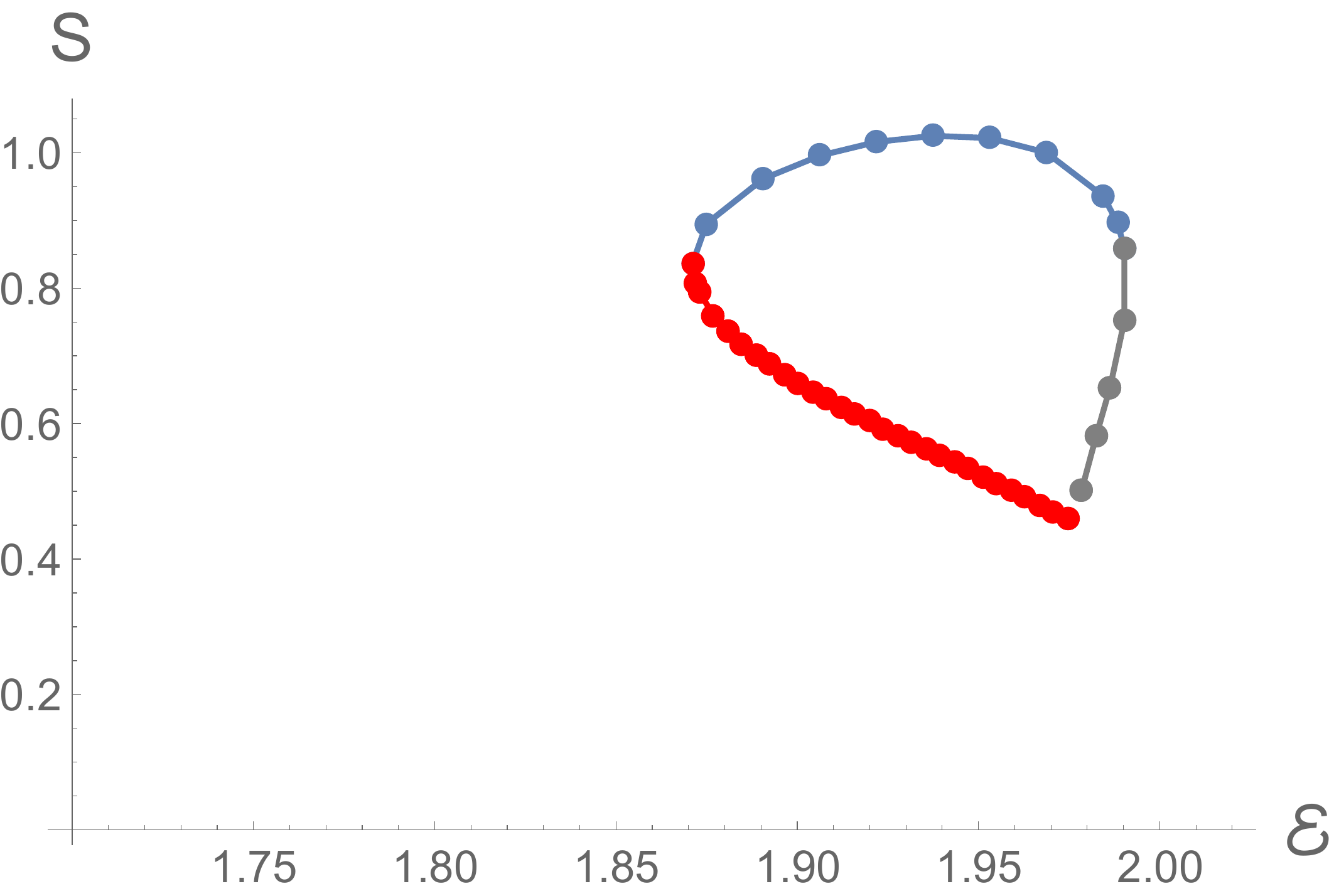}
}
\caption{Black hole entropy as function of the electric field for $T=1/\pi$, $T=0.275665$ and $T=0.24179$ ($G_N=1$). 
Plot (b) is for a temperature slightly above the minimal value allowed for $AdS$-Schwarzschild black holes, where the large and small black hole 
branches meet. Below this temperature neutral black holes only exist for a non-zero electric field.
\label{entropyABJM}}
\end{figure}

Figure \ref{entropyABJM} shows the area of the horizon by way of the Bekenstein-Hawking entropy 
\ba
S=\frac{\mathcal{A}}{4 G_N}=\frac{\pi y_{0}^{2}}{G_N}\int_0^{\frac{\pi}{2}}  d\theta\, \sin\theta\sqrt{C(0,\theta)B(0,\theta)}\,,
\ea
of the four black hole branches as a function of the electric field, for three values of the temperature that decreases in the plots from left to right. In this plot and those that follow, the blue, red, gray, and orange curves correspond to the L1, S1, L2, and S2 branches, respectively. The maximum value of 
$\mathcal{E}$ increases  as the temperature increases. For low enough temperatures, below the minimal value allowed for $AdS$-Schwarzschild black holes, 
black holes only exist above a minimal value of ${\cal E}$.

\begin{figure}[b!]
\centering
\includegraphics[width=80mm]{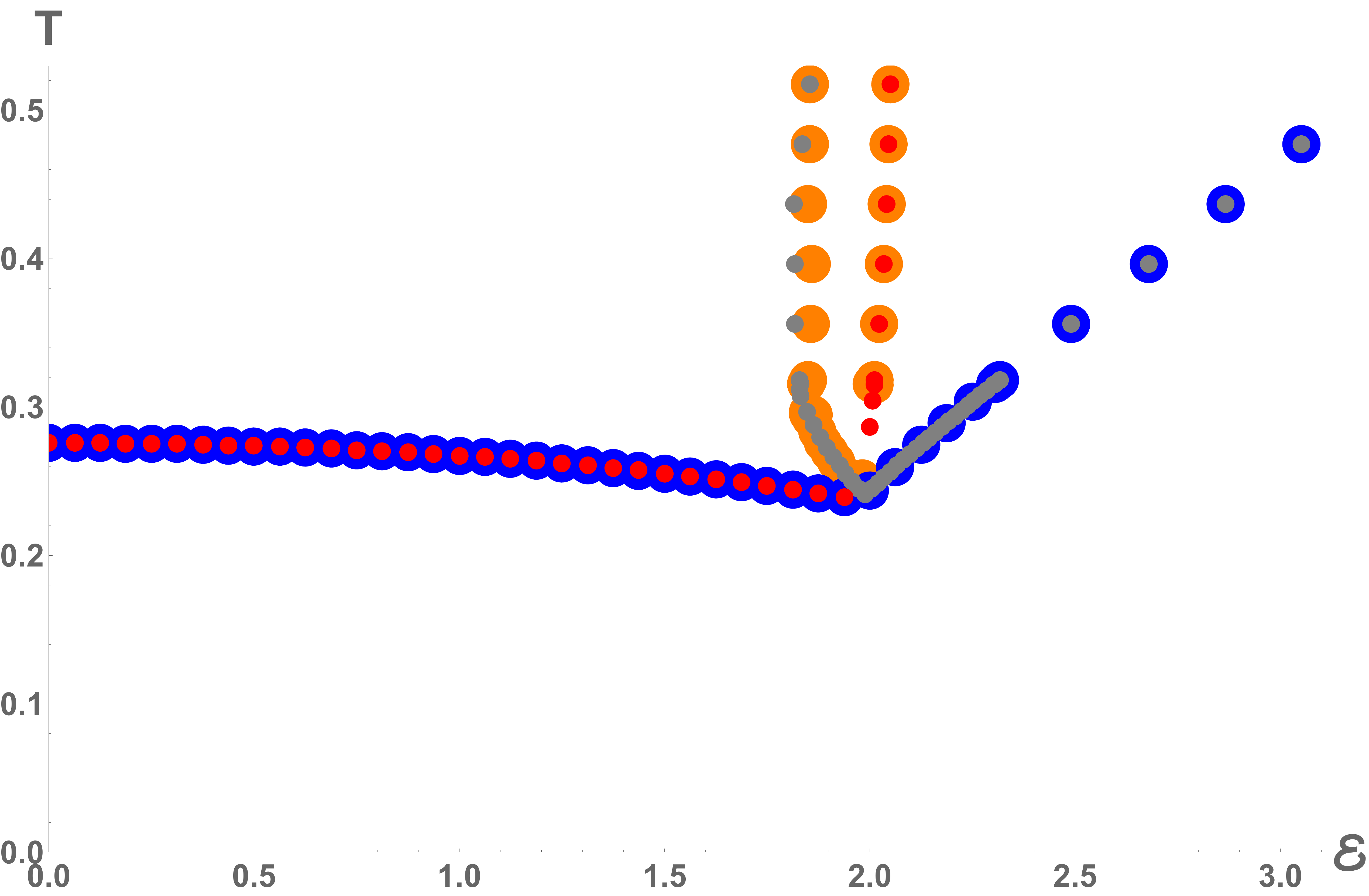}
\caption{\label{mintabABJM}The minimum temperatures of the L1 (blue), S1 (red), L2 (gray), and S2 (orange) curves as a function of $\mathcal{E}$.}
\end{figure}
\begin{figure}[t!]
        \centering
        \subfloat[]{
        \includegraphics[height=50mm]{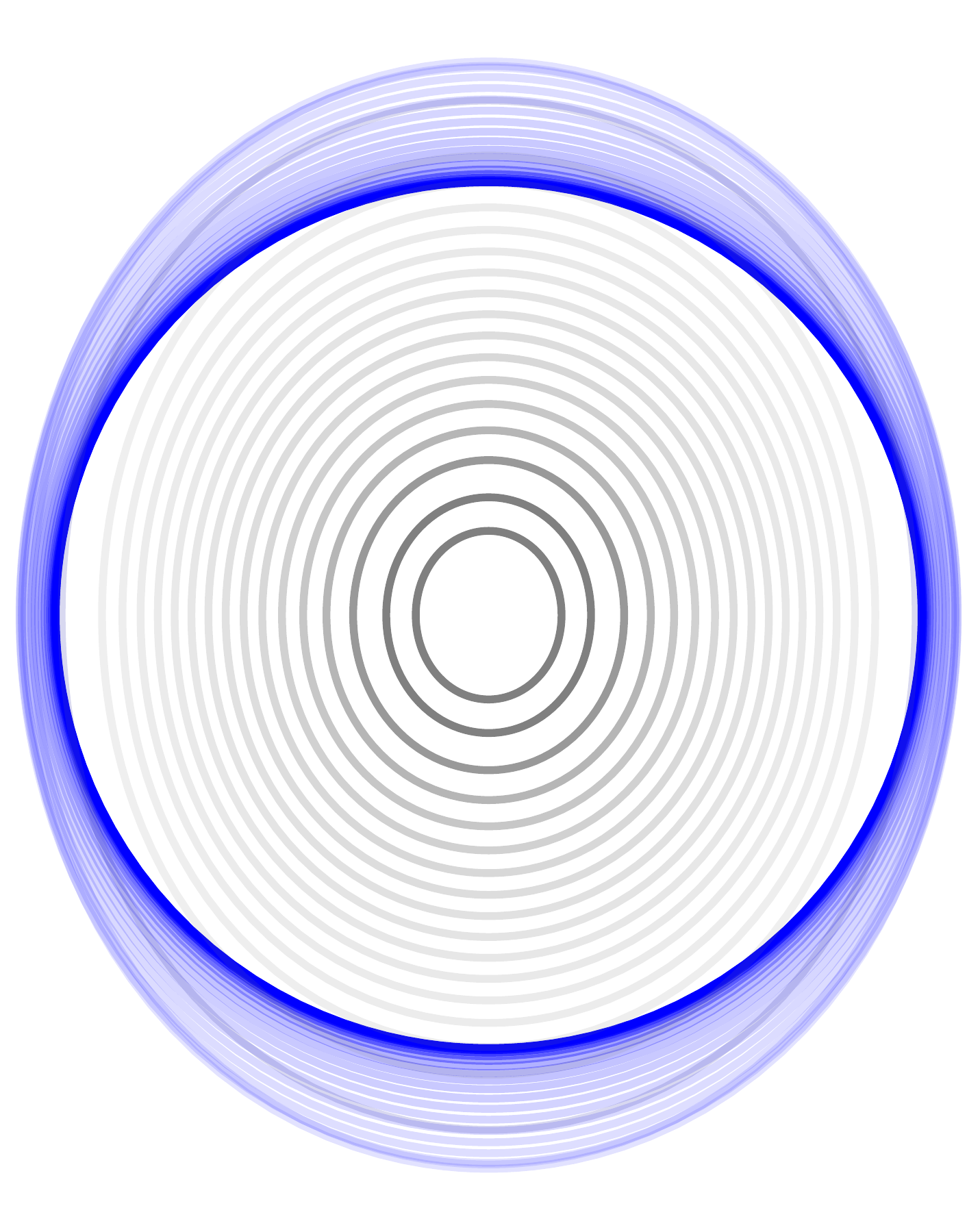}
        }
\ \ \ \ \ \ \ \ \ \ \ \ \ \ \ \ \ \ 
       \subfloat[]{
        \includegraphics[height=50mm]{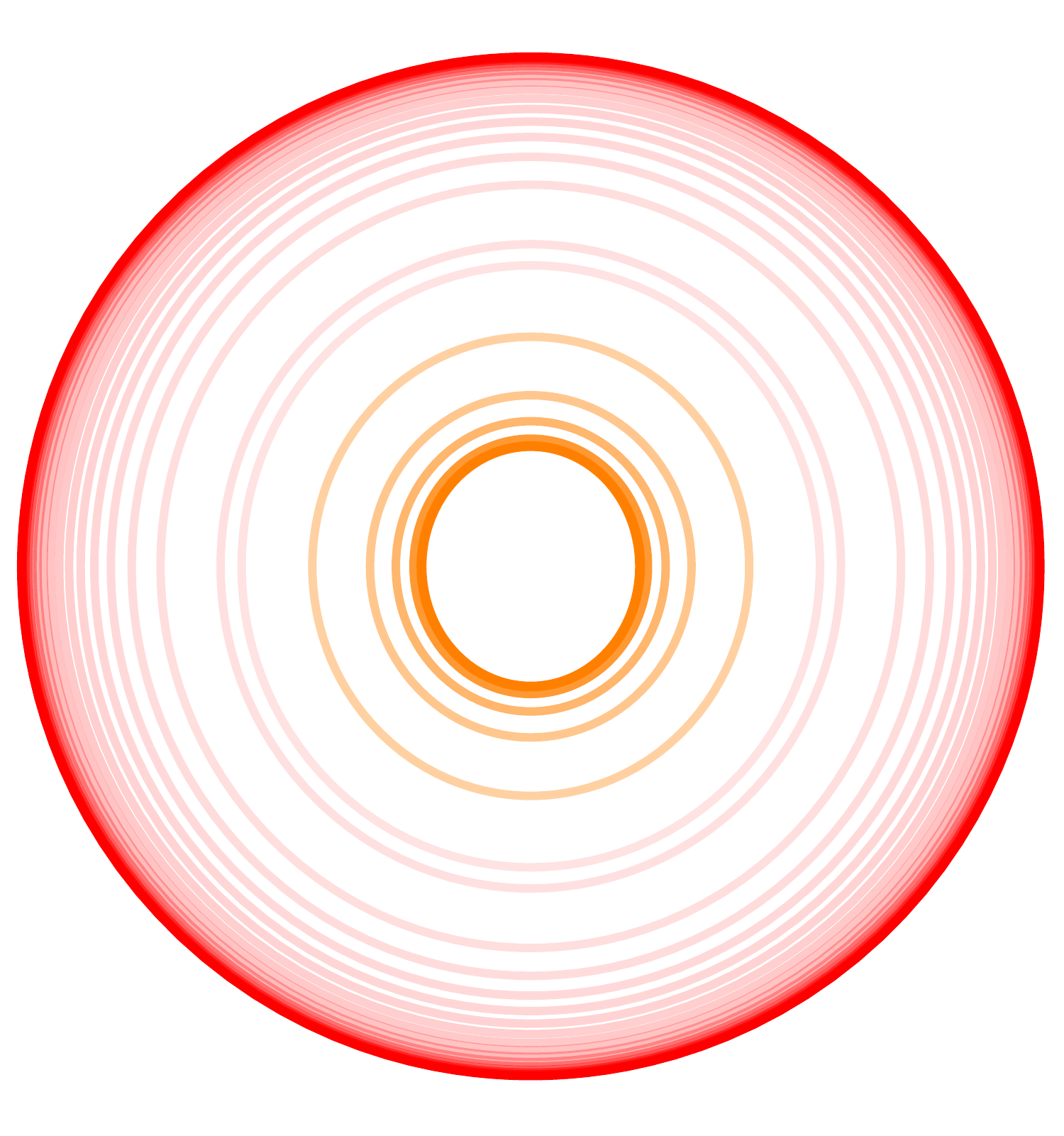}
        }
       
\caption{Isometric embeddings of  black hole horizons at fixed $T=1/\pi$. The curves for the L1, L2, S1, and S2 black holes are blue, gray, red, and orange, respectively.}
 \label{bigvsmall}
\end{figure}
Before looking at plots of other observables, it is useful to spend some time discussing the space of solutions. In figure \ref{mintabABJM}, we plot the extremal values of ${\mathcal{E},T}$ for the L1, L2, S1, and S2 black hole branches drawn as blue, gray, red, and orange curves, respectively. We see that each curve has a minimum and maximum value of the electric field that depends on the temperature. For small $\mathcal{E}$, the L1 and S1 black holes have the same minima, as we expect from $AdS$-Schwarzschild. Around $\mathcal{E}=2$, there are four black hole solutions. The electric field minima for the S2 and L2 fall along the same curve, with any discrepancies in this plot arising from the numerical difficulty in finding S2 black hole solutions for a certain range in temperatures.  The plot also shows that the maxima for the L1 and L2, as well as S1 and S2, branches follow the same respective curves. Notice that below a certain temperature we do not find any black holes solution, irrespectively of the value of the electric field.

\begin{figure}[b!]
        \centering
        \subfloat[]{
        \includegraphics[height=50mm]{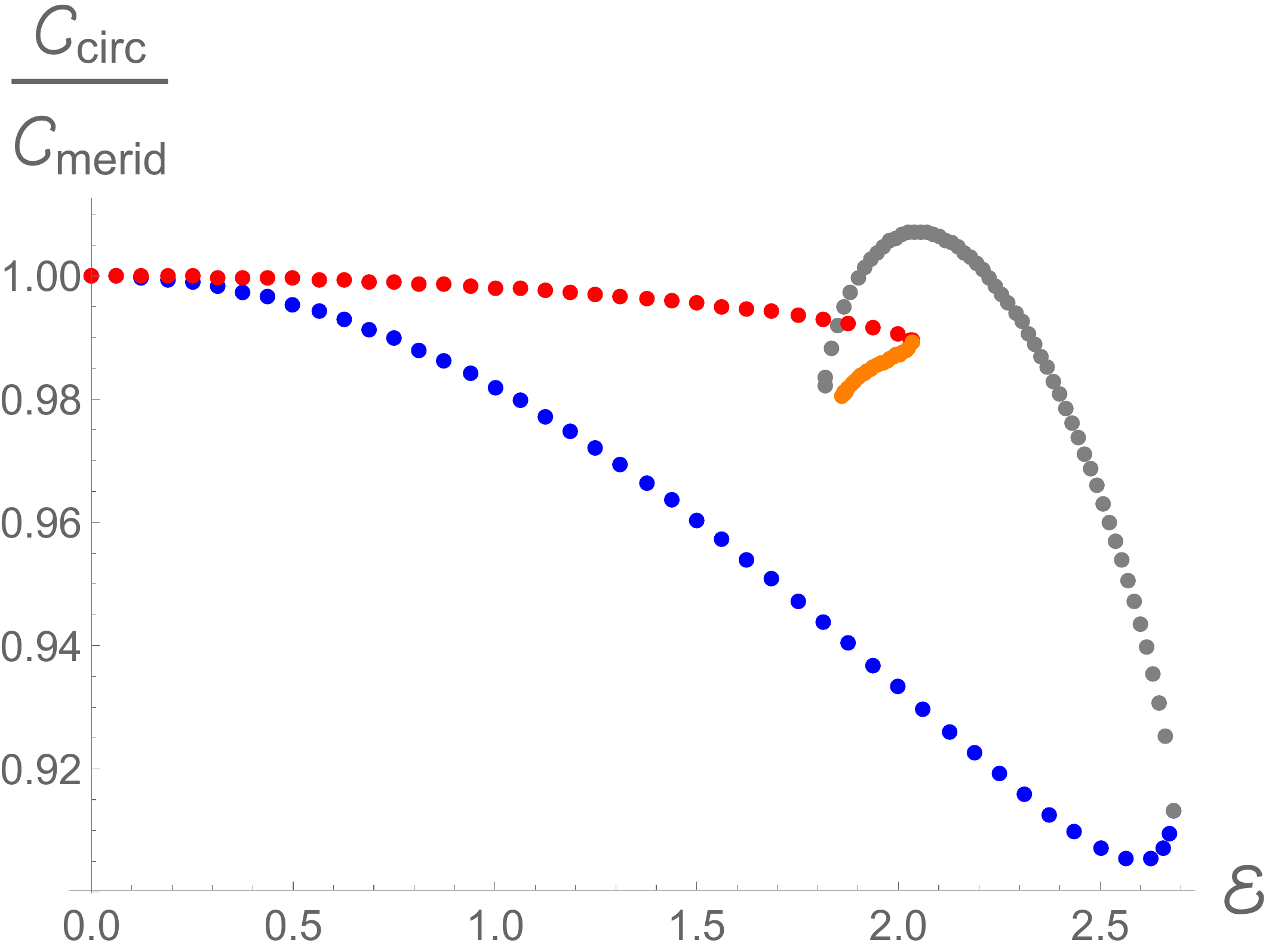}
        }
\caption{\label{circrat}The ratio of the circumferences at the horizon equator and a meridian at fixed $T=.396$. }
\end{figure}

By computing isometric embeddings of the horizon geometry in Euclidean space, we can monitor the shape of the horizon. These are plotted in figure \ref{bigvsmall}  for the four black hole branches at a fixed temperature of $T=1/\pi$. In (a), the L1 and L2 black holes are plotted in blue and gray, respectively. In (b), the S1 curve is red and the S2 curve orange. More transparent curves correspond to larger values of the electric field with the faintest curves of L1 and L2 or S1 and S2 corresponding to the same value of 
$\mathcal{E}$. We can see that the black holes have the same shape when these branches meet at the extremal values of the electric field. These plots can be summarized by looking at the ratio of the horizon circumference at the equator to that of a meridian, as shown in figure \ref{circrat}. The L1 and S1 black holes start as round spheres at $\mathcal{E}=0$, after which the L1 black hole deforms much more than the S1 black hole. The L1 and L2 curves meet at the maximum value of $\mathcal{E}$, and the L2 black hole also deforms much more than the S2 curve.

As the electric field is increased, the black holes becomes slightly deformed but do not pinch at the equator as in \cite{PBH}. This is one of the main qualitative differences from coupling the polarized black holes to a scalar field.

\begin{figure}[t!]
\centering
\subfloat[ ]{
\includegraphics[width=60mm]{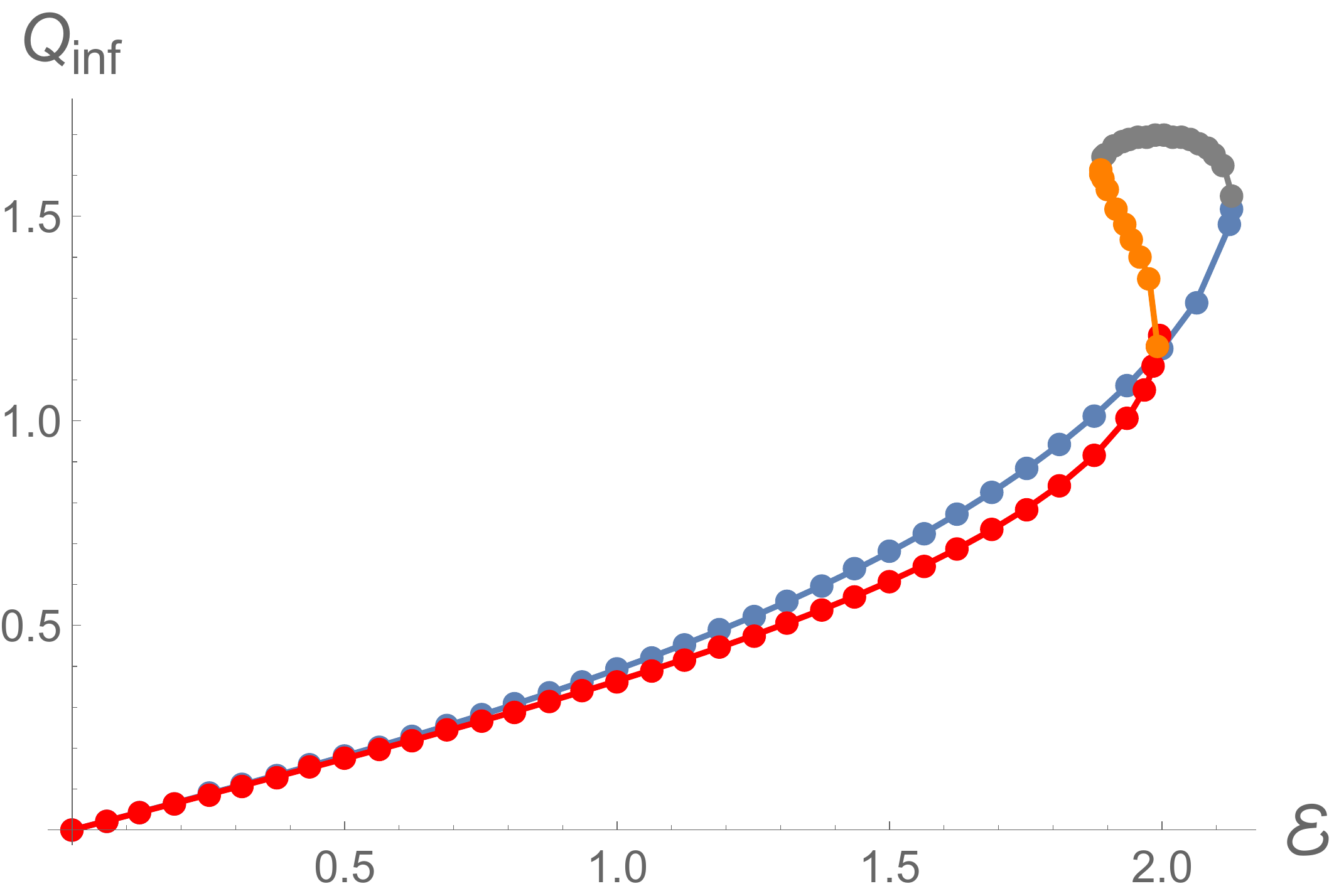}
\label{totalQa}
}
\subfloat[ ]{
\includegraphics[width=60mm]{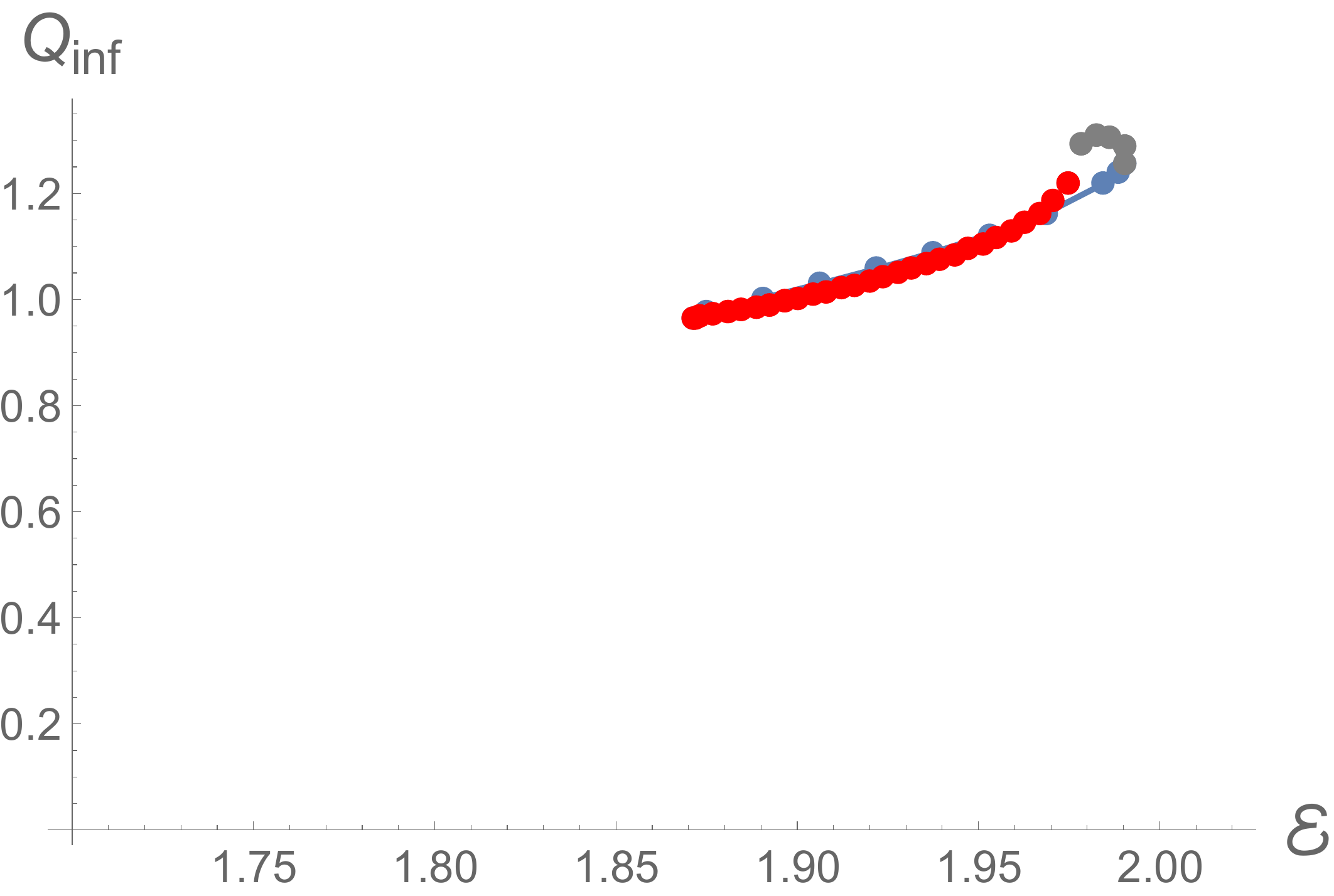}
\label{totalQb}}
\\
\centering
\subfloat[ ]{
\includegraphics[width=60mm]{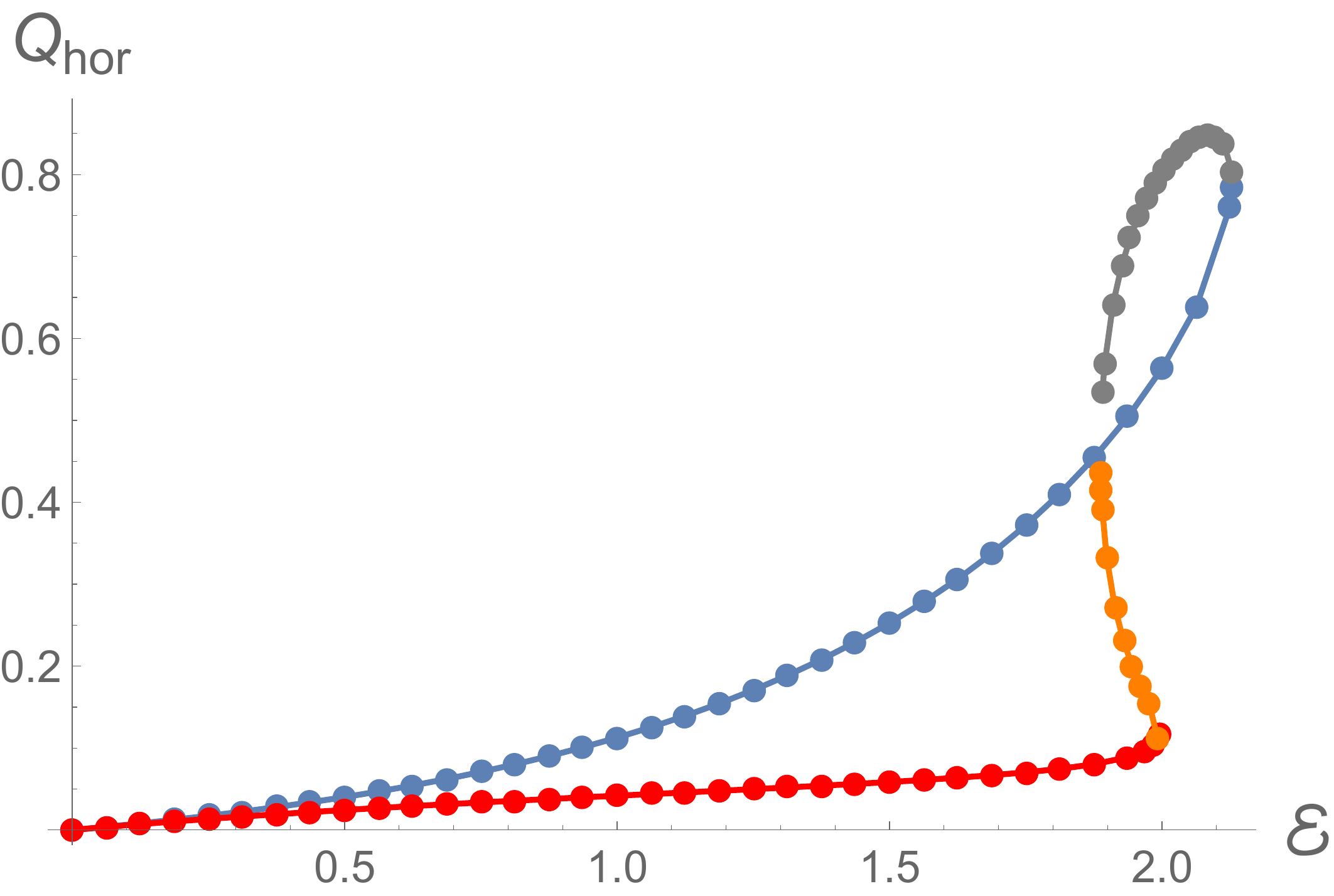}
\label{totalQc}
}
\subfloat[ ]{
\includegraphics[width=60mm]{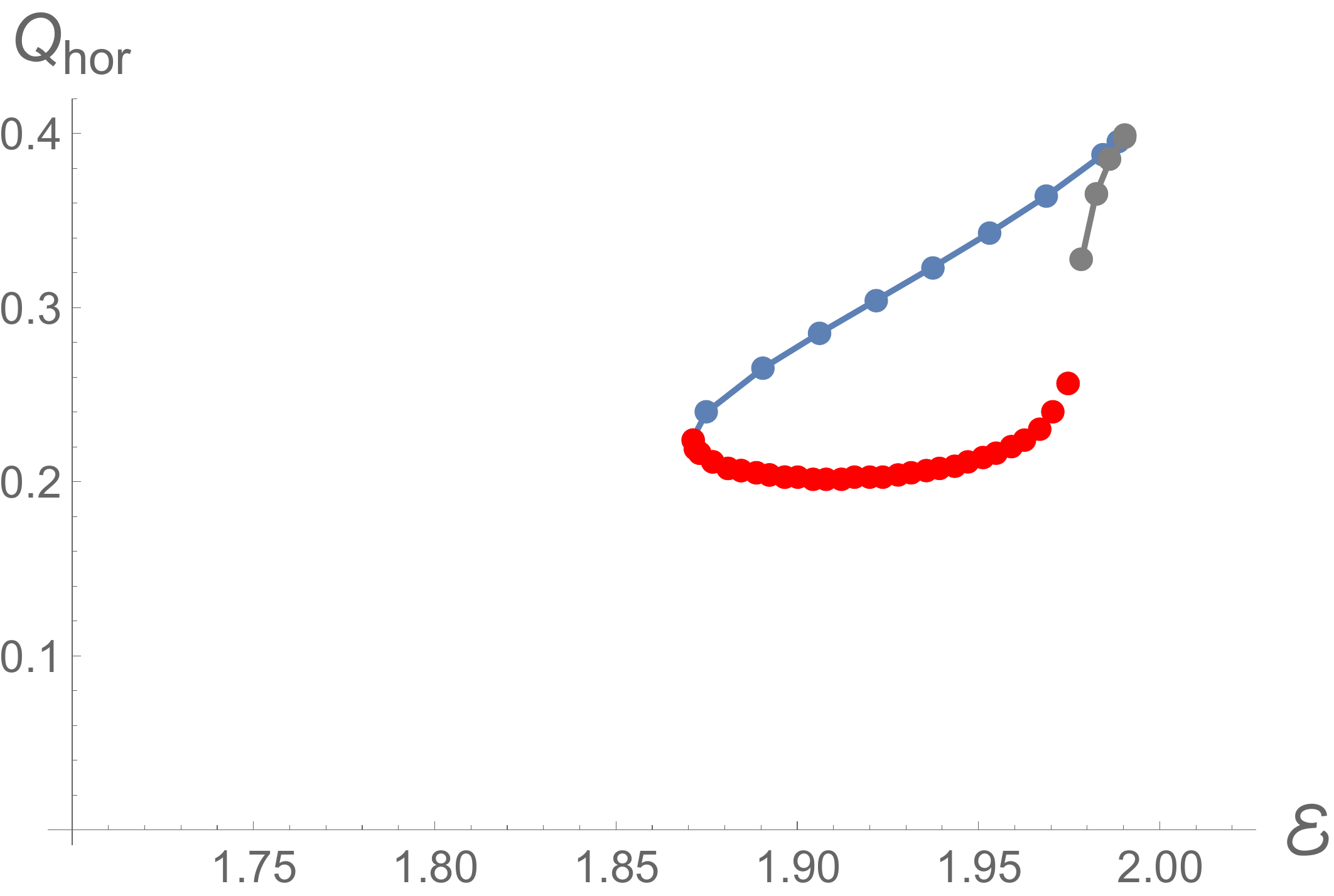}
\label{totalQd}
}
\caption{ (a)-(b) Total charge in one hemisphere  at  the $AdS$ boundary; (c)-(d) and at the black hole for $G_N=1$. The first column is at $T=0.275665$ and the second at $T=0.24179$. 
}
\end{figure}

In contrast to the $AdS$ soliton, the black hole has two surfaces which accumulate charge: the boundary and the horizon. We can therefore look at the total charge contained in one hemisphere, by integrating the electric flux through each of these surfaces as defined in (\ref{eq:carge_dens}).  Figures \ref{totalQa} and \ref{totalQb} show the total boundary charge for two values of the temperature, while figures \ref{totalQc} and \ref{totalQd} show the total charge in one hemisphere of the black hole horizon for the same two values of temperature. In this case, conservation of charge requires that the difference between the hemispherical charges at the boundary and horizon is equal to the electric flux (\ref{fluxeq}) through the equator. We checked that this is true to within $10^{-4}$.

\begin{figure}[t!]
\centering
\subfloat[ ]{
\includegraphics[width=50mm]{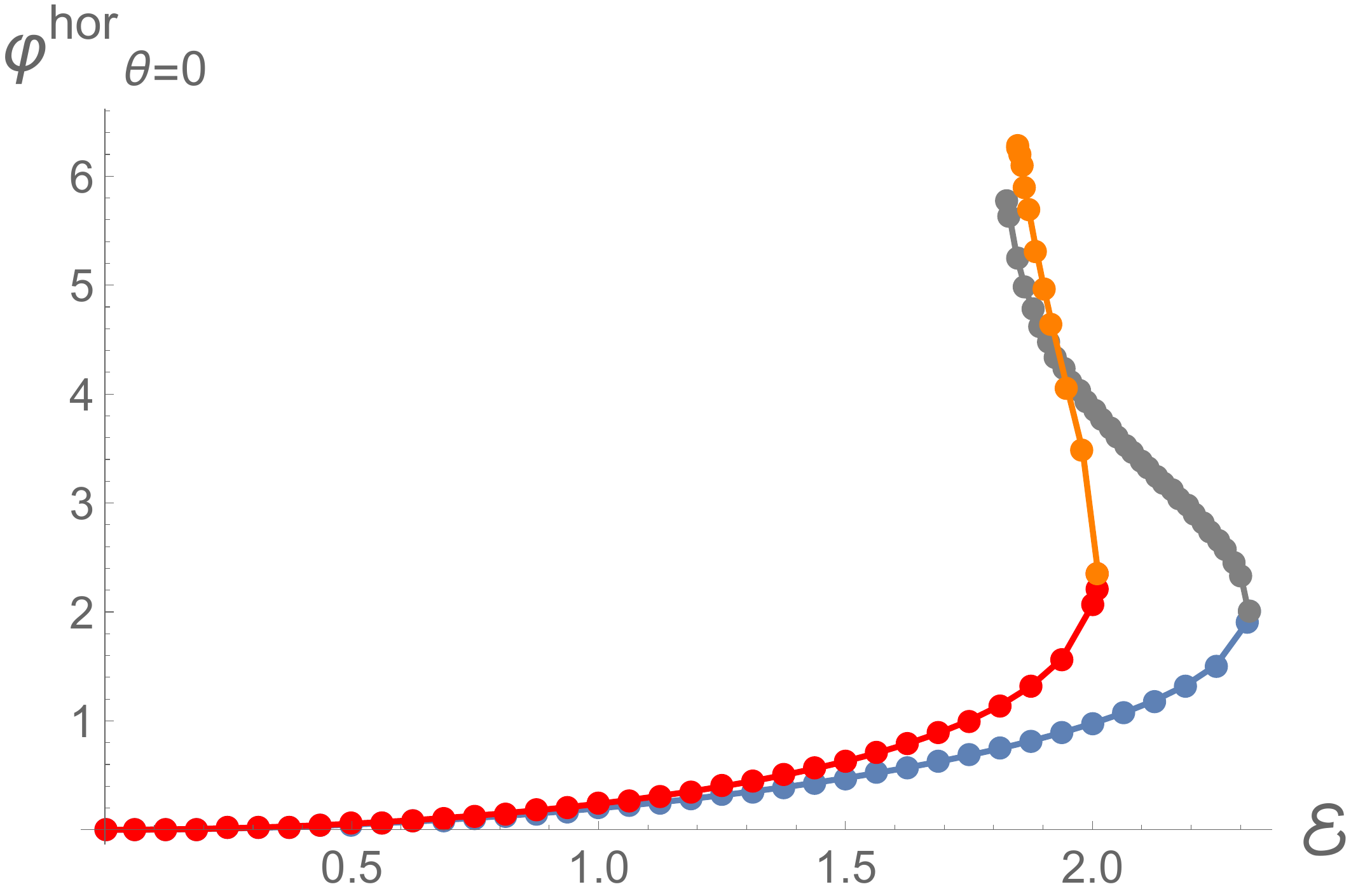}
}
\subfloat[ ]{
\includegraphics[width=50mm]{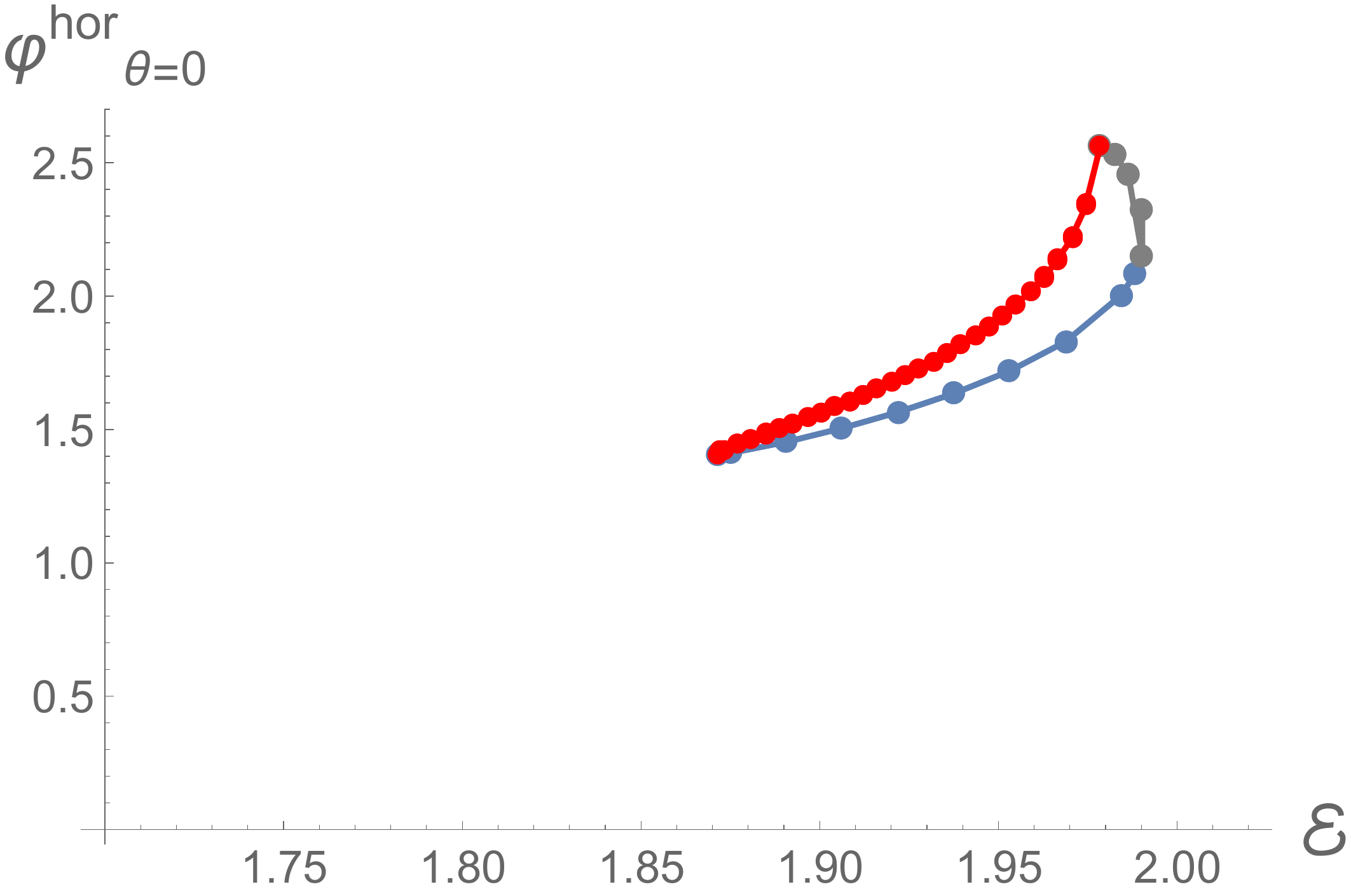}
}
\caption{ \label{scalar}The value of the scalar field at the horizon for $T=1/\pi$ and $T=0.24179$, from left to right. 
}
\end{figure}

The value of the scalar field at the pole of the horizon is shown in figure \ref{scalar}. It shows that while the value doesn't change much with temperature, the maximum electric field allowed at that temperature for the large black hole branch, which is the right-most curve on the plot, does.  

\begin{figure}[b!]
\centering
\subfloat[]{
\includegraphics[width=50mm]{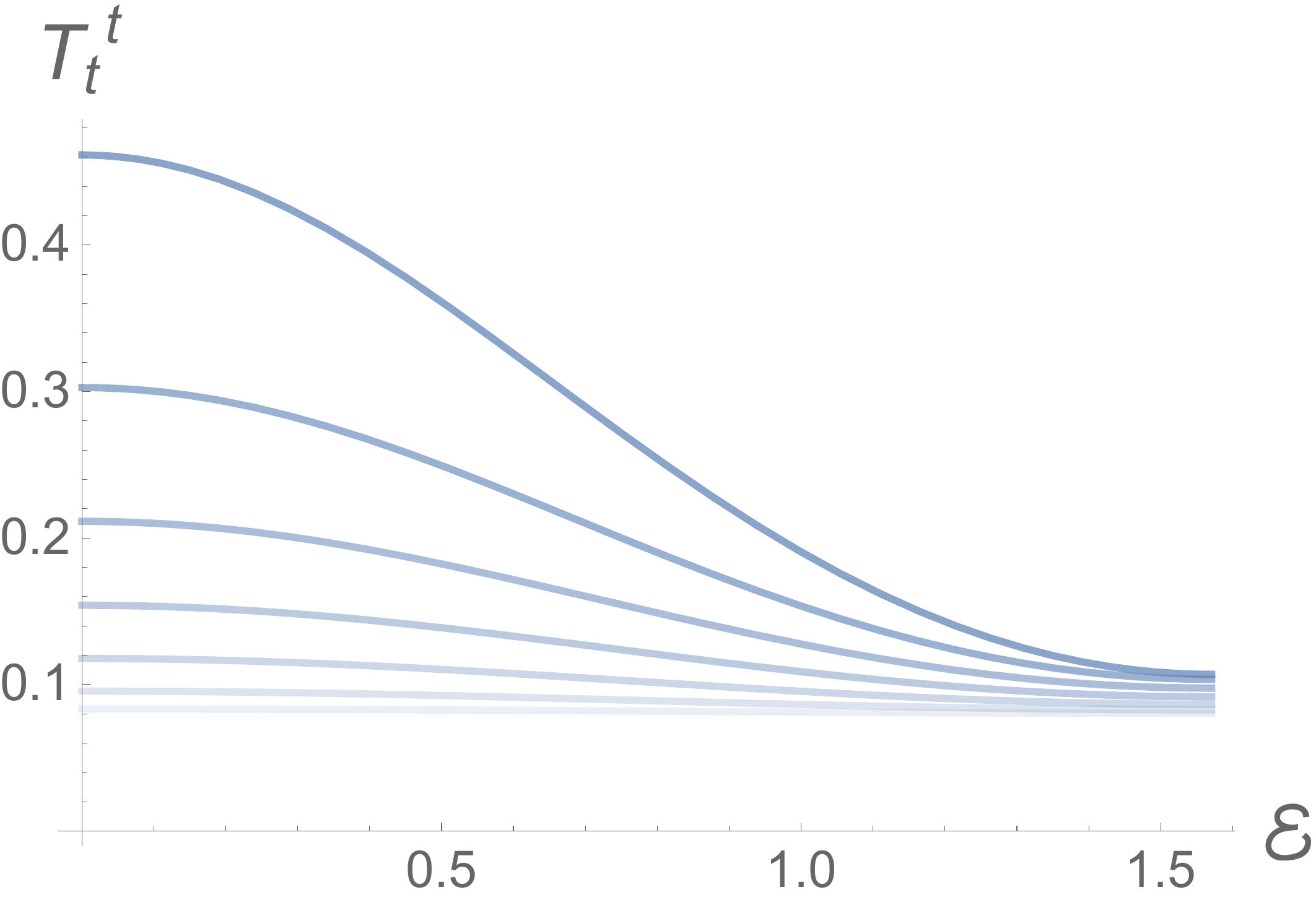}
}
\subfloat[]{
\includegraphics[width=50mm]{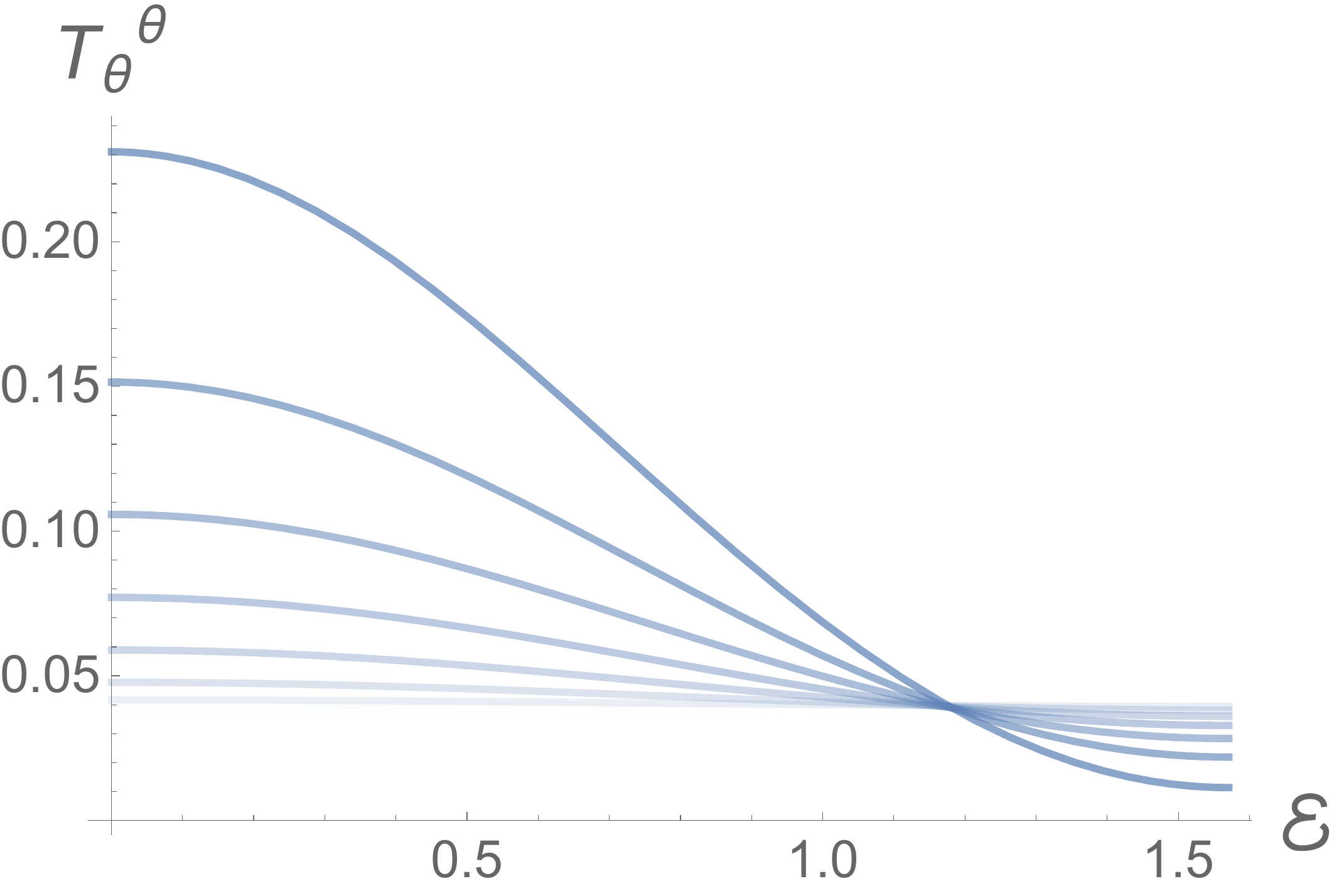}
}
\subfloat[]{
\includegraphics[width=50mm]{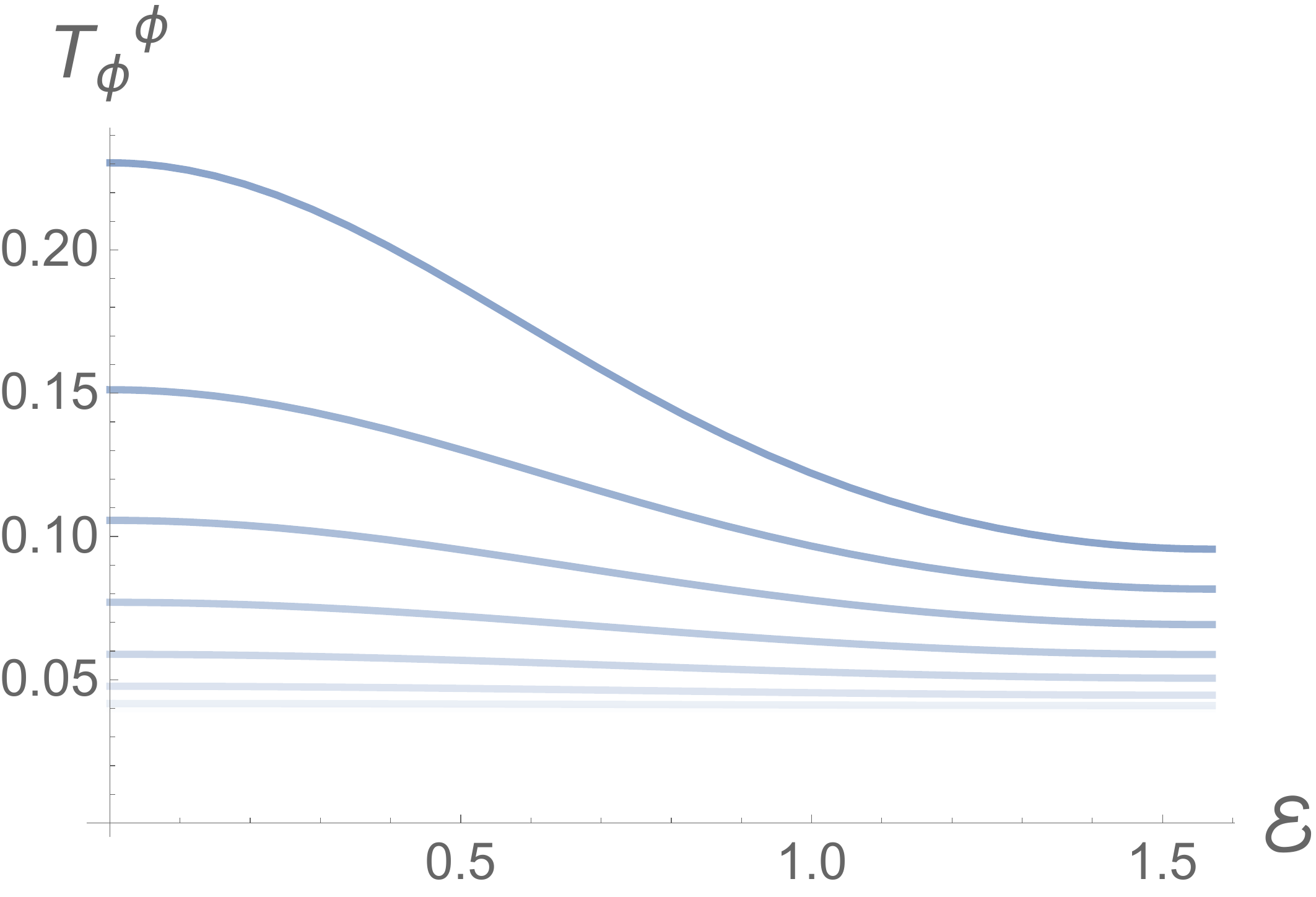}
}
\caption{ (a) Boundary energy density, (b) the $\theta\theta$ component and (c) the $\phi\phi$  component of the boundary stress tensor at $T=1/\pi$. Fainter curves correspond to lower values of the electric field. (Setting $G_N=1$.)
\label{StressTensorBH}}
\end{figure}

The non-vanishing components of the boundary energy-momentum tensor of the dual to the black hole geometry are
\ba
T_\tau^\tau&=&\frac{-y_0}{256\pi G_N}\left(16(1+Q^2+y_0^2)-y_0^2\left(6\alpha_3-3 \phi_0^2\right)\right) ,\notag\\
T_\theta^\theta&=&\frac{y_0}{256\pi G_N}\left(16(1+Q^2+y_0^2)+y_0^2\left(6\chi_3-3 \phi_0^2\right)\right),  \\
T_\phi^\phi&=&\frac{y_0}{128\pi G_N}\left(8(1+Q^2+y_0^2)-3y_0^2\left(\alpha_3+\chi_3- \phi_0^2\right)\right),\notag
\ea
where, as before, $\alpha_i$, $\chi_i$, and $\phi_i$ are the $i$-th order power-law modes associated to the functions $A$, $C$, and $\varphi$, respectively.  Angular profiles for each component are plotted in figure \ref{StressTensorBH} for the large black hole at several values of electric field magnitude and temperature. Above ${\cal E}=0$, the energy density and spatial components of the stress tensor are all maximal at the pole and minimal at the equator, as for the $AdS$ soliton. The only qualitative difference is that the $T_{\theta \theta}$ component does not becomes negative for the large black hole. This means that there is no expansion in the fluid on the boundary around the equator, as for the $AdS$ soliton case and the polarized black holes described in \cite{PBH}. We also checked that our numerical solutions obey the conservation equation (\ref{conservationT}) with a $1\%$  precision with respect to $T_\phi^{\,\phi}$.


\begin{figure}[h!]
\centering
\subfloat[]{
\includegraphics[width=52mm]{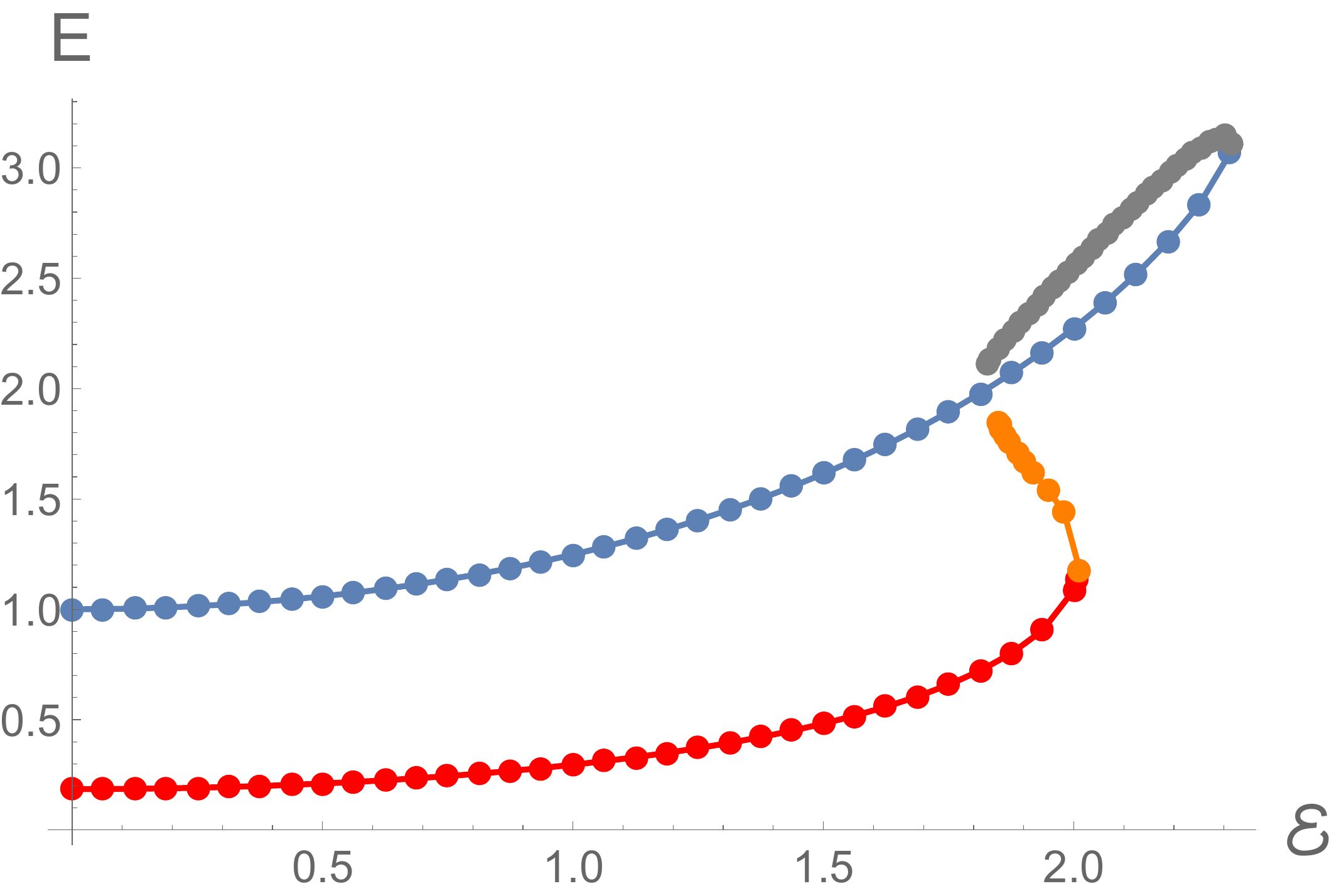}
}
\subfloat[ ]{
\includegraphics[width=52mm]{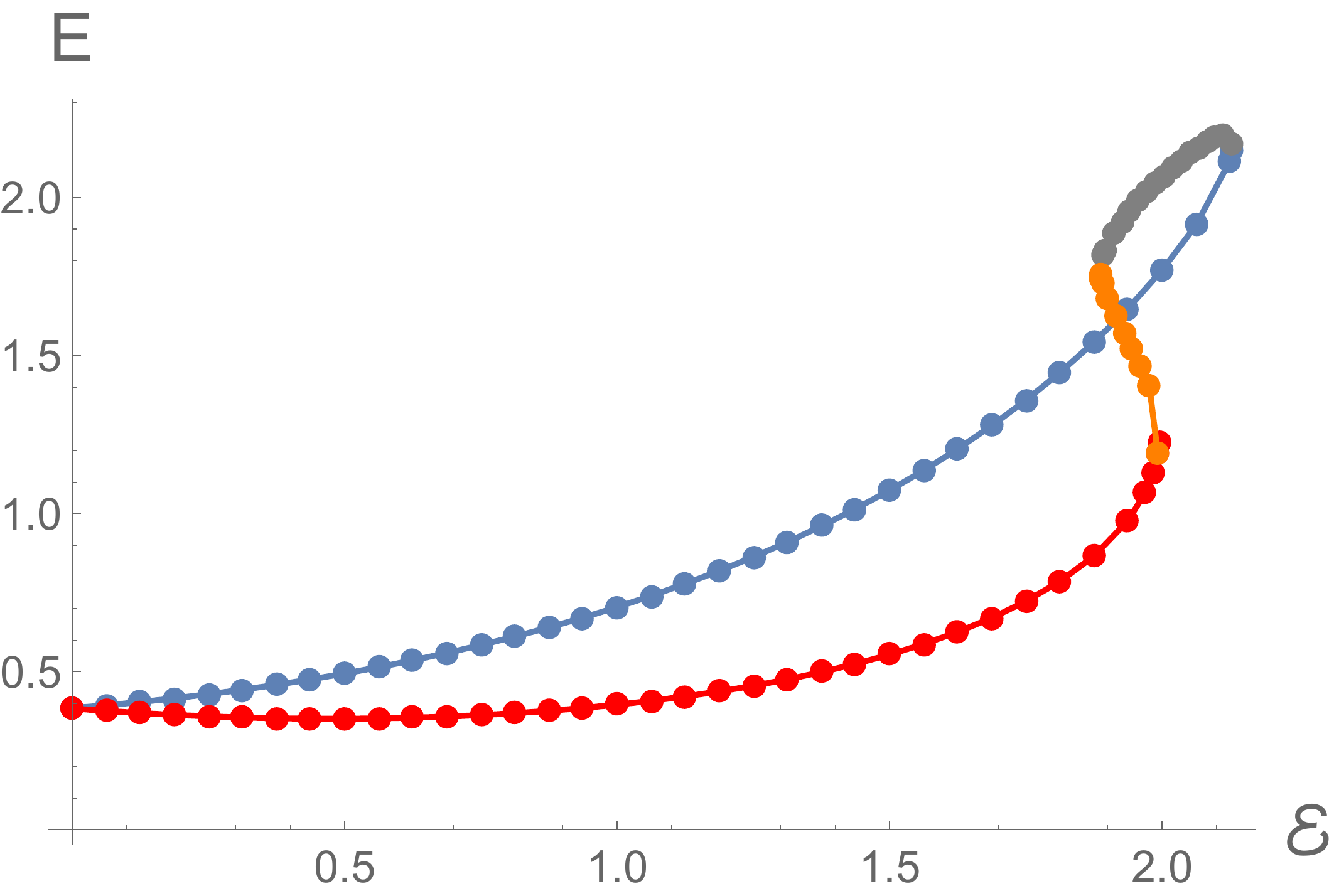}
}
\subfloat[ ]{
\includegraphics[width=52mm]{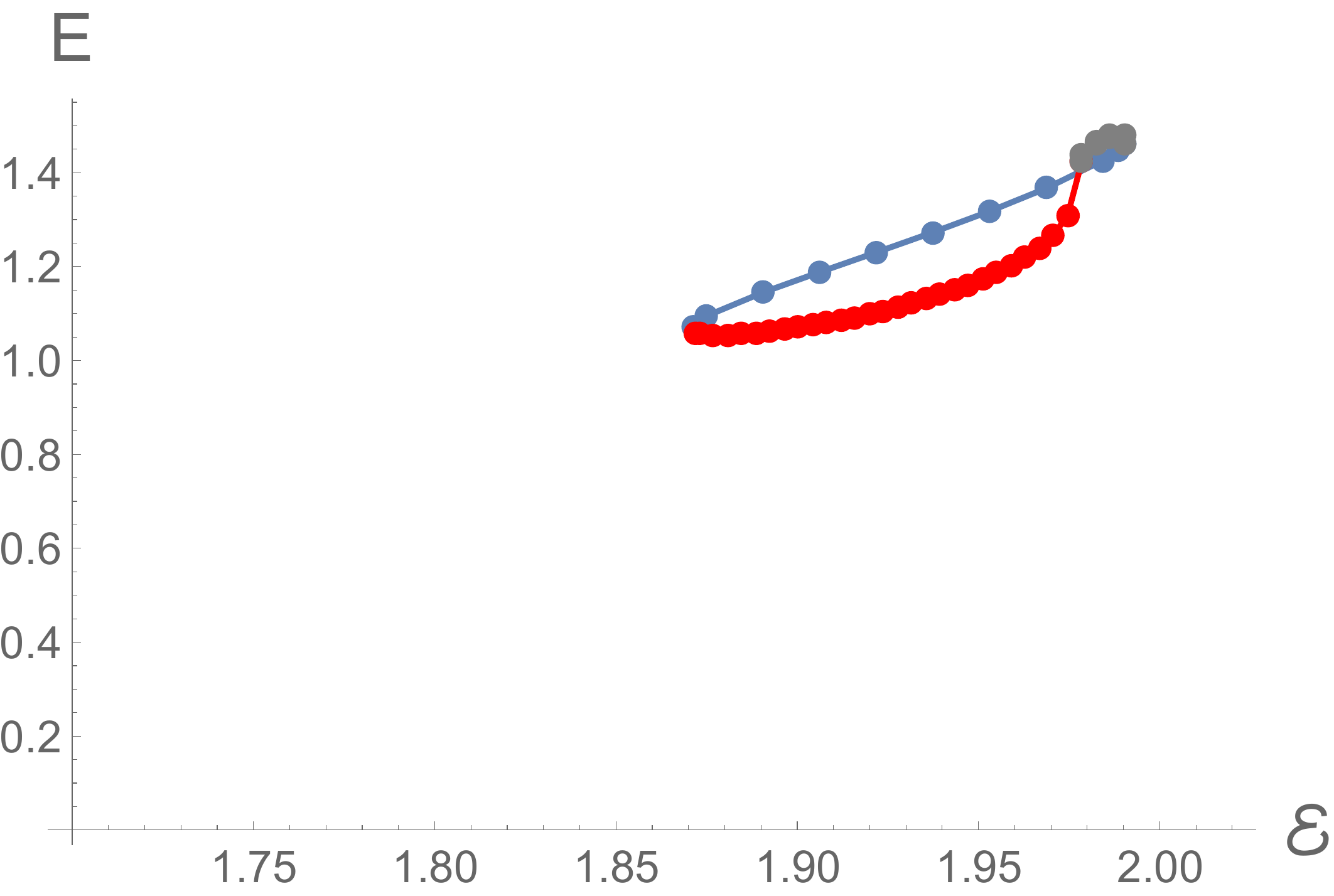}
}
\caption{Total energy at the boundary for the dual state of the black hole  for $G_N=1$ at $T=1/\pi$, $T=0.275665$ and $T=0.24179$ from left to right.
\label{totalE}}
\end{figure}
By integrating the energy density we obtain the total energy measured at infinity. This is plotted in figure \ref{totalE}. 

\section{Thermodynamics}

Defined with Euclidean signature, the $AdS$ soliton and black hole geometries of the previous two sections are already primed to study the thermodynamic properties of the boundary theory. The black hole geometries correspond to phases that depend on electric field magnitude, and on temperature that fixes the periodicity of the thermal circle, so that the solutions are regular at the Euclidean horizon. The soliton solution exists for any temperature and depends only on $\mathcal{E}$. These phases are in thermodynamic competition with each other, and by comparing their free energies we will be able to draw the corresponding phase diagram. 

\begin{figure}[t!]
\centering
\subfloat[]{
\includegraphics[width=50mm]{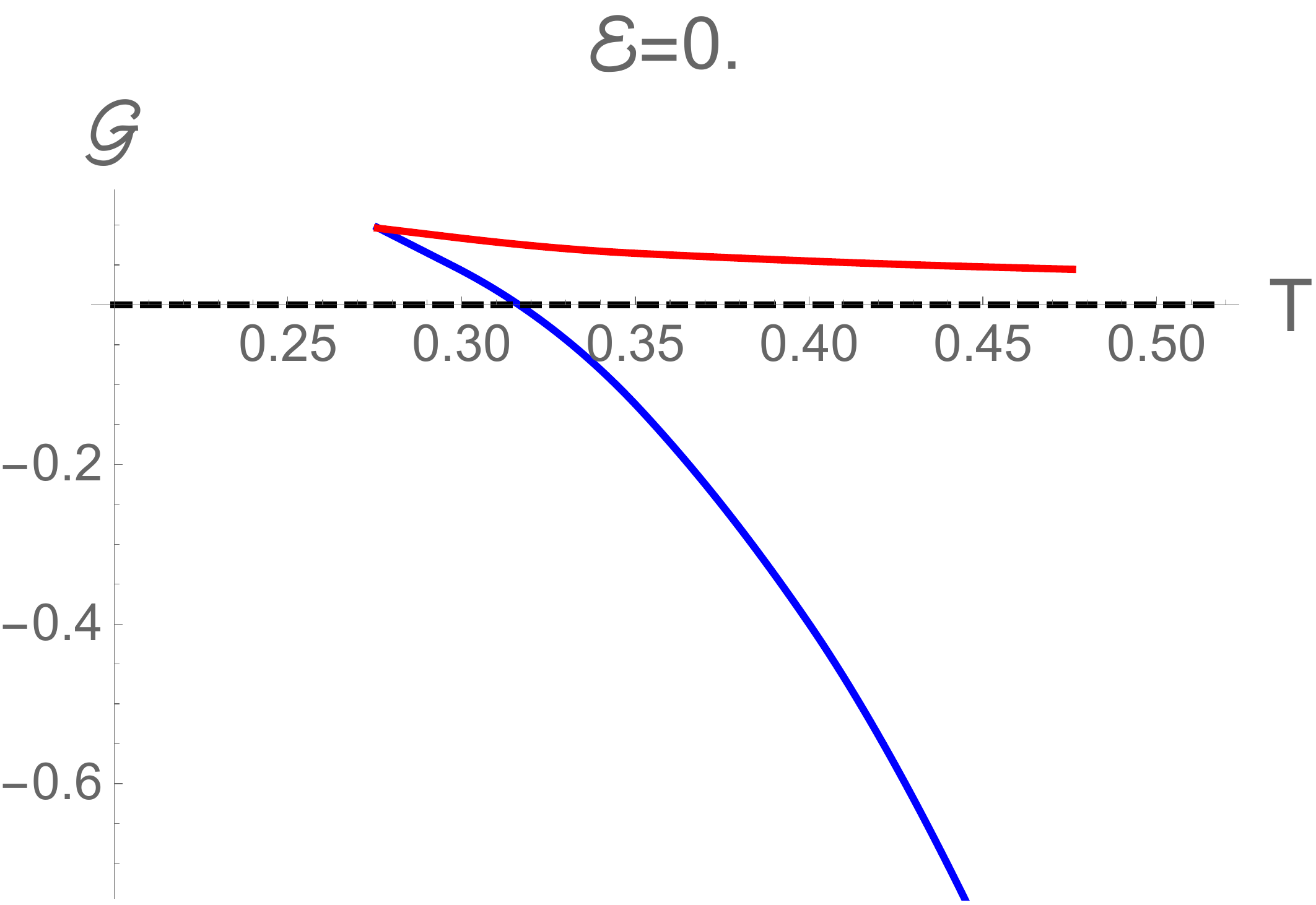}
}
\subfloat[]{
\includegraphics[width=50mm]{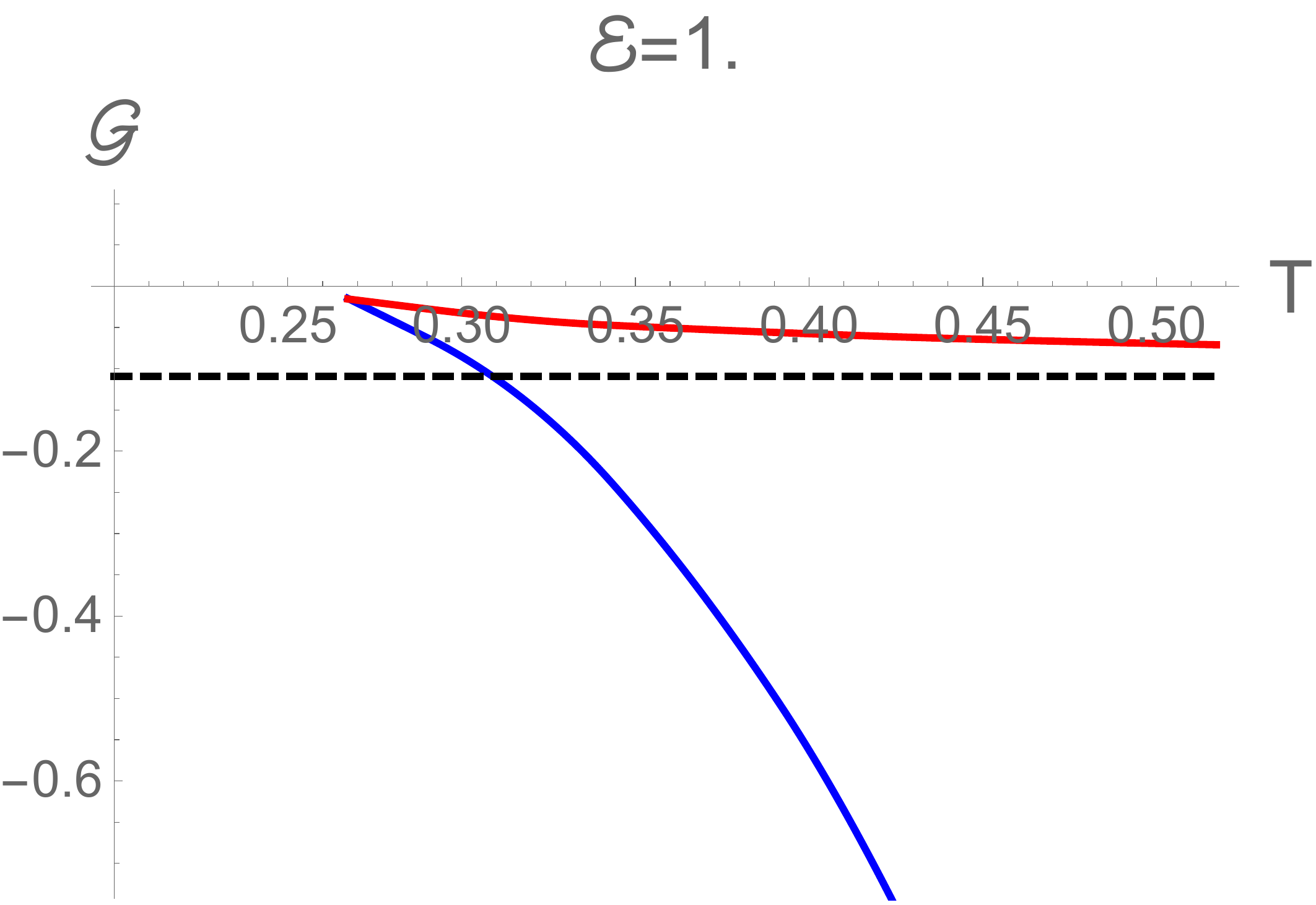}
}
\subfloat[]{
\includegraphics[width=50mm]{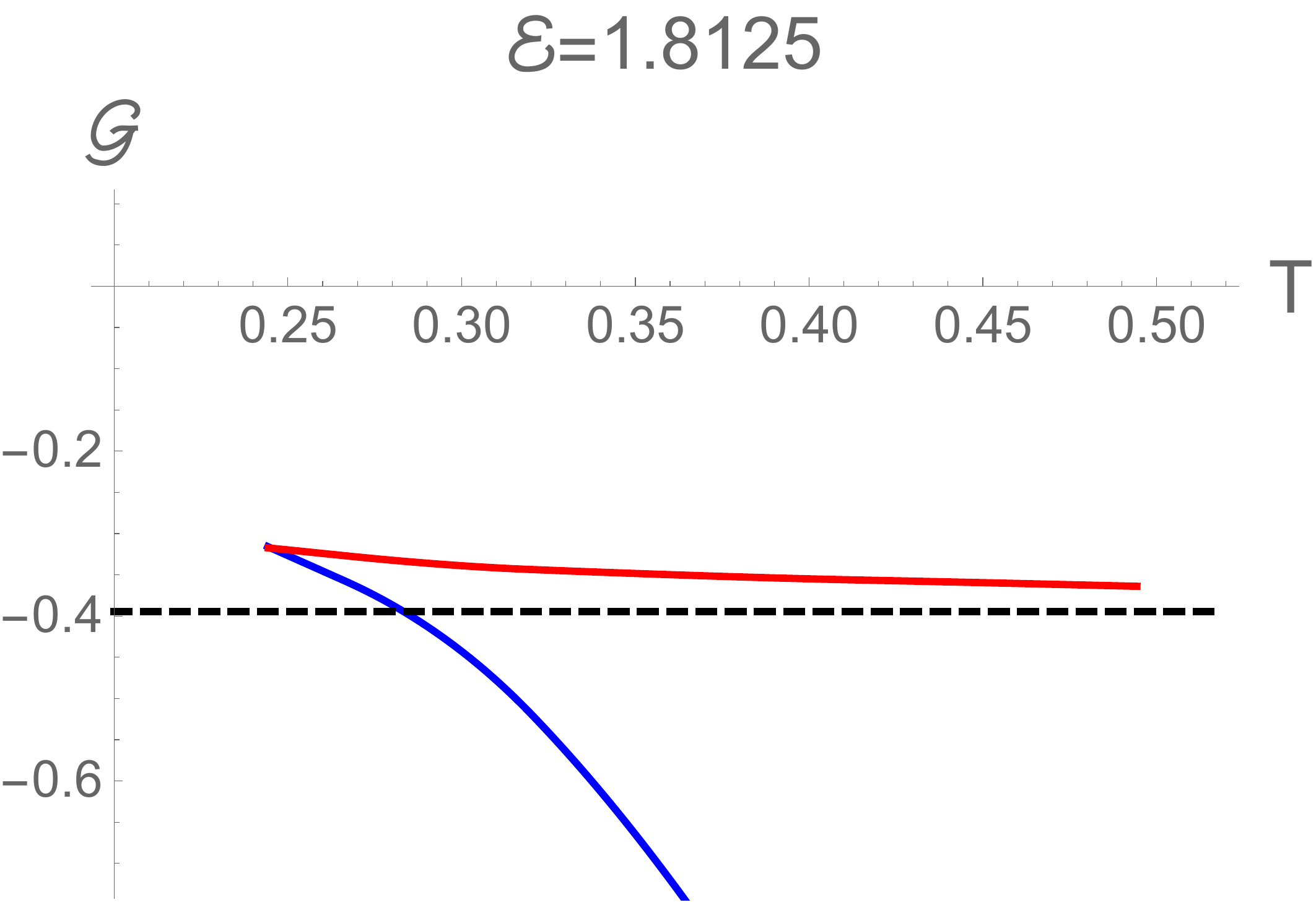}
}
\\
\subfloat[]{
\includegraphics[width=50mm]{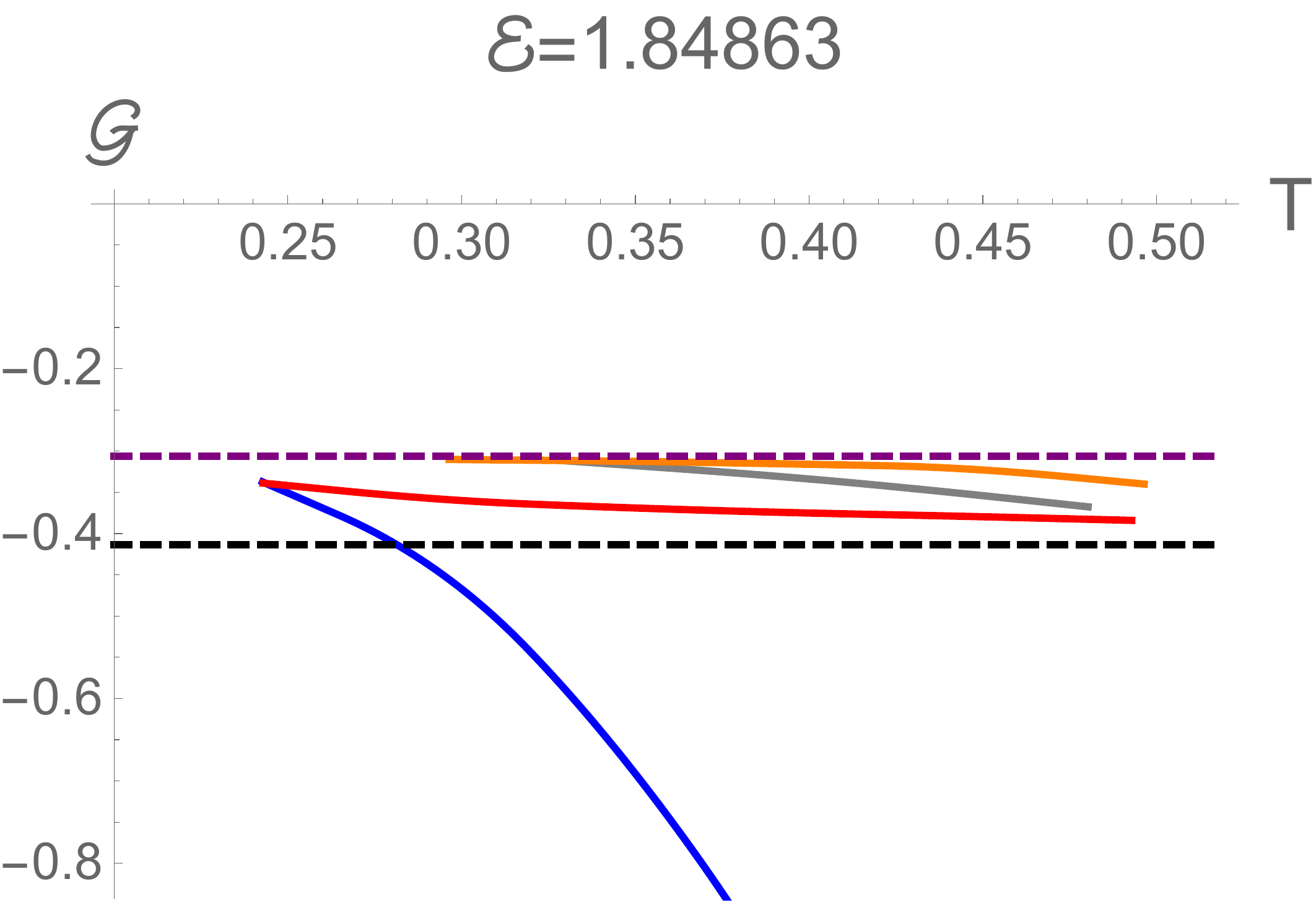}
}
\subfloat[ ]{
\includegraphics[width=50mm]{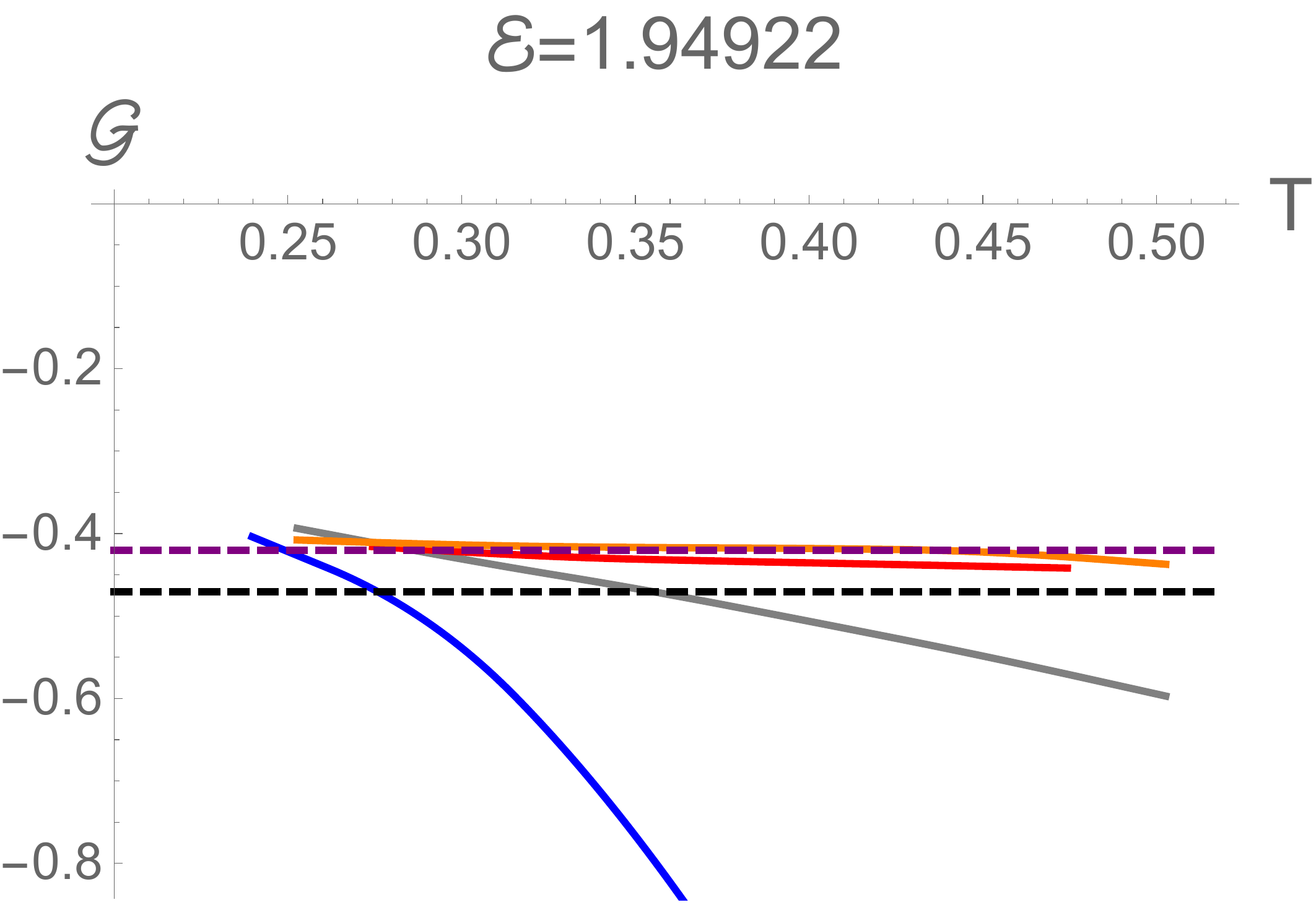}
}
\subfloat[ ]{
\includegraphics[width=50mm]{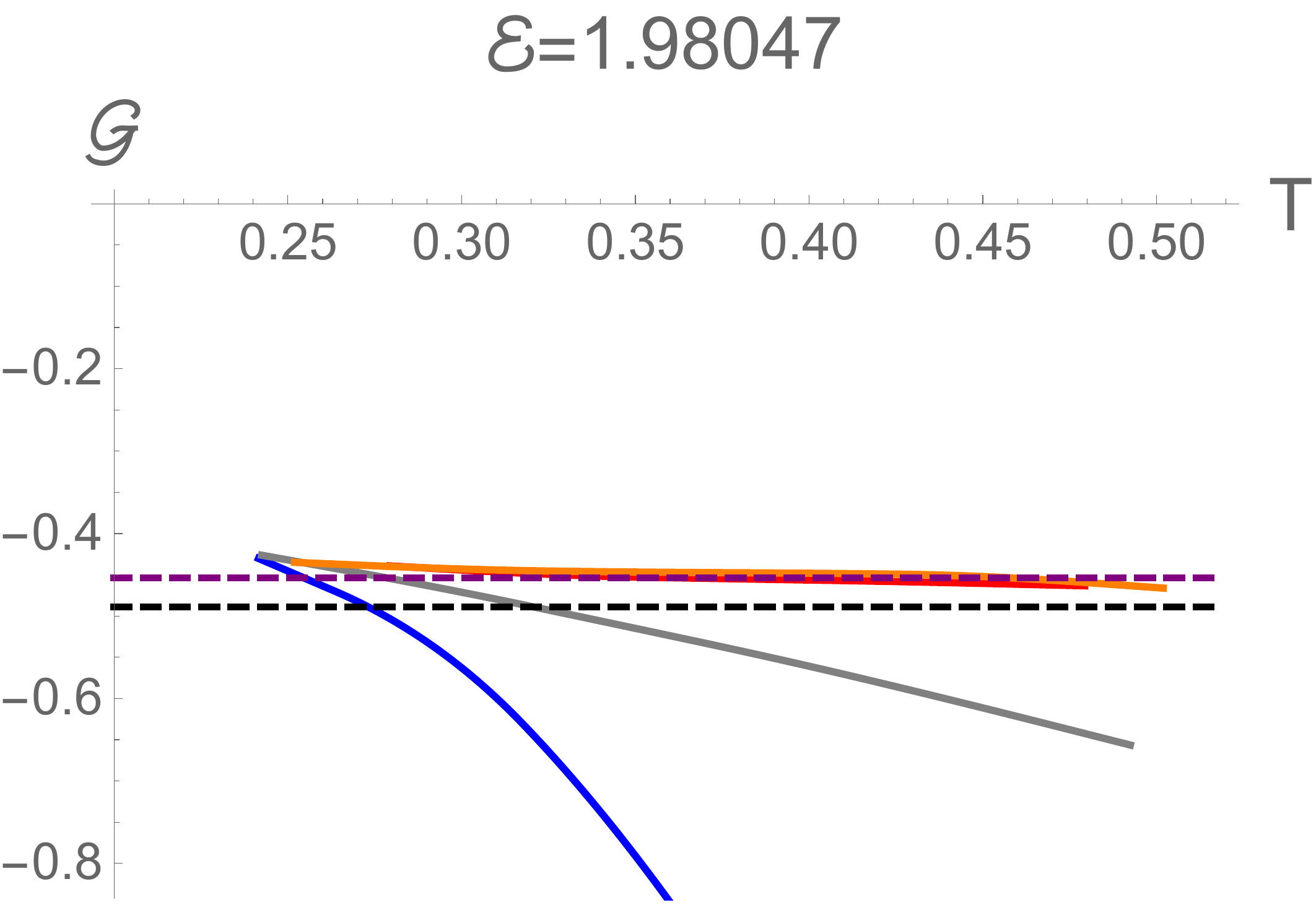}
}
\\
\subfloat[]{
\includegraphics[width=50mm]{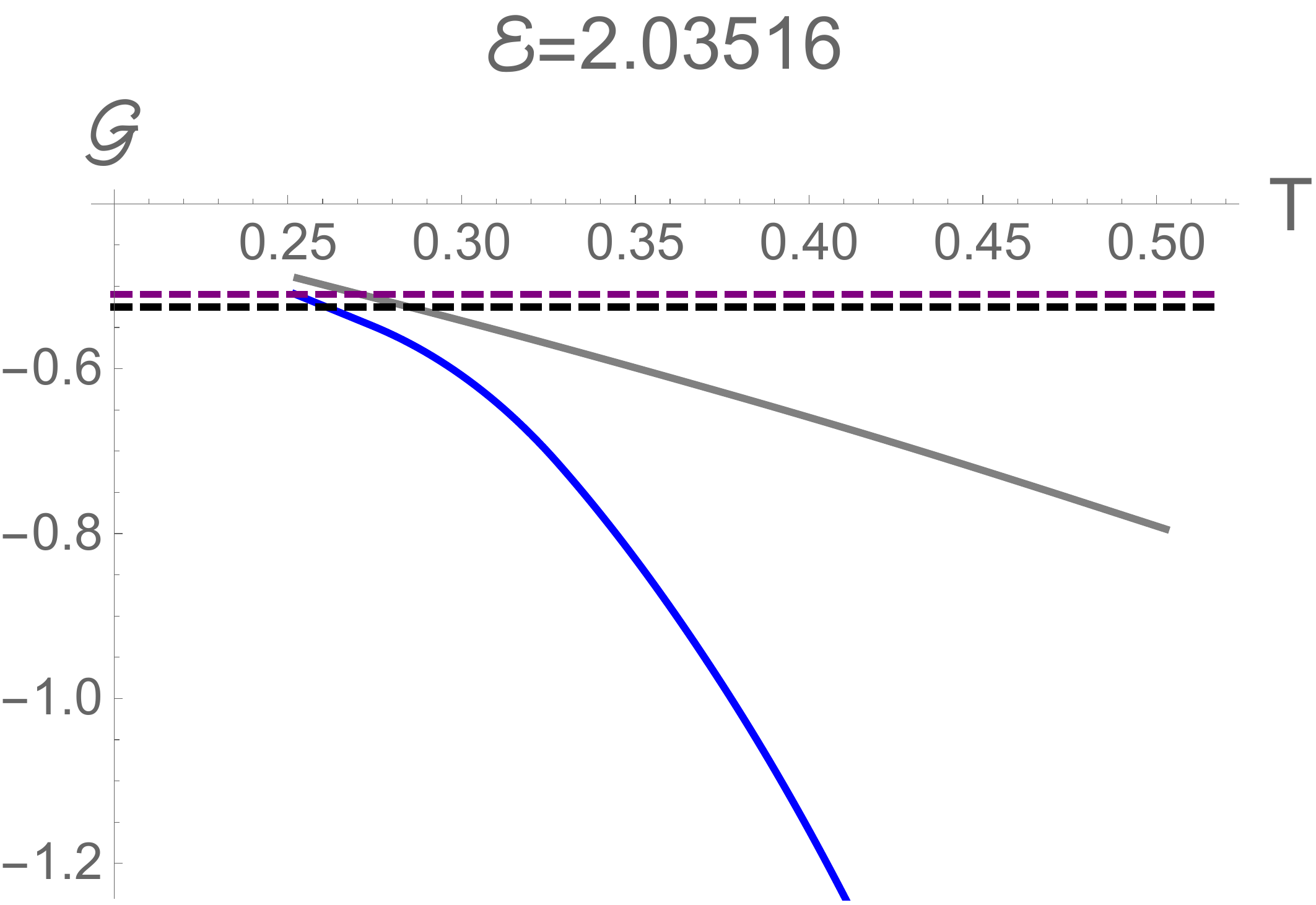}
}
\subfloat[ ]{
\includegraphics[width=50mm]{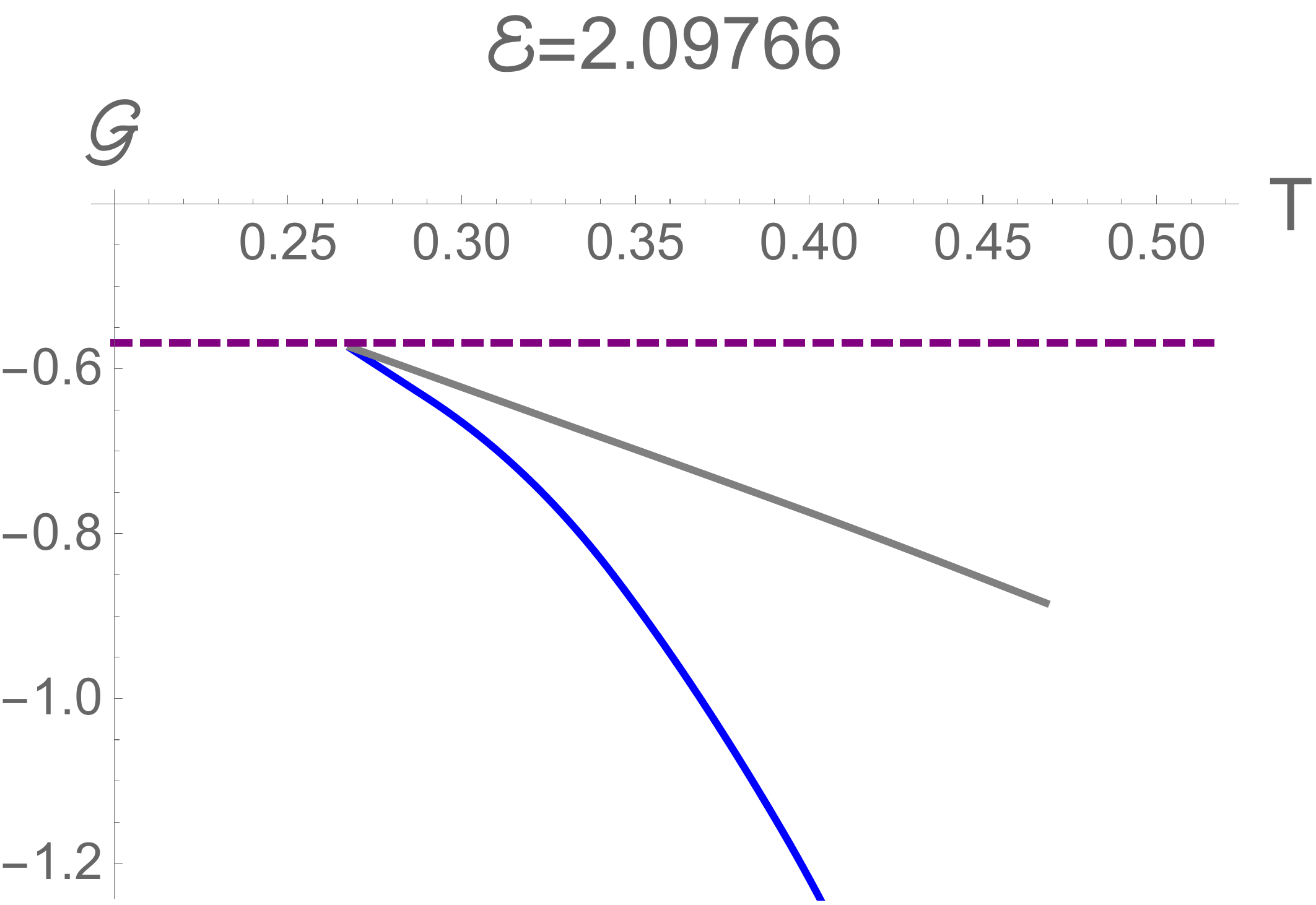}
}
\subfloat[ ]{
\includegraphics[width=50mm]{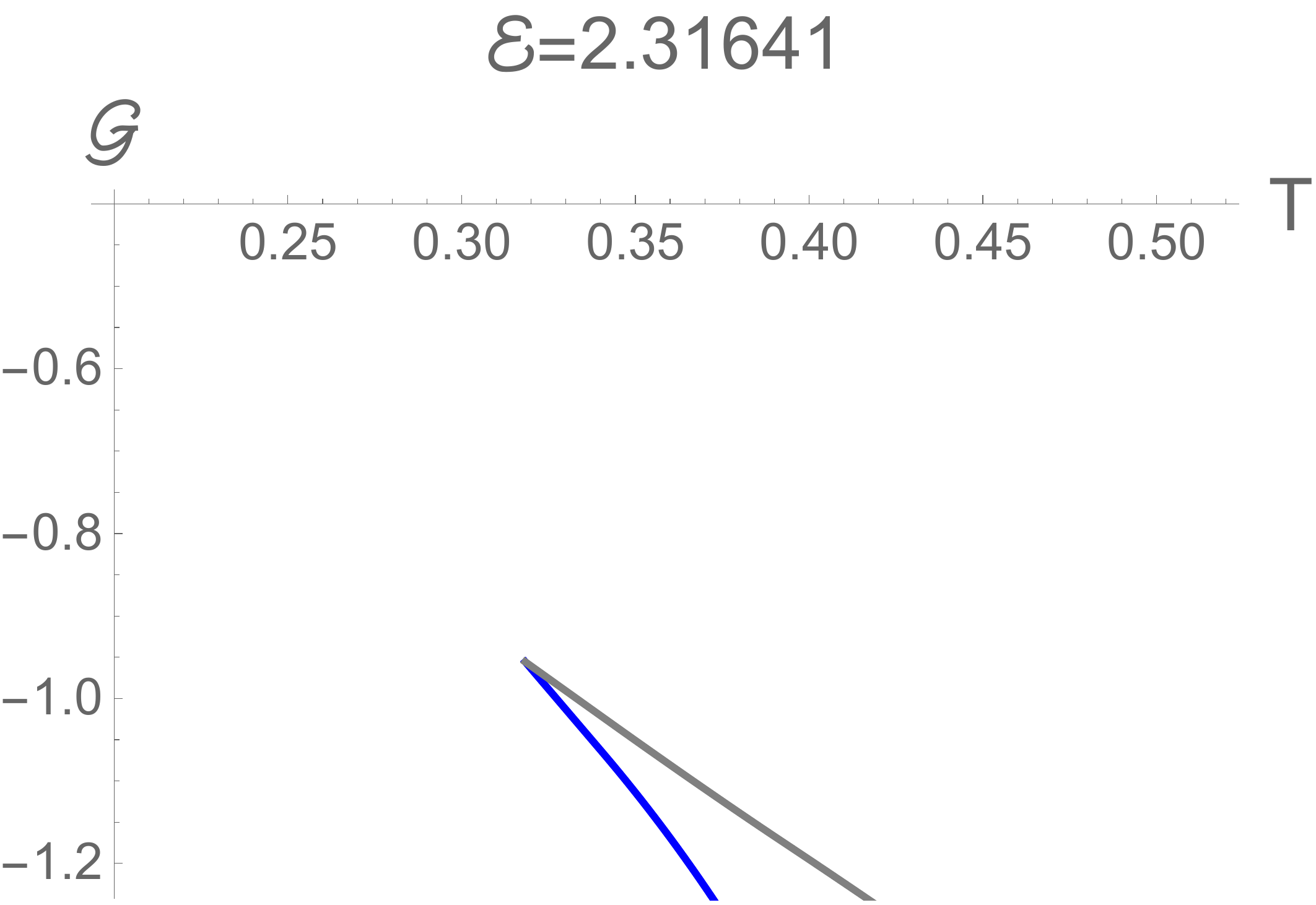}
}
\caption{Gibbs free energy for L1, L2, S1, and S2 black hole branches (blue, gray, red, and orange curves), and $AdS$ soliton (black and purple dashed line) for several values of the electric field as a function of the temperature. In these plots we set $G_N=1$.
\label{free2dABJM}}
\end{figure}

The Gibbs free energy associated to a geometry with a dipolar electrostatic source at the boundary is 
\be
{\cal G}= E-TS- \pi \int_0^{\pi}  d\theta \sin \theta\,\rho(\theta) \, \mathcal{E}\cos\theta \,,
\ee
where $E$ is the energy, $S$ is the Bekenstein-Hawking entropy and $\rho(\theta)$ is the charge density at the boundary. For the $AdS$ soliton solutions without horizon, the entropy term vanishes and the free energy is independent of the temperature. The free energy is shown in figure \ref{free2dABJM}. The L1, L2, S1, and S2 black holes are depicted by blue, gray, red, and orange curves, respectively. The two soliton branches are shown by black and purple dotted lines. It is clear from these plots that the L1 and black soliton branches are the two with lowest free energy and are therefore the only two that matter for the phase diagram. We will therefore restrict our attention to these phases. 

\begin{figure}[t!]
\centering
\includegraphics[width=80mm]{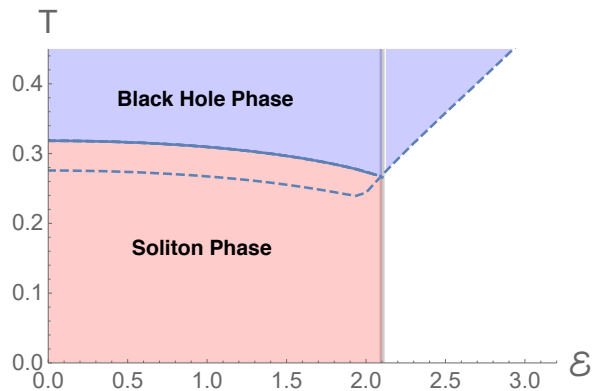}
\caption{Phase diagram with the critical temperature (solid blue curve) above which the black hole phase is thermodynamically favoured. The dashed line shows the minimum temperature of the L1 black hole, and the vertical gray line marks the maximum electric field for the soliton.
 \label{phasediagramABJM}}
\end{figure}

Below the maximum value of the electric field for the soliton $\mathcal{E}_{c}^{Sol}$, the phase transition occurs when the blue and black curves of figure  \ref{free2dABJM} cross. This begins with the Hawking-Page phase transition at $T=1/\pi$ for $\mathcal{E}=0$ and $T_c(\mathcal{E})$ decreases with $\mathcal{E}$ until $\mathcal{E}=\mathcal{E}_{c}^{Sol}$. Beyond this value, the L1 black hole is the black hole phase with lowest free energy, and no soliton solution exists. These results are summarised in the phase diagram of figure \ref{phasediagramABJM}. The blue region corresponds to the black hole phase while the red region corresponds to the $AdS$ soliton phase. The solid blue line marks the phase transition up to $\mathcal{E}_{c}^{Sol}$, shown here as a vertical gray line. The blue dashed curve shows the minimum temperature for the L1 black hole as a function of the electric field. The black hole phase technically exists and is thermodynamically stable for $T>T_{min}$, even for $\mathcal{E}$ greater than the maximum value for the soliton. 
The situation is very similar to the case of charged black holes dual to SYM with an R-charge chemical potential \cite{Yamada:2007gb}.
We expect that the black holes are  metastable for $\mathcal{E}>\mathcal{E}_{c}^{Sol}$ and that strictly speaking the canonical ensemble is only well defined for $\mathcal{E}<\mathcal{E}_{c}^{Sol}$. This phase diagram is qualitatively similar to those produced in \cite{Basu:2016mol} for the case of spherically symmetric Einstein-Maxwell-dilaton gravity in global $AdS$.

%

In order to verify our numerical solutions, we can derive a first law of thermodynamics that encodes the conservation of energy of the black hole and soliton systems. 
 For the soliton, the energy responds to small changes in $\mathcal{E}$. We have
\be
\delta_\mathcal{E}E- \pi\, \mathcal{E} \int_0^{\pi}  d\theta \sin \theta\, \cos\theta\,\delta_\mathcal{E}\rho(\theta) =0\,.
\ee
This is satisfied to $10^{-2}$ on our numerical solutions and analytically to fourth order in $\mathcal{E}$. 
For the black hole, the energy at infinity, entropy, and charge density all respond to variation in the temperature. At fixed electric field, the first law for the black hole can be written
\be
\delta_T E- T\delta_T S -\pi\, \mathcal{E} \int_0^{\pi}  d\theta \sin \theta\, \cos\theta\,\delta_T\rho(\theta)=0\,.
\ee
This is also satisfied to $10^{-2}$ on our numerical solutions.
\subsection{Free ABJM with external electric field} \label{ABJM}

The ABJM theory has global $SU(4)\times U(1)$ $R$-symmetry, which is dual to a local symmetry in the gravity description. The truncation of this gravity theory to $U(1)^4$  considers the three $U(1)$'s inside the $SU(4)$. In particular, the  $U(1)$ gauge field that we turned on in the bulk is dual to a global current associated to one of the $U(1)$'s of the $SU(4)$ R-symmetry. To understand the effect of turning on 
a source for this $U(1)$ global current, recall that the ABJM theory has four complex scalars and four Majorana spinors, which transform in the fundamental of the $SU(4)$ R-symmetry.
Our choice of $U(1)$ corresponds to the generator $Q={\rm diag}(1,-1,0,0)$ of $SU(4)$.
Thus we  have one scalar and one fermion with charge $+1$, one scalar and one fermion with charge $-1$, two neutral scalars and two neutral fermions.
We conclude that, at zero coupling and finite external source, the problem of computing the partition functions reduces to that of analysing one complex 
scalar field and one Majorana fermion with conformal coupling on $\mathbb{R}\times S^2$, minimally coupled to the external electric field. 
The full partition function can then be computed by considering the others scalars and fermions that do not couple to the electric field and by considering that all
fields are in the fundamental times anti-fundamental of the gauge symmetry $U(N)\times U(N)$.
Then, the gauge-invariant states can be written as products of traces of products of pairs of elementary fields.

The single particle states of a charged scalar on the two-sphere in the presence of the dipolar electrostatic potential were computed in \cite{PBH}. 
%
Using a basis of spherical harmonics the hamiltonian 
is diagonal in the azimuthal quantum number $m$ but it becomes an infinite tridiagonal matrix in the quantum number $l\ge |m|$,
\be
\langle l',m'|H|l,m\rangle=\delta_{m,m'}\left\{ \delta_{l,l'}\left(l+\frac{1}{2} \right)+{\cal E}\left[\delta_{l,l'-1}\,\sqrt{\frac{(l+1-m)(l+1+m)}{(2l+1)(2l+3)}}+(l\leftrightarrow l')\right]\right\} .
\label{Hamiltonian}
\ee
 

We shall now study the case of a free charged fermion on a two sphere with a dipolar potential, following the logic of \cite{lapsphere}.
The Dirac equation reads
\be
\left(i\hat{\nabla}-M \right)\psi =\of{ i e^\alpha_a \gamma^a \Big(D_\alpha+\frac{i}{4}\omega_\alpha^{ij}\sigma_{ij}\Big)-M}\psi=0\,,
\label{eq:Dirac}
\ee
where $e^\alpha_a={\rm Diag}\left( 1, 1, 1/\sin\theta  \right)$ is the zweibein on the two-sphere and 
$\omega$ is the spin connection.  
Here the covariant derivative 
includes the gauge field as $D_{\alpha}=\partial_{\alpha}-iC_{\alpha}$.
The irreducible representation of Majorana spinors in $SO(1,2)$ is two-dimensional, so the $\gamma$-matrices reduce to Pauli matrices. With Lorentzian $(-,+,+)$ signature, we choose ($\{ \gamma^a,\gamma^b\} =2\eta^{ab}$) 
\ba
\gamma^0&=&i \sigma_3\, ,\quad\gamma^1=\sigma_1\, ,\quad\gamma^2=\sigma_2\, ,\quad\\
\sigma_{ab}&=&-\frac{i}{2}\sqof{\gamma^a,\gamma^b},
\ea 
and we use latin indices in the tangent space. 

Next we consider the particular case of an external electric field for which $C=\mathcal{E} \cos\theta dt$. In this case 
the Dirac operator becomes
\be
\hat{\nabla} = \gamma^0 \left( \partial_t - i  \mathcal{E} \cos\theta\right) + \gamma^1 \left(  \partial_\theta +\frac{\cos\theta}{2\sin\theta} \right) 
+ \frac{\gamma^2}{\sin\theta} \,\partial_\phi\,.
\ee
Doing the usual Fourier decomposition of the spinor components
\be
\psi = e^{-i(\omega t - m \phi)}  \left(
\begin{array}{c}
\psi_+ \\
\psi_-
\end{array}
\right) ,
\ee
for a half integer $m$ and where 
$\psi_\pm$ are functions of $\theta$, the Dirac equation  (\ref{eq:Dirac})  becomes
\be
\big( M\mp i(\omega + \mathcal{E} \cos\theta) \big) \psi_\pm =i \left(  \partial_\theta +\frac{\cos\theta\pm2m}{2\sin\theta} \right) \psi_\mp  \,.
\ee
As usual, it is convenient to square the Dirac equation by multiplying (\ref{eq:Dirac}) by $(i\hat{\nabla}+M)$. This looks like 
a Klein Gordon equation 
$(\hat{\nabla}^2+M^2)\psi=0$, with
\be
\hat{\nabla}^2=
(\omega+  \mathcal{E} \cos\theta)^2  +i \gamma^2  \mathcal{E} \sin\theta 
+\frac{1}{\sin\theta} \left( \partial_\theta \sin\theta \partial_\theta\right) -\frac{1}{4} 
-\frac{1}{4\sin^2\theta} \left(1 +4m^2 +i \gamma^0 4m \cos\theta  \right) .
\ee
Notice that the second term in this operator has a $\gamma^2$ matrix and therefore is not diagonal.

Let us first consider the case of zero electric field. In this case the Klein-Gordon equation is diagonal.
In terms of the coordinate $x=\cos \theta$, the functions $\psi_\pm(x)$ satisfy the differential equation
\be
\left(  \partial_x \left( (1-x^2) \partial_x\right) - \frac{1}{4(1-x^2)}\left( 1+4m^2 \mp 4m x \right) 
+ \omega^2 +M^2 -\frac{1}{4} \right) \psi_\pm(x)=0\,.\label{EOMsq}
\ee
Notice that the equations for $\psi_+$ and $\psi_-$ can be interchanged by sending $x\rightarrow -x$, and that equation (\ref{EOMsq}) is singular at the poles $x=\pm1$. After the redefinition
\ba
\psi_\pm=(1\mp x)^\frac{\alpha}{2}(1\pm x)^\frac{\beta}{2} Y_\pm\,,
\ea
with $\alpha$ and $\beta$ greater than zero for regularity, the equation can be written in hypergeometric form 
\be
(1-x^2)Y_\pm''+\big(\pm \sgn(m)-2(1+|m|)x\big)Y_\pm'+l(l+2|m|+1)Y_\pm=0\,,
\ee
for $\alpha=|m-1/2|$, $\beta=|m+1/2|$, and $\omega^2+M^2=(l+|m|+1/2)^2$. Again, we will use the conformal value for the fermion mass, which  is $M=0$.  The latter constraint on $\omega$ ensures the square integrability of solutions on the interval $x \in [-1,1]$.  Solutions to this equation are Jacobi polynomials of order $l\ge0$
\ba
Y_+&=&a_{l,m}P_l^{(\alpha,\beta)}(x)\,,
\\
Y_-&=&b_{l,m}P_l^{(\beta,\alpha)}(x)\,,
\ea
where the coefficients $a_{l,m}$ and $b_{l,m}$ are related by the Dirac equation and fixed by the normalisation condition 
\be
\int_0^{2\pi}d\phi \int_{-1}^1dx \,\psi^\dagger \psi=1\,.
\ee
This gives 
\be
b_{l,m}=- \sgn(m)\, a_{l,m}\, \quad |a_{l,m}|=\frac{ \sqrt{l! \,\Gamma (l+2 |m|+1)}}{2^{|m|+1}\sqrt{\pi}\,\Gamma \left(l+|m|+\frac{1}{2}\right)}.
\ee

\begin{figure}[t!]
\centering
\subfloat[]{
\includegraphics[width=60mm]{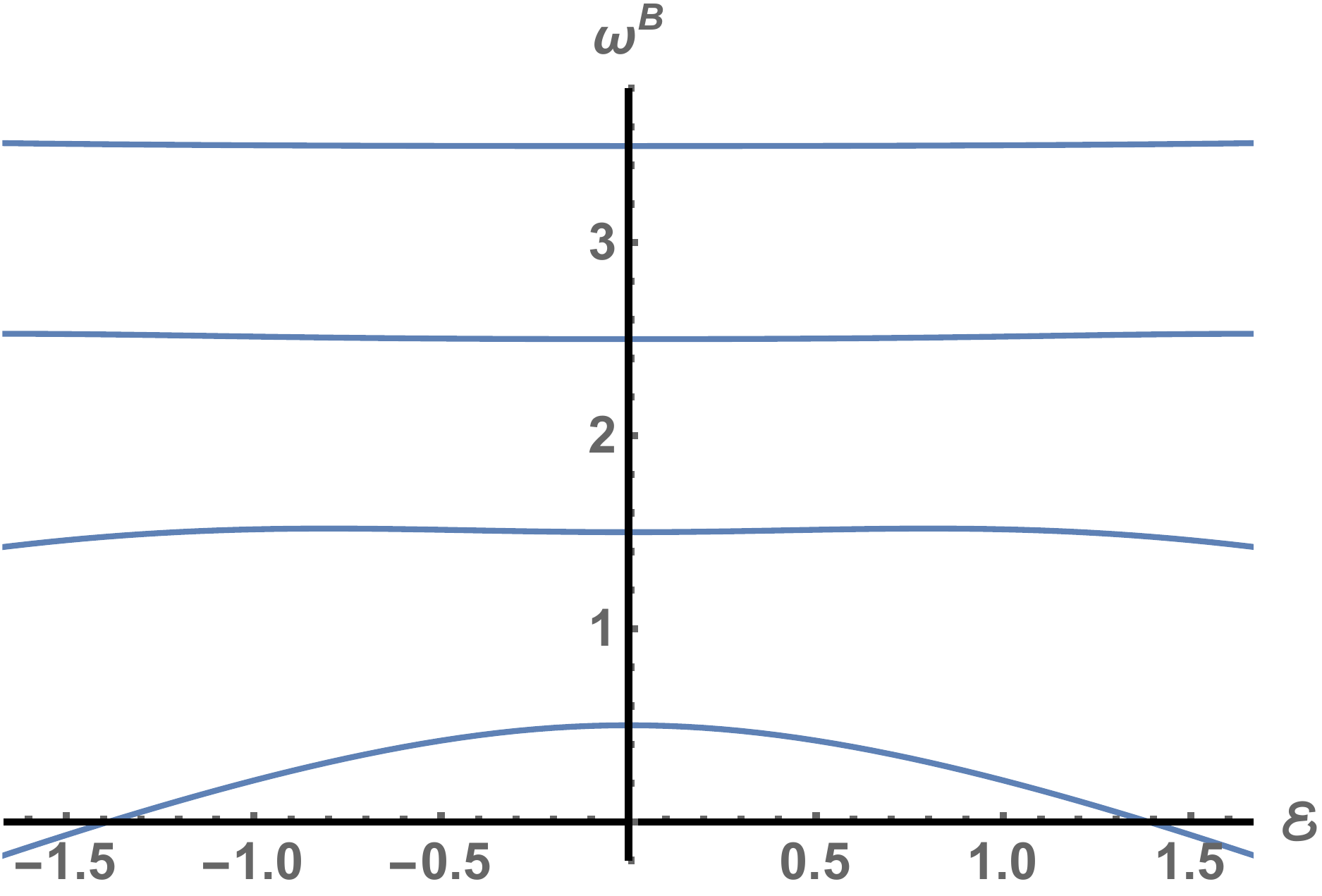}
}
\hspace{2mm}
\subfloat[]{
\includegraphics[width=60mm]{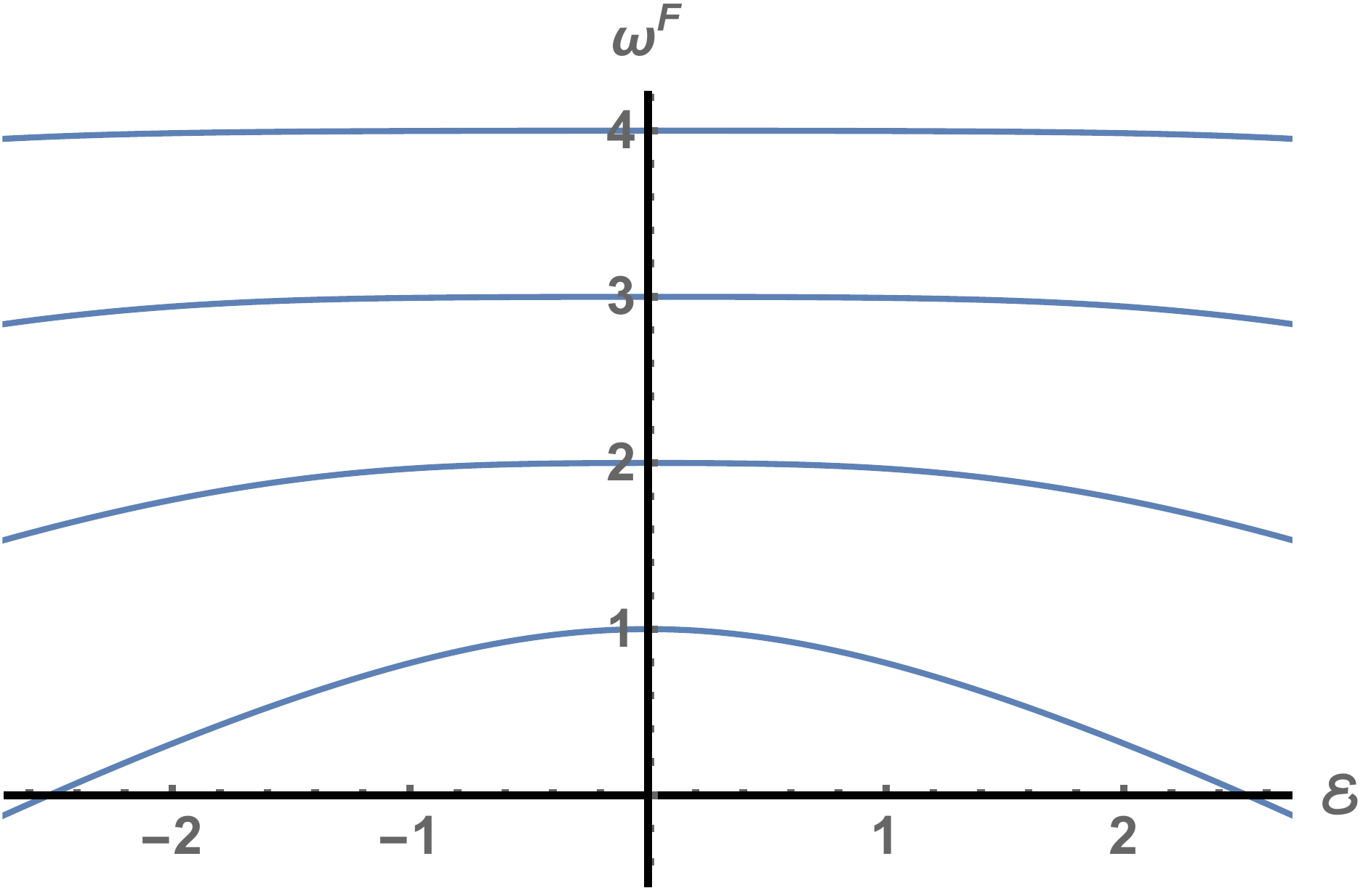}
}
\caption{The first four energy levels for (a) the $m=0$ sector of the free boson, and (b) the $m=1/2$ sector of the free fermion, as a function of the dipolar potential $\cal E$.
 \label{fig:Fermion}}
\end{figure}

We can now consider the Hamiltonian 
\be
\hat{H}=i\partial_t=-\mathcal{E} x +i \gamma^0\hat{\nabla}_{S^2}
\ee
where $\hat{\nabla}_{S^2}$ is the Dirac operator on the two-sphere. The eigenvalues for this operator were found above, and are given by $\omega_{l,m}=l+|m|+1/2$, with $l=0,1,2,\dots$ and $m\in \mathbb{Z}+1/2$.
On a energy eigenstate for zero electric field $\psi_{l,m}$, the Hamiltonian acts 
\be
\hat{H}\psi_{l,m} = -\mathcal{E} x \,\psi_{l,m} + \omega_{l,m}\psi_{l,m}\,.
\ee
We can compute the matrix elements of $\hat{H}$ in this basis
\begin{align}
\langle  l',m'|\hat{H}|l,m\rangle &=     \int_0^{2\pi}d\phi\int_{-1}^1 dx\, \psi_{l',m'}^\dagger \hat{H} \psi_{l,m}\\
&=  \delta_{m,m'}\left[\omega_{l,m}\delta_{l,l'}-2\pi\mathcal{E}\left(\frac{2^{2 \left| m\right| } \Gamma \left(l+\left| m\right| +\frac{1}{2}\right) \Gamma \left(l+\left| m\right| +\frac{3}{2}\right)}{l! (2 \left| m\right| +l )! (\left| m\right| +l+1) }\right. \delta_{l',l+1}\right.\\
&\qquad \qquad + \left. \left.
\frac{2^{2 \left| m\right| } \Gamma \left(l+\left| m\right| -\frac{1}{2}\right) \Gamma \left(l+\left| m\right| +\frac{1}{2}\right)}{(l-1)! (2 \left| m\right| +l-1 )! (\left| m\right| +l) }
 \delta_{l',l-1}\right)a_{l,m}a_{l',m}\right],
\end{align}
which simplifies to
\be
\langle l',m'|\hat{H}|l,m\rangle=\delta_{m,m'}\left\{ \delta_{l,l'}\left(l+|m|+\frac{1}{2} \right)-{\cal E}\left[\delta_{l,l'-1}\,\frac{\sqrt{(l+1)(l+2|m|+1)}}{2(l+|m|+1)}+(l\leftrightarrow l')\right]\right\} .
\nonumber
\label{HamiltonianFermion}
\ee
We denote the eigenvalues of this hamiltonian by $\omega^F_{m,k}(\mathcal{E})$ with $m \in \mathbb{Z}+1/2$ and $k=1,2,\dots$. The resulting low energy spectra of the free boson for $m=0$ and free fermion for $m=1/2$ are shown in figure \ref{fig:Fermion}. Notice that in the bosonic sector, for ${\cal E}>{\cal E}^B_c\approx 1.3868$ the single particle ground state energy becomes negative, while for the fermion sector, this critical value is ${\cal E}^F_c\approx 2.5183$. 

We can now compute the single-particle boson and fermion partition functions. At finite temperature, these are defined by
\ba
z_B(x,{\cal E})= \sum_{m\in \mathbb{Z}} \sum_{k=1}^\infty e^{-\beta \omega^B_{m,k}({\cal E})}\,, 
\qquad  
z_F(x,{\cal E})= \sum_{m\in \mathbb{Z}+\frac{1}{2}} \sum_{k=1}^\infty e^{-\beta \omega^F_{m,k}({\cal E})}\,, 
\qquad  
x\equiv e^{-\beta}\,.
\ea
Following \cite{hep-th/9908001, hep-th/0110196, hep-th/0310285, Nishioka:2008gz} and using the relevant charges for our case, the Hagedorn temperature is determined by
the condition
\be
2z_B(x_H,{\cal E})+2z_F(x_H,{\cal E})+2z_B(x_H,0)+2z_F(x_H,0)=1\,,\qquad\qquad x_H=e^{-\beta_H}\,,
\ee
where we used the fact that the single particle partition functions are even functions of $\cal{E}$.

\begin{figure}[t!]
\centering
\subfloat[ ]{
\includegraphics[width=70mm]{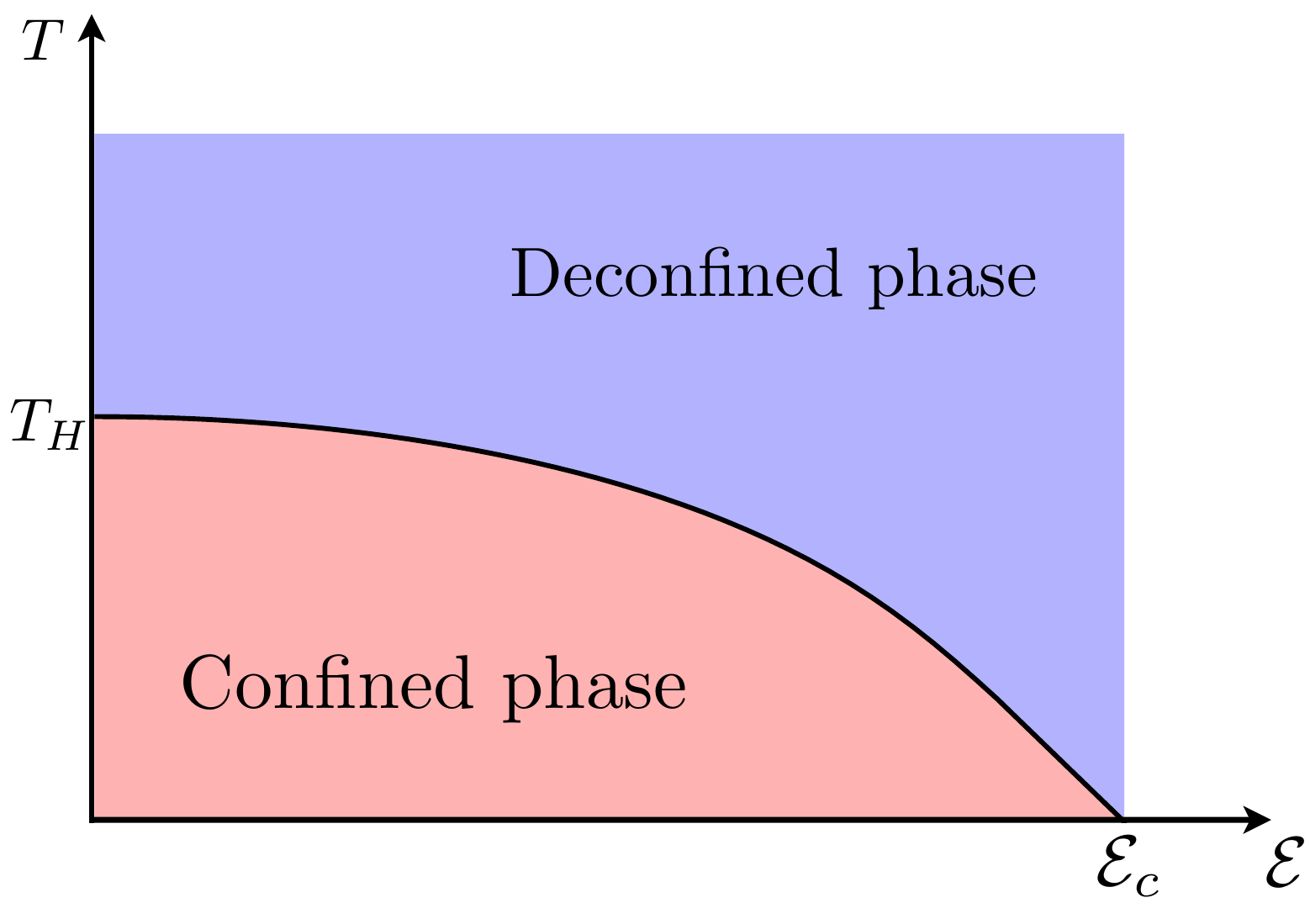}
}
\caption{ Large $N$ phase diagram of a theory with a free adjoint scalar and fermion field. The Hagedorn temperature decreases with   the dipolar potential ${\cal E}$ and goes to zero as ${\cal E}\to{\cal E}_c^B$.
 \label{fig:Hagedorn}}
\end{figure}
In Figure \ref{fig:Hagedorn}, we plot the Hagedorn temperature as a function of the electric field.
The black curve marks a phase transition.
As expected, there is a low temperature confined phase and a high temperature deconfined phase separated by a Hagedorn phase transition that starts at $T_c\approx 0.304836$ at $\mathcal{E}=0$ and goes to zero as ${\cal E}\to{\cal E}^B_c$, meaning that the addition of fermions does not change the value of the maximum electric field.
For ${\cal E}>{\cal E}^B_c$ the canonical ensemble does not exist.

\section{Conclusion}
We have shown that there exist asymptotically $AdS$ geometries coupled to a neutral scalar that are polarised by a dipolar electric field. There are two soliton and four black hole phases for a range of electric field values that depend on the temperature.

\begin{figure}[b!]
\centering
\subfloat[]{
\includegraphics[width=55mm]{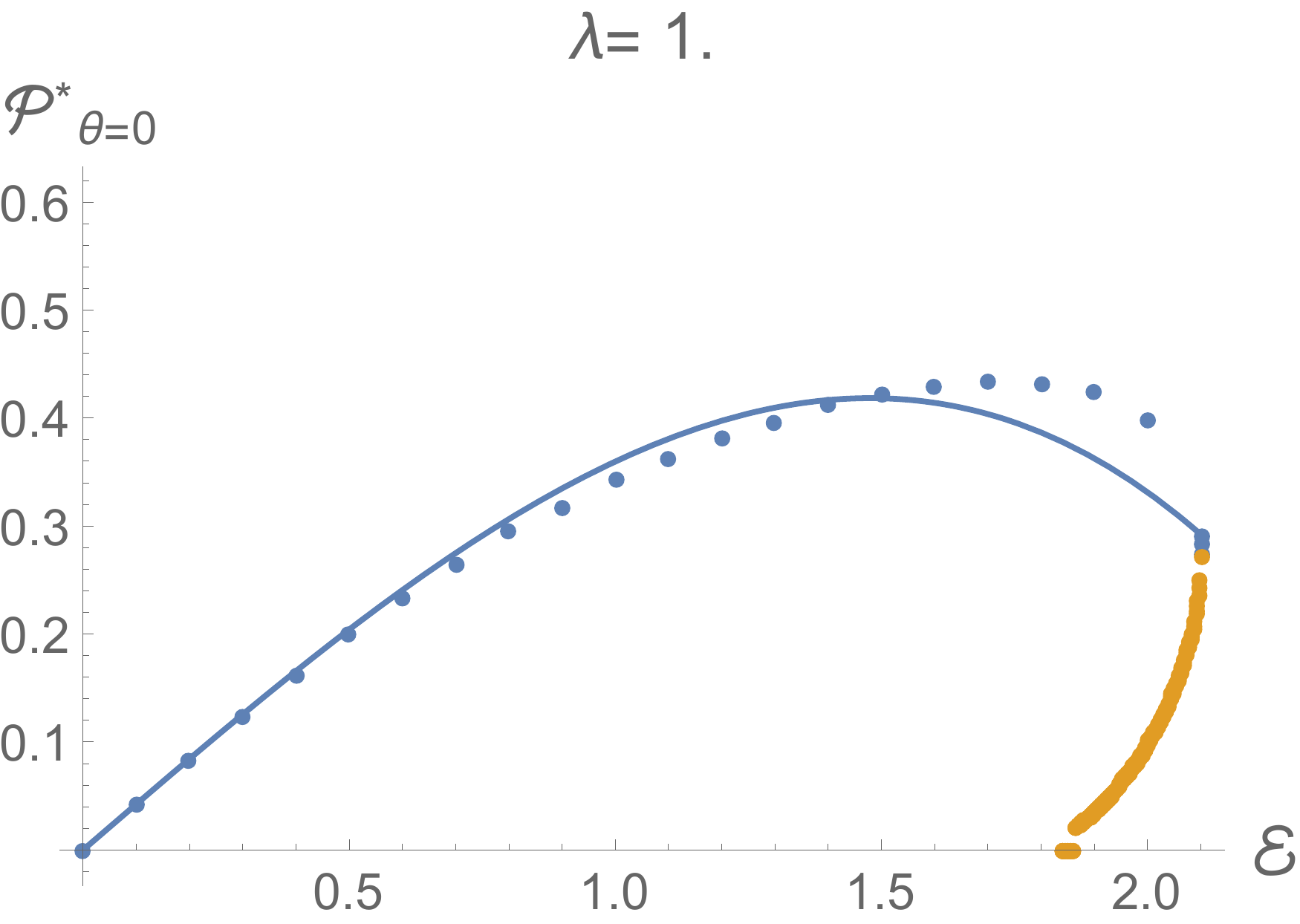}
}
\hspace{2mm}
\subfloat[]{
\includegraphics[width=55mm]{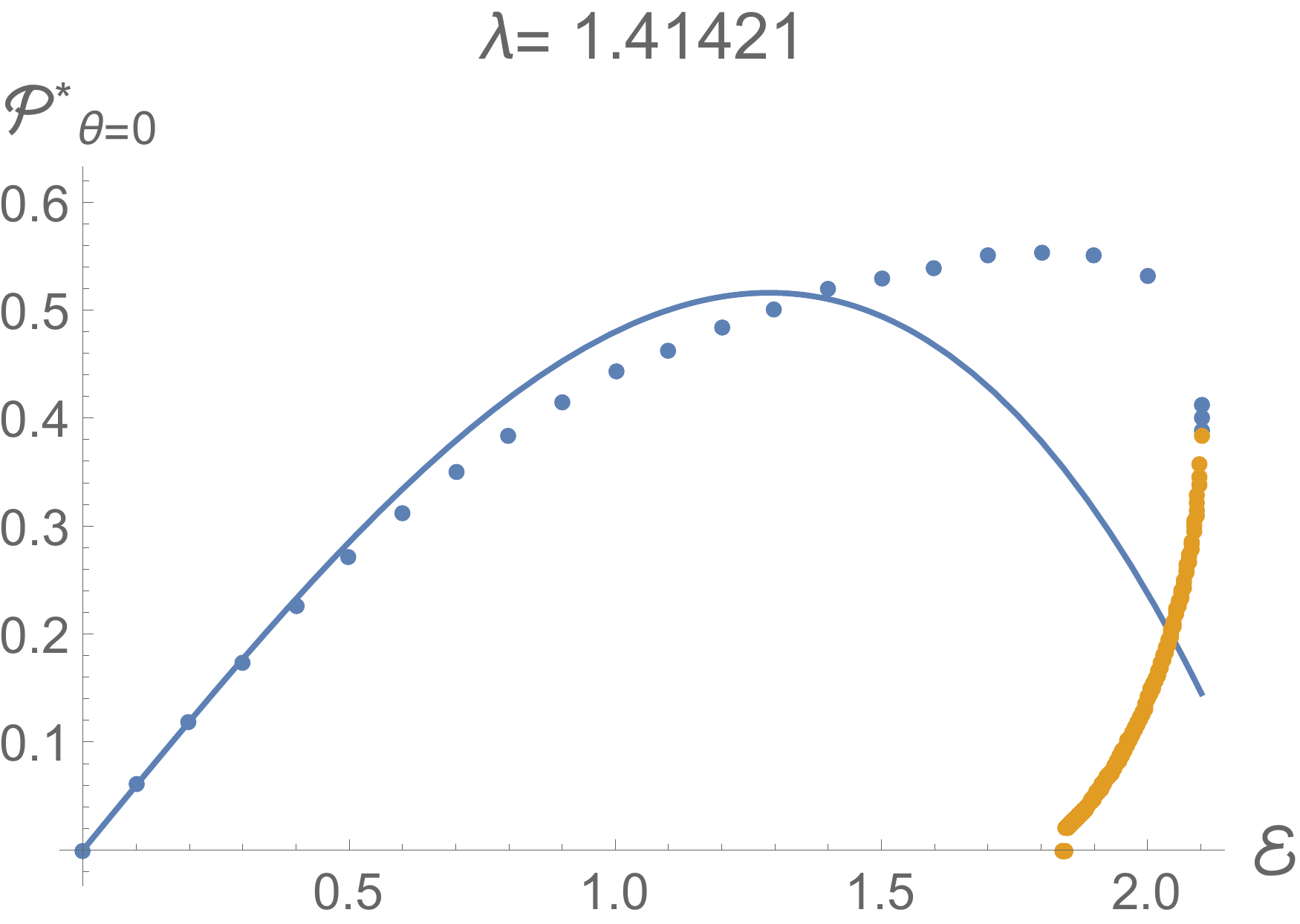}
}
\caption{Timelike static orbits of charged particles for two values of $\lambda=q/m$ as a function of the electric field. 
\label{ptq}}
\end{figure}

\begin{figure}[t!]
\centering
\subfloat[]{
\includegraphics[width=65mm]{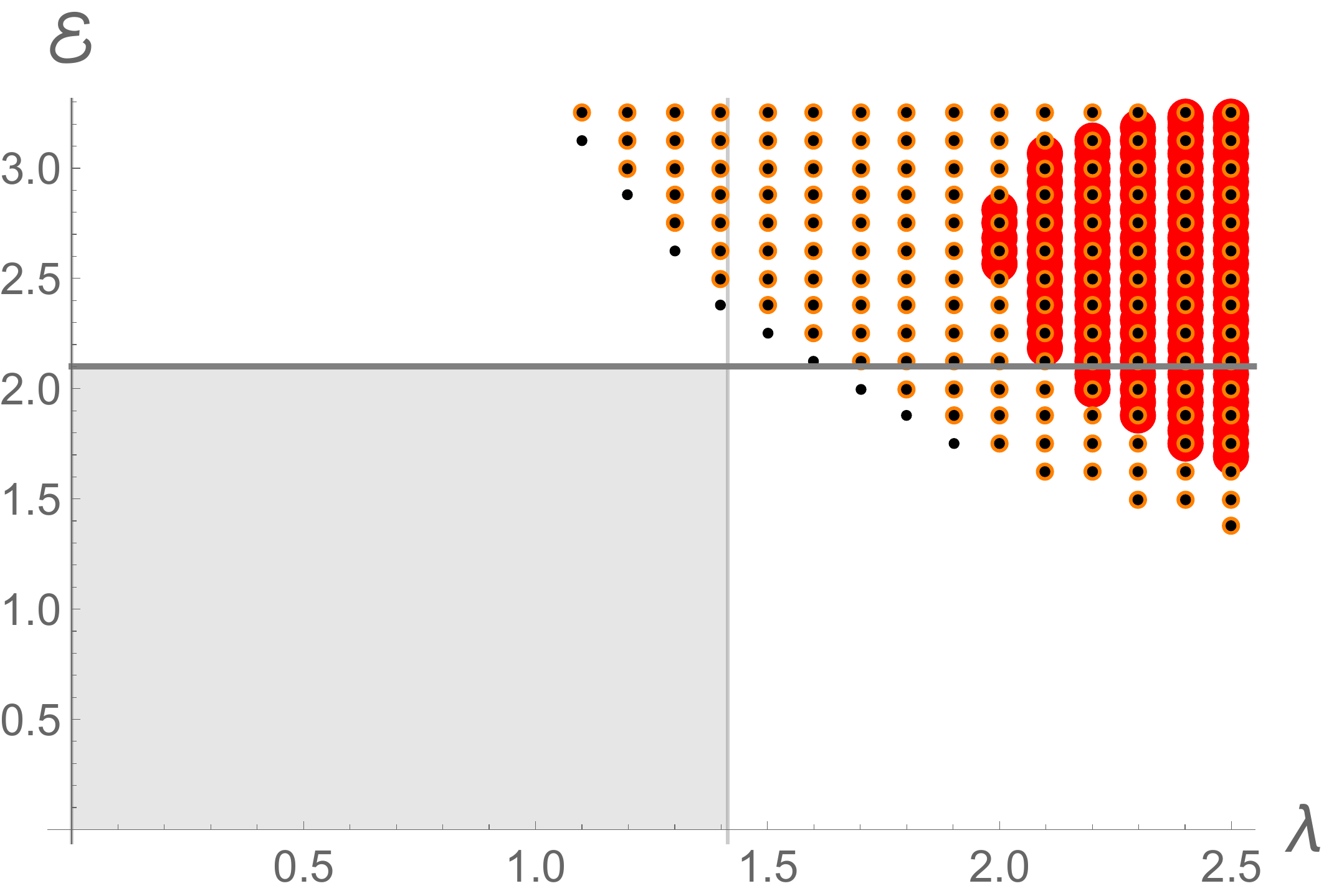}
}
\hspace{2mm}
\subfloat[]{
\includegraphics[width=65mm]{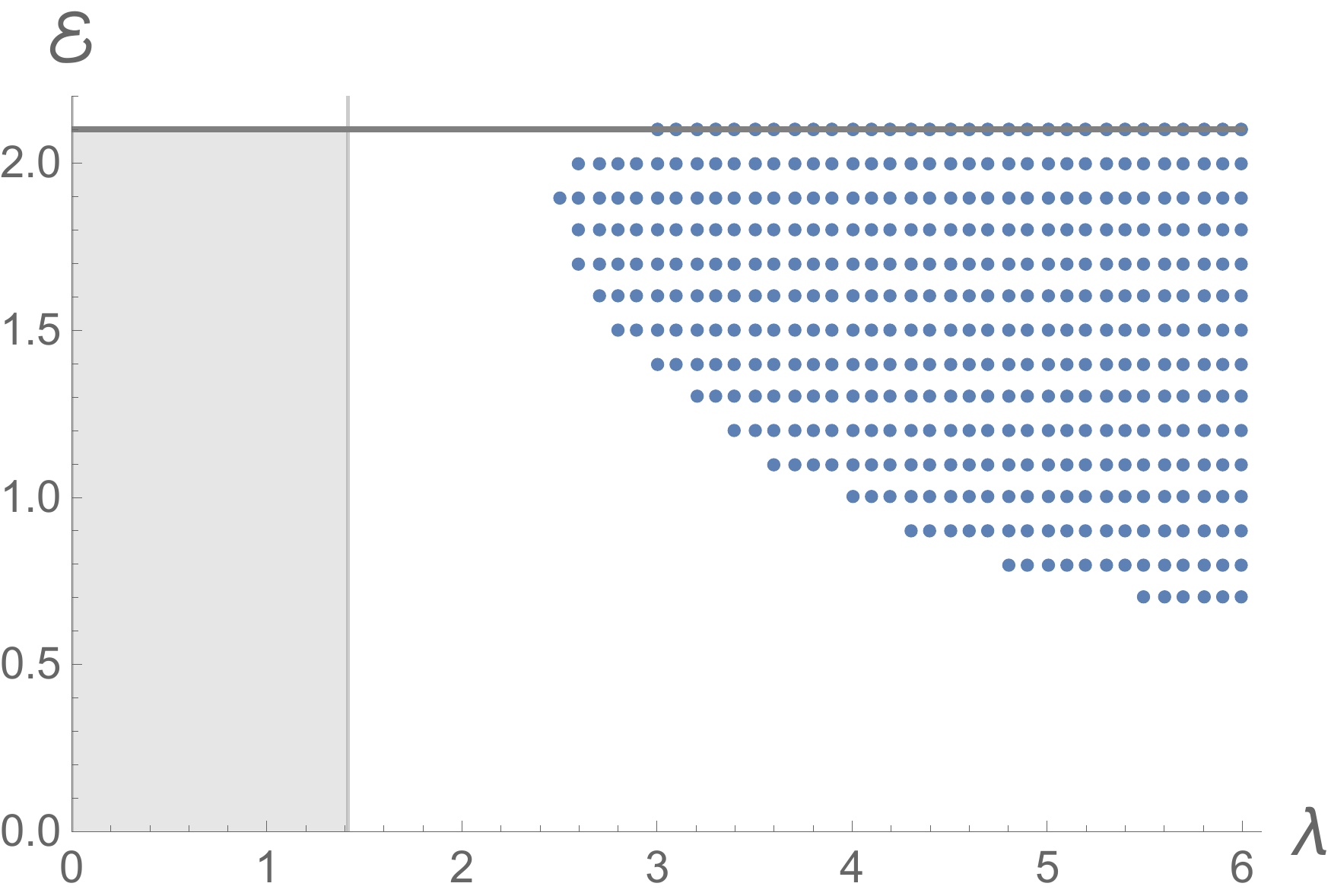}
}
\caption{Stable orbits of pointlike charged particles for (a) the black hole for
$T=0.517254$ (red), $T=1.91981$ (orange), and $T= 9.55129$ (gray) and (b) the soliton. A stable pair of particles that decrease the free energy of the black hole solution would fall in the gray region.
\label{pointlike}}
\end{figure}

It is interesting to ask if oppositely charged pairs of probe particles could exist in these polarised backgrounds. Such particles would sit in timelike static orbits at equilibrium 
positions located at the minima of the potential
\be
V=\sqrt{g_{\tau \tau}}-\frac{q}{m}A_\tau\,.
\ee
Note that the extremal limit of a black hole in flat space corresponds to $\lambda=|q/m|=\sqrt{2}$. 
Therefore, small probe black holes correspond to particles with $\lambda<\sqrt{2}$.
In Figure \ref{ptq} we plot the proper radius $\mathcal{P}^*_{\theta=0}$, along the $\theta=0$ pole, corresponding to stable orbits of charged massive particles in the soliton background, as a function of the electric field. These are shown for $\lambda$  less than and equal to $\sqrt{2}$. The solid blue curve is the analytic result found by the expansion to third order in $\mathcal{E}$ presented in Appendix \ref{App:pert}.
A contribution to the free energy for such point particles is 
\be
\delta \mathcal{G}=m V_{min}\,.
\ee
So adding charged particles becomes thermodynamically favorable for $V_{min}<0$. These exist for certain values of $\mathcal{E}$ and $\lambda$ as shown in figure \ref{pointlike} for the black hole and for the soliton. For the black hole, we plot this for three values of the temperature $T$. As the black hole temperature is increased, the boundary of these regions move toward smaller values of $\lambda$ until it crosses the vertical gray line corresponding to $\lambda=\sqrt{2}$, indicating that there exist probe black holes that form stable orbits in these polarized backgrounds. However, we suspect these to be metastable above $\mathcal{E}_{c}^{Sol}$, drawn as a horizontal gray line. A stable, thermodynamically preferable black hole would lie in the gray region. It is possible that these exist for sufficiently high temperatures, but we do not expect this to be the case. In the polarized soliton backgrounds, no such black hole orbits can exist.

The solutions we have constructed in this paper contain only one of the  neutral scalar fields of ABJM. In general, ABJM contains other fields, including massive charged 
scalars.\footnote{The mass and the charge of these scalars is not arbitrary, instead they are both function of the quantum numbers of the fields with respect to the round $\mathbb{CP}^3$. Furthermore, they can appear from a lower dimensional point of view as a set of complicated coupled equations which only effectively decouple close to the boundary. We will bypass this, and consider a massless charged scalar field as a proxy for the more complicated  cases.} Since the work of \cite{Hartnoll:2008vx,Hartnoll:2008kx,Dias:2010ma,Dias:2011tj,Markeviciute:2016ivy} we know that small and near extremal RN black holes in AdS can become unstable to perturbations governed by charged scalar fields. Motivated by these two facts, we decided to investigate whether charged scalar fields $\tilde{\phi}$ that are minimally coupled to gravity can become unstable. As such, we considered the following
\be
\mathcal{D}^a\mathcal{D}_a\tilde{\phi}=0\,,
\ee
with $\mathcal{D}=\nabla-i\,q\,A$. Since our background admits a Killing vector field $\partial/\partial t$ we can Fourier decompose our perturbations with respect to $t$
\be
\tilde{\phi}(t,r,\theta)=e^{i\omega t}\phi(r,\theta)\,.
\ee
Modes with $\mathrm{Im}(\omega)<0$ grow exponentially with time and are unstable, while modes with $\mathrm{Re}(\omega)>0$ are stable. In this paper we are not interested in the growth rate of these novel hairy solutions, instead we are interested to know where they connect in the moduli space with the ones we constructed. As such, we can set $\omega=0$, and search directly for zero-modes. These turn out to obey a rather simple equation of the form
\be
\nabla^2\phi=q^2A_t^2\phi\,,
\ee
where the metric connection and gauge field $A_t$ are the polarized black hole or soliton geometries constructed above. We solve this generalized eigenvalue problem numerically for $\phi$ and $q$. 

\begin{figure}[t!]
\centering
\subfloat[]{
\includegraphics[width=65mm]{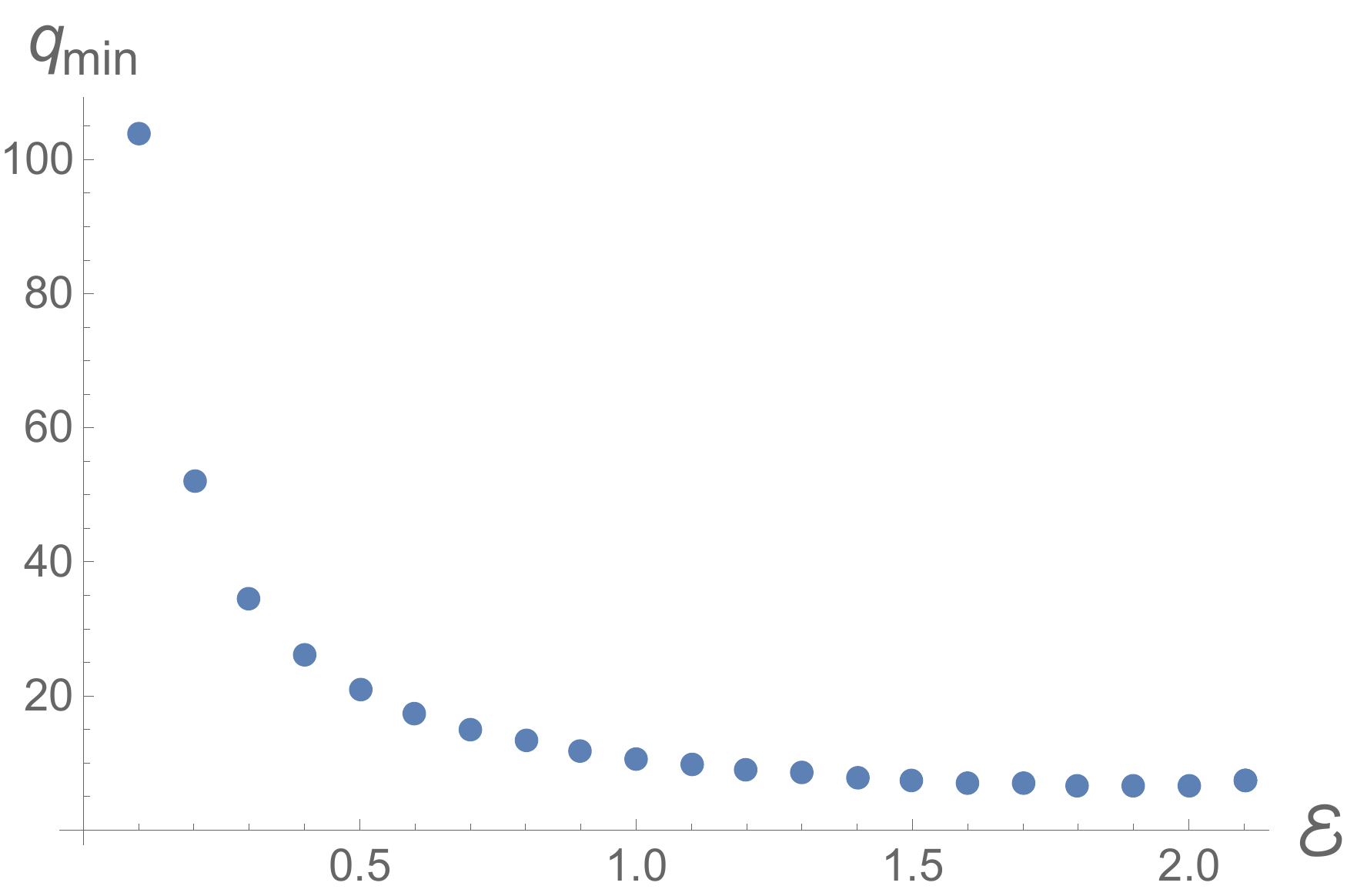}
}
\subfloat[]{
\includegraphics[width=65mm]{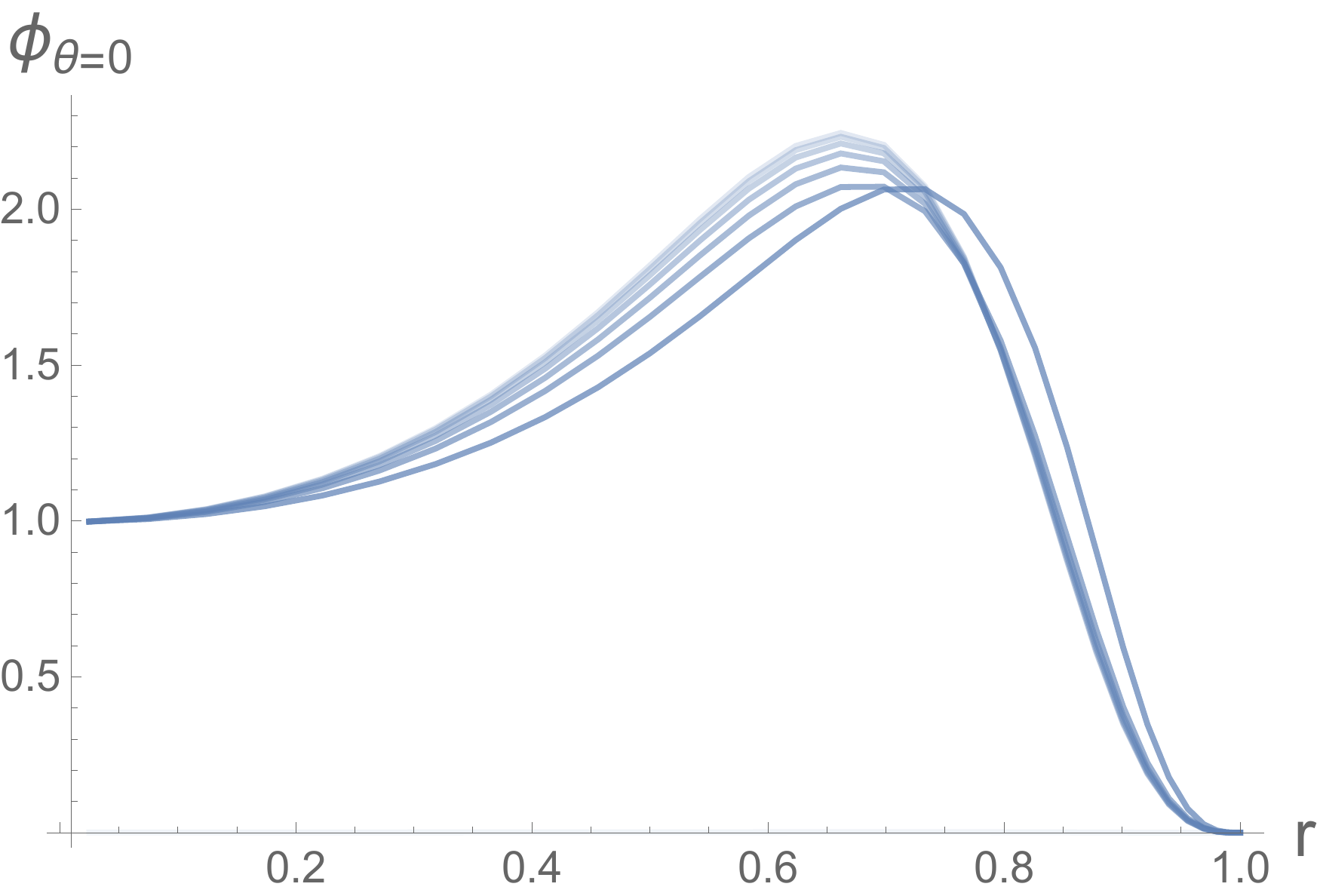}
}
\caption{Instability onset for a charged scalar in a polarized soliton background. The plot on the left shows the critical value of $q$ at which the instability occurs as a function of the electric field.  The plot on the right shows the scalar field profile at the pole as a function of the coordinate $r$ for a range of electric field values.
\label{supersol}}
\end{figure}

\begin{figure}[h!]
\centering
\subfloat[]{
\includegraphics[width=65mm]{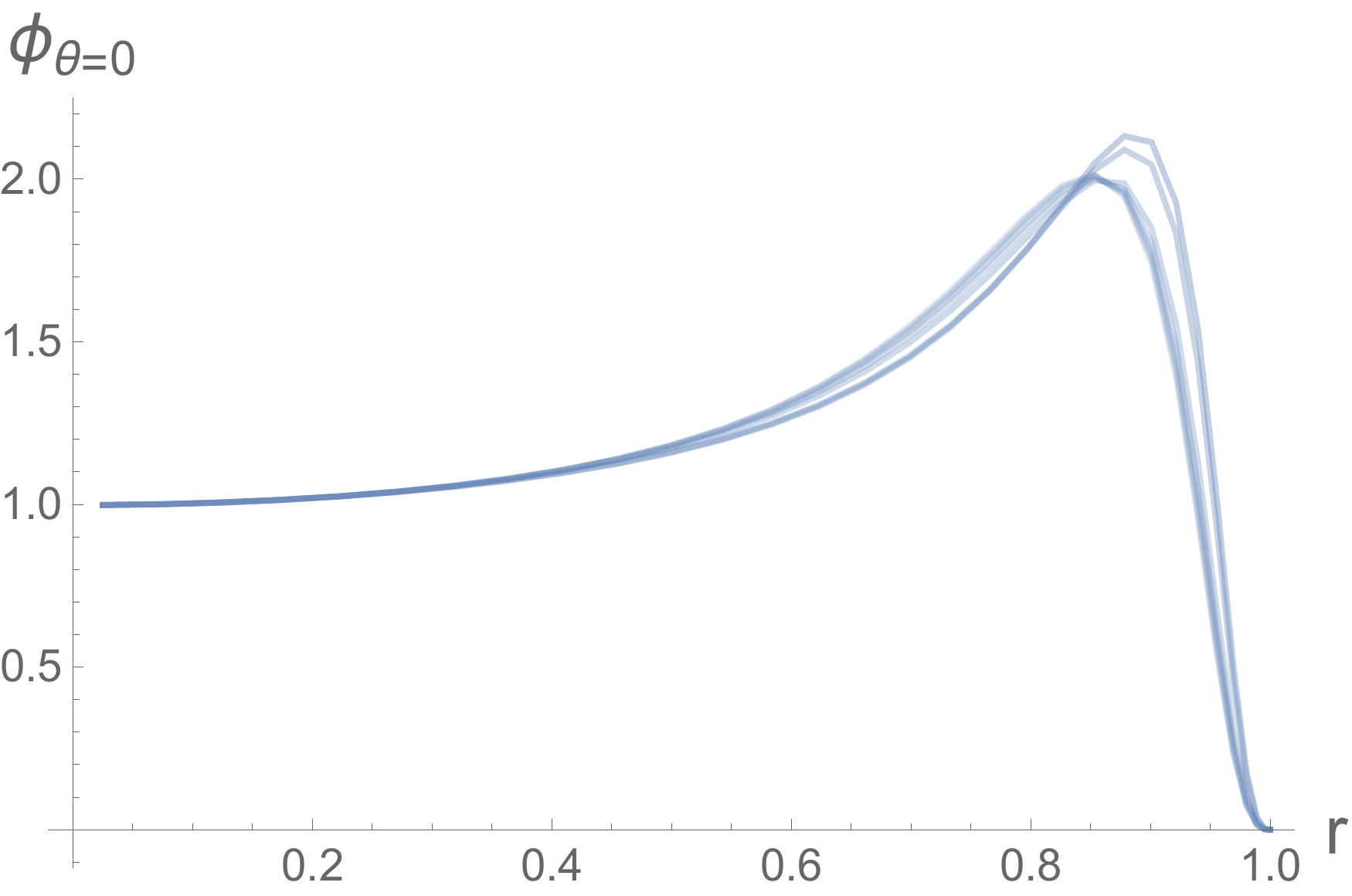}
}
\subfloat[]{
\includegraphics[width=65mm]{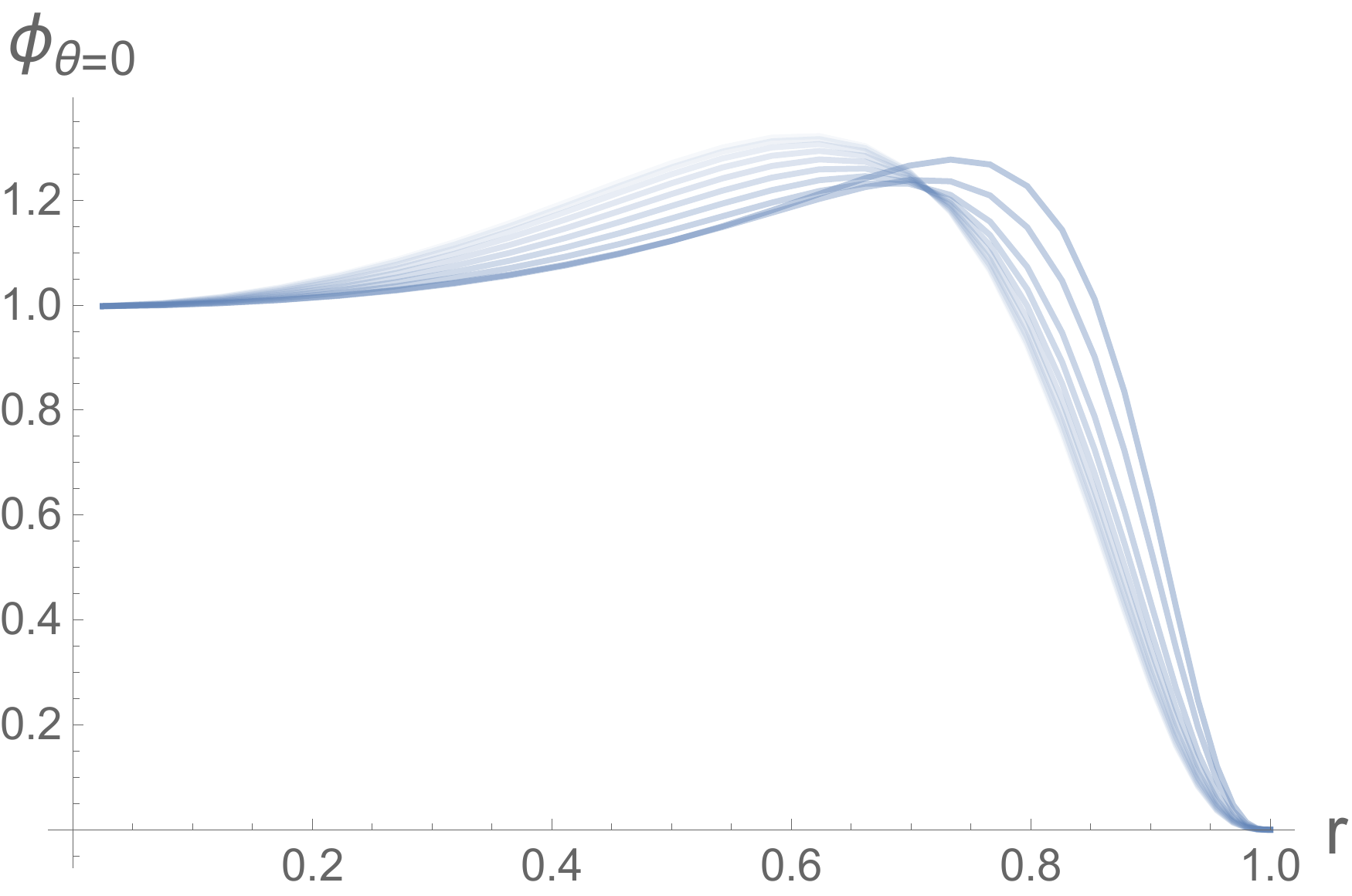}
}
\caption{Instability onset for a charged scalar in a polarized black hole background with a temperature $T=1/\pi$ for a range of electric field values. The curves get taller and fainter as the electric field decreases. The left and right plots show the scalar field profiles at the pole as a function of $r$ for the small and large black holes, respectively.
\label{superBH}}
\end{figure}

In figure \ref{supersol}, we plot the smallest eigenvalue $q$ corresponding to the minimum value at which an instability occurs in the soliton background as a function of the electric field. We also plot the associated scalar field profiles at the pole as a function of the radial coordinate for several values of the electric field. Similar scalar field profiles are plotted for the small and large black hole backgrounds in figure \ref{superBH} for a fixed temperature of $1/\pi$. In all backgrounds, the scalar field curves get taller and fainter as the electric field is decreased. We can see from these plots that there exist unstable modes at any value of the electric field. We also find such modes in the black hole background, for all values of temperature and electric field accessible to our solutions. To understand the relevance of these results for ABJM, one would have to determine the mass and charge of the scalar fields in ABJM, and see if they appear as minimally coupled fields from a 4D perspective. If a massless scalar field is found with sufficiently large $q>q_{\min}$, our results suggest that such a configuration will be unstable beyond a certain value of $\mathcal{E}(q_{\min})$.

The zero coupling phase diagram is very similar to the case of SYM with R-charge chemical potentials \cite{Yamada:2006rx}. This analogy suggests that at   weak coupling and high temperature a region of metastability also appears in our system for ${\cal E}>{\cal E}^B_c$. This would be qualitatively similar to the strong coupling phase diagram shown in figure \ref{phasediagramABJM}. However, since the spectrum depends on the coupling, ${\cal E}_c^B$ does not equal ${\cal E}_{c}^{Sol}$. It is possible that deformed ABJM theory can be compactified on the $S^2$ and simulated using Monte Carlo techniques. Such an undertaking could not only confirm our results at strong coupling, but also provide another test of the gauge/gravity duality. 
We leave these ideas for the future.

\section*{Acknowledgements}
The research leading to these results has received funding from the [European Union] Seventh Framework Programme [FP7-People-2010-IRSES] and [FP7/2007-2013] under grant agreements No 269217, 317089 and No 247252, and from the grant CERN/FIS-NUC/0045/2015. \emph{Centro de F\'isica do Porto} is partially funded by the Foundation for  Science and Technology of Portugal (FCT). J.P. is partially funded by SwissMAP. L.G. is funded by the FCT/IDPASC fellowship SFRH/BD/51983/2012. This research was supported in part by Perimeter Institute for Theoretical Physics.  Research at Perimeter Institute is supported by the Government of Canada through Industry Canada and by the Province of Ontario through the Ministry of Economic Development and Innovation. J. E. Santos is grateful to Siavash Golkar for countless discussions about ABJM.

\appendix

\section{Perturbative Analysis \label{App:pert}}

In powers of the electric field parameter $\mathcal{E}$, the expansion of the metric, gauge field, and scalar take the form 
\begin{equation}
g_{\mu\nu} = \bar{g}_{\mu\nu}+\sum_{j=1}^{+\infty} g^{(2j)}_{\mu\nu}{\cal E}^{2j}\, ,\quad
{\cal A}_{t}=\sum_{j=0}^{+\infty} a^{(2j+1)}_{t}{\cal E}^{2j+1}\, , \quad {\phi}=\sum_{j=0}^{+\infty}f^{(2j)}{\cal E}^{2j}.
\end{equation}
where $\bar{g}$ is $AdS_4$ in global coordinates and the back-reacted metric $g$ can be written in the quasi-spherical gauge
\begin{equation}
 ds^2 = -\left(1+\frac{y^2}{l^2}\right) Q_1(y,\theta) dt^2+Q_2(y,\theta)\,\frac{ dy^2}{1+\frac{y^2}{l^2}}+Q_3(y,\theta)\,y^2\left(d\theta^2+\sin^2\theta d\phi^2\right) .
\end{equation}

At linear order in $\cal{E}$ the equations of motion reduce to a second-order equations in $a_t^{(1)}$. This can be solved using separation of variables to get 
\ba
a^{(1)}_t(y,\theta) &=& \sum_{\ell=0}^{+\infty}a_\ell\,\frac{\Gamma\!\left(\frac{\ell+1}{2}\right)\Gamma\!\left(\frac{\ell+3}{2}\right)}{\sqrt{\pi}\Gamma\!\left(\ell+\frac{3}{2}\right)}\left(\frac{y}{l}\right)^\ell{}_2F_1\left(\frac{\ell}{2},\frac{\ell+1}{2},\ell+\frac{3}{2},-\frac{y^2}{l^2}\right) L_\ell\big(\!\cos(\theta)\big)\,,
\ea
where $L_\ell$ is the Legendre polynomial of degree $l$ and 
we imposed regularity at the center of $AdS$ and defined the constants $a_\ell$ to fix the boundary chemical potential. The factors of gamma functions were pulled out to impose that
\begin{equation}
\lim_{y\to+\infty}a^{(1)}_t(y,\theta) = \sum_{\ell=0}^{+\infty}a_\ell L_\ell\big(\!\cos(\theta)\big)\,.
\end{equation}
For the dipolar potential we are considering, we can therefore set $a_1=1$ and $a_\ell=0$ for all other $\ell \ne 1$. The solution for the gauge field to first order then takes the simple form
\begin{equation}
a^{(1)}_t(y,\theta)=\frac{2}{\pi  y^2}\left[\left(l^2+y^2\right) \arctan\left(\frac{y}{l}\right)-l y\right]\cos\theta\,.
\end{equation}

The backreaction of the metric subject to the linear source can be found at second order. In the quasi-spherical gauge, metric perturbations are gravitational modes of scalar-type labeled by spherical harmonics of degree $\ell$. The the metric components take the form
\begin{equation}
Q_i(r,\theta) = 1+\sum_{j=1}^{+\infty} q_i^{(2j)}(r,\theta)\,{\cal E}^{2j}\,.
\end{equation}
The only nontrivial contributions to the metric perturbations come from $\ell=0$ and $\ell=2$, admitting a composition of the form 
\begin{align}
q_i^{(2)} = \alpha_i(r) L_0(\theta)+\beta_i(r) L_2(\theta)\,, \qquad f^{(2)}=\alpha_5(r) L_0(\theta)+\beta_5(r) L_2(\theta)\,.
\end{align}
The residual gauge freedom can be fixed by setting $\alpha_3(r)=0$. Solving the equations of motion at this order, subject to normalizability at the conformal boundary and regularity at the $AdS$ center, we get the following analytical result
\begin{subequations}
\begin{multline}
\alpha_1(r) = \frac{2 \left(l^6-3 l^2 y^4\right)\left( \tan ^{-1}\left(\frac{y}{l}\right)\right)^2}{3 \pi ^2 y^4}+\frac{1}{3} \left(\frac{3 l^2}{2}+\frac{2 l^4 \left(\frac{1}{y^2}-\frac{4}{l^2+y^2}\right)}{\pi ^2}\right)
\\
-\frac{4 \left(l^7+3 l^5 y^2+3 l^3 y^4\right) \tan ^{-1}\left(\frac{y}{l}\right)}{3 \pi ^2 y^3 \left(l^2+y^2\right)}\,,
\qquad\qquad\qquad\qquad\qquad\qquad\qquad\qquad\quad
\end{multline}
\begin{multline}
\alpha_2(r) =-\frac{4 l^5 \left(l \tan ^{-1}\left(\frac{y}{l}\right)-y\right) \left(l^2 \tan ^{-1}\left(\frac{y}{l}\right)+y^2 \tan ^{-1}\left(\frac{y}{l}\right)-l y\right)}{3 \pi ^2 y^4 \left(l^2+y^2\right)}\,,
\\
\end{multline}
\begin{multline}
\beta_1(r) =-\frac{l^4 \left(\left(9 \pi ^2-8\right) l^2+5 \left(8+3 \pi ^2\right) y^2\right)}{24 \pi ^2 y^2 \left(l^2+y^2\right)}+\frac{4 l^4 \left(l^2+y^2\right) \tan ^{-1}\left(\frac{y}{l}\right)^2}{3 \pi ^2 y^4}\\
+\frac{l^3 \left(\left(9 \pi ^2-40\right) l^4+2 \left(9 \pi ^2-8\right) l^2 y^2+\left(9 \pi ^2-40\right) y^4\right) \tan ^{-1}\left(\frac{y}{l}\right)}{24 \pi ^2 y^3 \left(l^2+y^2\right)}\,,
\qquad\qquad\quad
\end{multline}
\begin{multline}
\beta_2(r) = \frac{l^4 \left(\left(56+9 \pi ^2\right) l^2+5 \left(8+3 \pi ^2\right) y^2\right)}{24 \pi ^2 y^2 \left(l^2+y^2\right)}+\frac{4 l^4 \left(l^2+y^2\right) \tan ^{-1}\left(\frac{y}{l}\right)^2}{3 \pi ^2 y^4}\\
-\frac{l^3 \left(\left(88+9 \pi ^2\right) l^4+2 \left(56+9 \pi ^2\right) l^2 y^2+\left(9 \pi ^2-40\right) y^4\right) \tan ^{-1}\left(\frac{y}{l}\right)}{24 \pi ^2 y^3 \left(l^2+y^2\right)}\,,
\qquad\qquad
\end{multline}
\begin{multline}
\beta_3(r) =\frac{1}{24} l^2 \left(\frac{\left(8+9 \pi ^2\right) l^2}{\pi ^2 y^2}-12\right)+\frac{l^3 \left(\left(8-9 \pi ^2\right) l^2+\left(56+9 \pi ^2\right) y^2\right) \tan ^{-1}\left(\frac{y}{l}\right)}{24 \pi ^2 y^3}\\
-\frac{2 \left(l^6+4 l^4 y^2-3 l^2 y^4\right) \tan ^{-1}\left(\frac{y}{l}\right)^2}{3 \pi ^2 y^4}\,,
\qquad\qquad\qquad\qquad\qquad\qquad\qquad\qquad\quad
\end{multline}
\begin{multline}
\alpha_5(r) = -\frac{l^3 \left(4 \left(l^3-l y^2\right) \tan ^{-1}\left(\frac{y}{l}\right)^2-\big(8 l^2 y+\left(\pi ^2-4\right) y^3\big) \tan ^{-1}\left(\frac{y}{l}\right)+4 l y^2\right)}{6 \pi ^2 y^4}\,,
\\
\end{multline}
\begin{multline}
\beta_5(r) =-\frac{l^3 }{6 \pi ^2 y^4}\left(8 \left(l^3+2 l y^2\right) \tan ^{-1}\left(\frac{y}{l}\right)^2\right.
\\
+y \left(\left(2+3 \pi ^2\right) l^2+\left(\pi ^2-10\right) y^2\right) \tan ^{-1}\left(\frac{y}{l}\right)+\left(10+3 \pi ^2\right) (-l) y^2\bigg).
\qquad\qquad
\end{multline}
\end{subequations}

At third order, the gauge field can be written as
\begin{equation}
a^{(3)}_t = f_1(r) L_1(\chi)+f_3(r) L_3(\chi)\,,
\end{equation}
Regularity at the origin and normalisability impose that these are real functions given by
\begin{subequations}
\begin{multline}
f_1(y)=\frac{8 l^2 \tan ^{-1}\left(\frac{y}{l}\right)^3}{525 \pi ^3 y^6}  \Big(65 l^6 + 176 l^4 y^2 + 144 I l y^5 + 2 l^2 y^4 \big(-115 + 72 \log(2)\big) \\ 
+ y^6 \big(-197 + 144 \log(2)\big) + 144 y^4 \big(l^2 + y^2\big) \big(\log(y) - \log(-I l + y)\big)\Big)\\
+\frac{l^2 \tan ^{-1}\left(\frac{y}{l}\right)^2}{700 \pi ^3 y^5}\bigg(4608 i y^3 \left(l^2+y^2\right) \text{Li}_2\left(\frac{l-i y}{l+i y}\right)+l \Big(5 \left(49 \pi ^2-192\right) l^4 \\
 -8 \left(594+35 \pi ^2\right) l^2 y^2+4608 y^4 \big(\log (y)-\log (y-i l)\big)+y^4 \left(-525 \pi ^2-6032+4608 \log (2)\right)\Big)\bigg)
\nonumber
\end{multline}
\begin{multline}
+\frac{l^2}{10500 \pi ^3 y^3}\left[3675 \pi ^2 l^3+6400 l^3-69120 i y \left(l^2+y^2\right) \text{Li}_4\left(\frac{l-i y}{l+i y}\right)+768 i \pi ^4 l^2 y+2620 \pi ^2 l y^2\right.\\
\left.
+34560 l y^2 \text{Li}_3\left(\frac{l-i y}{l+i y}\right)
-77760 l y^2 \zeta (3)+143520 l y^2+5760 \pi ^2 l y^2 \log (2)+768 i \pi ^4 y^3\right]\\
-\frac{l^2 \tan ^{-1}\left(\frac{y}{l}\right)}{1050 \pi ^3 y^4}\left(735 \pi ^2 l^4+240 l^4-10368 y^2 \left(l^2+y^2\right) \text{Li}_3\left(\frac{l-i y}{l+i y}\right)-7776 l^2 y^2 \zeta (3)\right.\\
\left. -158 \pi ^2 l^2 y^2+10040 l^2 y^2+576 \pi ^2 l^2 y^2 \log (2)-6912 i l y^3 \text{Li}_2\left(\frac{l-i y}{l+i y}\right)-7776 y^4 \zeta (3)\right.\\
\left.-788 \pi ^2 y^4+576 \pi ^2 y^4 \log (2)\right),
\qquad\qquad\qquad\qquad\qquad\qquad\qquad\qquad\qquad\qquad
\end{multline}
\begin{multline}
f_3(r) = \frac{4 l^2 \tan ^{-1}\left(\frac{y}{l}\right)^3}{525 \pi ^3 y^6}
\Big(245 l^6+558 l^4 y^2-240 i l^3 y^3+405 l^2 y^4\\
+48 y^2 \left(l^2+y^2\right) \left(5 l^2+y^2\right) \big(-\log (2 y)+\log (y-i l)\big)-208 i l y^5+124 y^6\Big)\\
l^2 +\frac{1}{1400 \pi ^3 y^5} \tan ^{-1}\left(\frac{y}{l}\right)^2\bigg[l \Big(35 \left(21 \pi ^2-128\right) l^4+2 \left(2672+1155 \pi ^2\right) l^2 y^2 \\
\left. +512 y^2 \left(15 l^2+13 y^2\right) \big(-\log (2 y)+\log (y-i l)\big)+\left(3104+1575 \pi ^2\right) y^4\Big)\right. \\
-1536 i y \left(l^2+y^2\right) \left(5 l^2+y^2\right) \text{Li}_2\left(\frac{l-i y}{l+i y}\right)\bigg]\\
+\frac{l^2 }{63000 \pi ^3 y^4}\bigg(-3840 i \pi ^4 l^4-11520 \left(15 l^3 y+13 l y^3\right) \text{Li}_3\left(\frac{l-i y}{l+i y}\right)+388800 l^3 y \zeta (3)\\
\left. -191775 \pi ^2 l^3 y-520800 l^3 y-28800 \pi ^2 l^3 y \log (2)-4608 i \pi ^4 l^2 y^2\right.\\
\left.+69120 i \left(5 l^4+6 l^2 y^2+y^4\right) \text{Li}_4\left(\frac{l-i y}{l+i y}\right)+336960 l y^3 \zeta (3)-135020 \pi ^2 l y^3\right.\\
-270720 l y^3-24960 \pi ^2 l y^3 \log (2)-768 i \pi ^4 y^4\bigg)
\\
 +\frac{l^2}{1050 \pi ^3 y^4} \tan ^{-1}\left(\frac{y}{l}\right)\bigg[-6480 l^4 \zeta (3)+2645 \pi ^2 l^4+10080 l^4+480 \pi ^2 l^4 \log (2)+576 \pi ^2 l^2 y^2 \log (2)\\
 \left.-384 i \left(15 l^3 y+13 l y^3\right) \text{Li}_2\left(\frac{l-i y}{l+i y}\right)-7776 l^2 y^2 \zeta (3)+1767 \pi ^2 l^2 y^2+1680 l^2 y^2\right.\\
 -1728 \left(5 l^4+6 l^2 y^2+y^4\right) \text{Li}_3\left(\frac{l-i y}{l+i y}\right)-1296 y^4 \zeta (3)-248 \pi ^2 y^4+96 \pi ^2 y^4 \log (2)\bigg]
\,,
\end{multline}
\end{subequations}
where $\text{Li}_k(x)$ is a polylogarithm function of order $k$, and $\zeta(x)$ is the Riemann zeta function. 

We can use the third order expansion to compute quantities presented in the text. For example, the location of point-like particles with charge to mass ratio $\lambda=|q/m|$
in terms of the proper distance along the $\theta=0$ axis is 
\be
\mathcal{P}_{\theta=0}^*=\frac{4 \lambda}{3 \pi }\,\mathcal{E}
+\lambda \frac{  20 \big(224 \lambda ^2-972 \zeta (3)-435\big)+9 \pi ^2 \big(271+288 \log (2)\big)}{14175 \pi ^3}\,\mathcal{E}^3\,.
\ee

\bibliographystyle{utphys}
\bibliography{ABJM}

\end{document}